\documentclass[%
twocolumn, onecolumn,
 aip,
 amsmath,amssymb,
 reprint,%
]{revtex4-1}

\usepackage{graphicx}
\usepackage{dcolumn}
\usepackage{bm}
\usepackage{float}
\usepackage{amsmath}
\usepackage{algorithm}
\usepackage{algorithmic}
\usepackage{xcolor}
\usepackage[utf8]{inputenc}
\usepackage[T1]{fontenc}
\usepackage{mathptmx}
\usepackage{multirow}
\usepackage{gensymb}

\usepackage{natbib}
\usepackage{lipsum}

\begin{document}
\preprint{AIP/123-QED}
\title{Structure Prediction of Epitaxial Organic Interfaces with Ogre, Demonstrated for TCNQ on TTF}

\author{Saeed Moayedpour}
\affiliation{Department of Chemistry, Carnegie Mellon University, Pittsburgh, PA 15213, USA}
\author{Imaneul Bier}
\affiliation{Department of Materials Science and Engineering, Carnegie Mellon University, Pittsburgh, PA 15213, USA}
\author{Wen Wen}
\affiliation{Department of Chemistry, Carnegie Mellon University, Pittsburgh, PA 15213, USA}
\author{Derek Dardzinski}
\affiliation{Department of Materials Science and Engineering, Carnegie Mellon University, Pittsburgh, PA 15213, USA}
\author{Olexandr Isayev}%
\affiliation{Department of Chemistry, Carnegie Mellon University, Pittsburgh, PA 15213, USA}
\author{Noa Marom}%
\email{nmarom@andrew.cmu.edu}
\affiliation{Department of Materials Science and Engineering, Carnegie Mellon University, Pittsburgh, PA 15213, USA}
\affiliation{Department of Chemistry, Carnegie Mellon University, Pittsburgh, PA 15213, USA}
\affiliation{Department of Physics, Carnegie Mellon University, Pittsburgh, PA 15213, USA}

\date{\today}

\begin{abstract}

Highly ordered epitaxial interfaces between organic semiconductors are considered as a promising avenue for enhancing the performance of organic electronic devices including solar cells, light emitting diodes, and transistors, thanks to their well-controlled, uniform electronic properties and high carrier mobilities. Although the phenomenon of organic epitaxy has been known for decades, computational methods for structure prediction of epitaxial organic interfaces have lagged far behind the existing methods for their inorganic counterparts. We present a method for structure prediction of epitaxial organic interfaces based on lattice matching followed by surface matching, implemented in the open-source Python package, Ogre. The lattice matching step produces domain-matched interfaces, where commensurability is achieved with different integer multiples of the substrate and film unit cells. In the surface matching step, Bayesian optimization (BO) is used to find the interfacial distance and registry between the substrate and film. The BO objective function is based on dispersion corrected deep neural network interatomic potentials, shown to be in excellent agreement with density functional theory (DFT). The application of Ogre is demonstrated for an epitaxial interface of 7,7,8,8-tetracyanoquinodimethane (TCNQ) on tetrathiafulvalene (TTF), whose electronic structure has been probed by ultraviolet photoemission spectroscopy (UPS), but whose structure had been hitherto unknown [Organic Electronics 48, 371 (2017)]. We find that TCNQ(001) on top of TTF(100) is the most stable interface configuration, closely followed by TCNQ(010) on top of TTF(100). The density of states, calculated using DFT, is in excellent agreement with UPS, including the presence of an interface charge transfer state.

\end{abstract}

\maketitle

\section{\label{sec:level1}Introduction}

Organic-organic heterojunctions, i.e., interfaces between different organic materials, are at the heart of organic electronic devices, including organic photovoltaics (OPV),\cite{Mazzio2014,2009Kippelen} organic light emitting diodes (OLEDs),\cite{Rieneke2013, liu2018NatRevMater, Brutting2001, Armstrong2009} and organic field effect transistors (OFETs).\cite{Mas-Torrent2011, Chen2020ChemRev,Sirringhaus2014, Braga2009, wang2022organic} Essential device functionalities take place at interfaces. In organic solar cells, charge separation of bound excitons into free charge carriers occurs at donor-acceptor interfaces, where the holes go into the highest unoccupied molecular orbital (HOMO) of the donor and the electrons go into the lowest unoccupied molecular orbital (LUMO) of the acceptor. Conversely, in OLEDs, electrons and holes recombine at a donor-acceptor interface, leading to photon emission.\cite{Armstrong2009} In more sophisticated multilayer OLED designs, additional functions may be separated into different materials. A multilayer OLED may include a hole-injection layer, a hole-transport layer, an electron-blocking layer, an emission layer, a hole-blocking layer, an electron-transport layer, and an electron-injection layer, forming multiple active interfaces.\cite{Rieneke2013} In OFETs, the gate stack contains, at the minimum, an organic semiconductor interfaced with the gate dielectric and the source and drain electrodes. In some cases, more than one organic semiconductor is included in the gate stack to create an $n$-channel and a $p$-channel. Additional organic layers may be included, e.g., to control the morphology of the organic semiconductor(s) and the contact resistance at the electrode interfaces.\cite{Chen2020ChemRev} Hence, the structure and electronic properties of organic interfaces are critical to the performance of organic electronic devices.        

Often, the films used in organic devices are amorphous and the interfaces between them are disordered. Disorder may lead to non-uniform electronic and optical properties that are averaged over multiple configurations.\cite{Khan2021, Khan2019} In addition, various defects, as well as grain boundaries in polycrystalline films,\cite{Choi2020} may form traps and provide scattering sites for charge carriers. This is detrimental to charge transport and lowers the carrier mobility, which is a key performance parameter for electronic devices. Molecular order produces well-controlled properties and higher mobilities, which may lead to improved device performance.\cite{Gershenson2006, Coropceanu2007, Nakayama2020,  Podzorov2013,Li2010, Reese2007, Fusella2018, Raimondo2011, wang2022organic} In analogy with inorganic materials, high-quality crystalline organic heterojunctions may be grown by molecular beam epitaxy. \cite{Schreiber2004,Forrest1994,Moret2011,Sassella2013,Koma1995,Koma1999,Fenter1997,Forrest1994a,Hooks2001, Yang2009ChemSocRev, Yang2015} In contrast to inorganic epitaxy, molecular materials are bound by weak dispersion interactions and organic substrates do not have dangling bonds on the surface. Moreover, there is competition between adsorbate-substrate and adsorbate-adsorbate intermolecular interactions. This leads to less strict lattice-matching requirements between the substrate and film for organic epitaxial growth.\cite{Koma1995, Forrest1997, Yang2009ChemSocRev, Yang2015}  Epitaxial interfaces between various organic semiconductors have been grown, including $\alpha$-quaterthiophene on rubrene\cite{Campione2009}, pentacene on $C_{60}$\cite{Al-Mahboob2009}, rubrene on tetracene\cite{Moret2011}, $C_{60}$ on pentacene\cite{Nakayama2016, Nakayama2018, Conrad2009, Huttner2019, Iwasawa2020} or rubrene\cite{Fusella2018, Mitsuta2017}, perfluoropentacene on pentacene\cite{Nakayama2019} or diindenoperylene (DIP) \cite{Hinderhofer2011}, DIP on copper-hexadecafluorophthalocyanine (F$_{16}$CuPc) \cite{Barrena2006}, bis(trifluoromethyl)dimethylrubrene on rubrene \cite{Takahashi2021}, tetraazanaphthacene on pentacene,\cite{Gunjo2021} and  7,7,8,8-tetracyanoquinodimethane (TCNQ) on tetrathiafulvalene (TTF) \cite{Kattel2017}. Band transport has been observed in organic epitaxial interfaces.\cite{Nakayama2019, Fusella2018} Moreover, epitaxial interfaces with a well-defined structure and uniform properties are conducive to spectroscopic characterization of the interface band alignment, interface charge transfer (CT) states, and exciton dynamics.\cite{Iwasawa2020,Takahashi2021, Kattel2017, Conrad2009, Gunjo2021, Doring2019} 

Computer simulations can help explore the vast configuration space of possible substrate and film combinations to direct fabrication efforts to systems likely to result in robust epitaxial growth of high-quality films. Furthermore, simulations can help the characterization of epitaxial organic interfaces by assigning spectroscopic signatures to putative interface structures. The structure prediction capabilities for organic interfaces lag far behind the existing methods for inorganic interfaces. Several computational tools and codes exist for constructing models of inorganic epitaxial interfaces.\cite{Dardzinski2022, Moayedpour2021, Mathew2016, Ding2016, Raclariu2015,Sun2013, Gao2019} Because these tools are designed to work with spherical atoms, not molecules, which are significantly more complex, they do not work out of the box for organic interfaces. Recently, progress has been achieved in the development of algorithms for structure prediction of organic monolayers on inorganic substrates.\cite{Mannsfeld2011, Packwood2017, Packwood2017a, Obersteiner2017, Scherbela2018}. Organic-organic interfaces have been investigated mainly by molecular dynamics simulations, based on classical force fields, which enable studying the effect of disorder.\cite{Han2015, Fu2014, Fu2013, Yoo2019} However, to our knowledge, to date, no  computational tools have been developed for structure prediction of epitaxial organic-organic interfaces.

Here, we introduce a new version of the open source Python package, Ogre, with new functionality of predicting the structure of epitaxial organic interfaces by lattice and surface matching. We note that Ogre generates ideal interfaces with no disorder and does not take into account growth conditions and kinetics.  The first version of Ogre was designed for modelling molecular crystal surfaces, calculating surface energies, and predicting Wulff shapes.\cite{yang2020ogre} In the  second version, the capability of performing structure prediction for epitaxial inorganic interfaces was added.\cite{Moayedpour2021} Similar to the previous version of Ogre, the workflow for organic interfaces begins by using Zur and McGill's lattice matching algorithm\cite{Zur1998} and proceeds to perform surface matching. The lattice matching step identifies potential domain-matched epitaxial interfaces, where commensurability is achieved with different integer numbers of the substrate and film unit cells. The lattice misfit tolerances for domain matching have been updated for the case of weak van der Waals epitaxy. In addition, a streamlined  calculation of the surface energies of different substrate facets may be performed to prioritize the most stable orientations for an interface Miller index search. In the surface matching step, Bayesian optimization (BO) is utilized to find the most stable interface configurations by exploring the three-dimensional space of interfacial distance and registry. 

 For surface matching, Ogre constructs commensurate surface unit cells of the substrate and film, ensuring that no molecules are broken when the surface is cleaved.\cite{yang2020ogre} The film is then moved with respect to the substrate in order to determine the optimal configuration. Owing to the large domain size and the number of atoms per molecule, even the smallest models of organic epitaxial interfaces, with only one layer of the substrate and the film, may contain hundreds to thousands of atoms. Therefore, it is imperative to use a computationally efficient method for evaluating the relative stability of interface configurations. The geometric score function we developed for inorganic interfaces\cite{Moayedpour2021} is not appropriate for organic interfaces because it is designed to work with spherical atoms, whose radii are determined by Hirshfeld partitioning\cite{Spackman2009} of the charge density of the bulk materials calculated with density functional theory (DFT). Instead, we use the Accurate Neural Network Engine for Molecular Energies (ANI) general-purpose machine learned interatomic potentials.\cite{Smith2017}  The ANI deep neural network potentials have been trained on a large and diverse database of first principles calculations for small molecule conformations\cite{Smith2020} using active learning\cite{Smith2018}. They have been shown to be transferable across chemical space, achieving a DFT level of accuracy on a large set of organic molecules while being six orders of magnitude faster. Because the ANI potentials were trained on isolated molecules they do not inherently contain a description of intermolecular dispersion (van der Waals) interactions. Therefore,  the Grimme D3 dispersion correction is added to the ANI energies.\cite{Smith2016, Grimme2011, Ehrlich2011} We show that ANI+D3 is in close agreement with DFT for the interfacial distance and registry. Preliminary ranking of the interface energies of the surface-matched configurations of all domain-matched interfaces is performed with ANI+D3. The ANI+D3 ranking is in reasonably good agreement with the DFT ranking and the most stable structures are  identified. In the final stage, DFT is used to calculate the electronic properties of a small subset of the most promising candidate structures. 

The application of Ogre is demonstrated for the epitaxial interface of TCNQ grown on top of TTF.\cite{Kattel2017} TTF and TCNQ are a quintessential donor/acceptor system. Both TTF and TCNQ are wide band gap semiconductors. On their own, they are insulators but when they are paired, electrons can transfer from the HOMO of TTF to the LUMO of TCNQ.\cite{Kattel2017, Murdey2005, BRAUN2010, Beltran2013, Atalla2016} Hence, TTF-TCNQ co-crystals and interfaces may exhibit metallic behavior. The  conducting layer, which forms at the interface, has been reported to exhibit a high carrier density, exceeding $10^{14}$ $cm^{-2}$.\cite{Kattel2017, Ando2008, Cohen1974, Ferraris2002}  The interface formed by TTF and TCNQ single crystals has been found to exhibit 2D metallic conductivity, which may be useful for organic transistors and sensing devices.\cite{Alves2013, Krupskaya2016, Krupskaya2016a, Takahashi2012, Mathis2012, Lezama2012, Calhoun2007, Alves2008} Kattel \textit{et al.} have grown an epitaxial interface of TCNQ on top of TTF.\cite{Kattel2017} The electronic structure of the interface was probed by ultraviolet photoemission spectroscopy (UPS) as a function of the thickness of the TCNQ film. A charge transfer (CT) state was observed at the interface, whose signature decayed when the TCNQ film thickness exceeded 2 nm. However, the interface was not structurally characterized, leaving the orientation of the TTF and TCNQ and their relative positions at the interface hitherto unknown. Using Ogre, we find that TCNQ(001) on top of TTF(100) is the most stable interface configuration, closely followed by TCNQ(010) on top of TTF(100). The resulting density of states (DOS) of the TTF substrate, a TCNQ film, and an interface with one layer of each material are in excellent agreement with the UPS experiment of Kattel \textit{et al.}. A CT state is found at the interface, consistent with their observations. 

\section{\label{sec:level1}Computational Details}
\subsection{\label{sec:level2}DFT}

The all-electron electronic structure code FHI-aims\cite{Blum2009} was used to perform DFT calculations. Computationally efficient light numerical settings and tier 1 basis sets were used throughout, owing to the large system sizes with several hundred atoms. Geometry optimizations, surface energy evaluations, reference DFT calculations for surface matching, and interface energy evaluations were conducted using the generalized gradient approximation of Perdew, Burke, and Ernzerhof (PBE)\cite{PhysRevLett.77.3865,PhysRevLett.78.1396} with the Tkatchenko-Scheffler (TS) pairwise dispersion method. \cite{tkatchenko2009accurate} Structural relaxations were performed for the bulk crystal structures of TTF and TCNQ until the residual force per atom was less than $10^{-2}$ ev/\AA. Relaxation was performed with fixed unit cell angles of $\beta$ = $101.43^\circ$ for TTF\cite{Batsanov2006} and $\beta$ = $98.351^\circ$ for TCNQ.\cite{Krupskaya2015} The number of k-points in each direction was determined by calculating the closest integer greater than 24 divided by the norm of the respective lattice parameter. The relaxed lattice parameters of TTF were $a$ = 7.44 \AA, $b$ = 4.28 \AA, and $c$ = 13.68 \AA, compared with the experimental values of $a$ = 7.35 \AA, $b$ = 4.01 \AA, and $c$= 13.901 \AA.\cite{Batsanov2006} The relaxed lattice parameters of TCNQ were $a$ = 8.98 \AA, $b$ = 7.31 \AA, and $c$ = 16.40 \AA, compared with the experimental values of $a$ = 8.8834 \AA, $b$ = 6.9539 \AA, and $c$ = 16.3948 \AA.\cite{Krupskaya2015} The interface models were constructed based on the relaxed bulk geometry with one layer of the TTF substrate, one layer of the TCNQ film, and a vacuum region of 40 \AA.  The electronic structure calculations for the the final lowest energy interfaces were performed using the range-separated hybrid functional of Heyd–Scuseria–Ernzerhof (HSE).\cite{heyd2003hybrid,ge2006erratum} We note that the size of the systems calculated here is very large. For example, the TCNQ(001)/TTF(100) interface model with a single layer of each material contained 1298 atoms.

\subsection{\label{sec:level2}ANI neural network potentials}

The ANAKIN-ME (ANI)\cite{Smith2017} transferable neural network molecular potentials are used for calculations of organic interfaces. The representation used by the ANI-2x model\cite{Devereux2020} is the Smith symmetry functions (SSFs), $\overrightarrow{G_i}^Z$, a modified version of the Behler and Parrinello symmetry functions. A separate neural network model is used for each element. The SSFs encode N inter-atomic distances between the $i^{th}$ atom and its neighbors within a cutoff radius $R_c$ into an invariant fixed-length atomic environment vector (AEV), $\overrightarrow{G_i}^Z = \{ G_1, G_2, G_3, ..., G_M \}$. The elements $G_M$ encodes specific regions of an individual atom's radial and angular chemical environment. Each $\overrightarrow{G_i}^Z$ for the $i^{th}$ atom with atomic number $Z$ is then used as input into a single neural network (NNP). With an invariant AEV, $\overrightarrow{G_i}^Z$, the total energy of a molecule $m$ is expressed as:
\begin{equation}
E_{total}(m) = \sum_i^{all\: atoms} NNP_{Z_i} \:(\overrightarrow{G_i}^Z)
\label{eq:ANI}
\end{equation}
The ANI-2x potentials have been trained on a large and diverse dataset of first-principles calculations for small organic molecules selected by an active learning procedure.\cite{Smith2018} Test cases have demonstrated the ANI potentials to be chemically accurate compared to reference DFT calculations. The Grimme D3 dispersion correction\cite{Grimme2010} is added \emph{post hoc} to ANI-2x calculations to account for intermolecular van der Waals interactions. All functionality is implemented in the open source TorchANI\cite{Gao2020} package (https://github.com/aiqm/torchani), which is written in python using the PyTorch library. TorchANI is designed to take advantage of the modularity and simplicity of PyTorch to provide a framework for fast iterative improvements in architecture and training schedules for ML models, while also allowing for experimentation and easy integration into existing software, such as Ogre. 

\section{\label{sec:level1}Results and Discussion}

\subsection{\label{workflow} Workflow Overview}
 
Ogre is written in Python 3. It employs several modules from the Python Materials Genomics (pymatgen) \cite{ong2013python} and the Atomic Simulation Environment (ASE) \cite{larsen2017atomic} libraries. The package can be downloaded from www.noamarom.com under the terms of a BSD-3 license. The inputs of Ogre are the bulk structures of the substrate and film materials, as well as a configuration file with user-specified settings for directing the workflow of Ogre. Standard input structure file formats such as the POSCAR format of the Vienna \emph{ab initio} Simulations Package's (VASP) \cite{Joubert1999,Kresse1996,Kresse1CI9C6a,Kresse1993,Kresse1994}, crystallographic information files (CIF), and the geometry.in format of the FHI-aims code \cite{Blum2009} are supported by Ogre. Figure \ref{fig:lattice} shows an overview of the workflow of interface structure prediction with Ogre. Ogre's workflow includes four main steps: surface selection (Section \ref{sec:surf_select}), lattice matching (Section \ref{sec:lattice}), surface matching (Section \ref{sec:surf_match}), and ranking (Section \ref{sec:ranking}). Finally, the electronic properties may be calculated for the interface(s) predicted to be the most stable (Section \ref{sec:Electronic})

\begin{figure}[htbp]
\centering
\includegraphics[scale = 0.35]{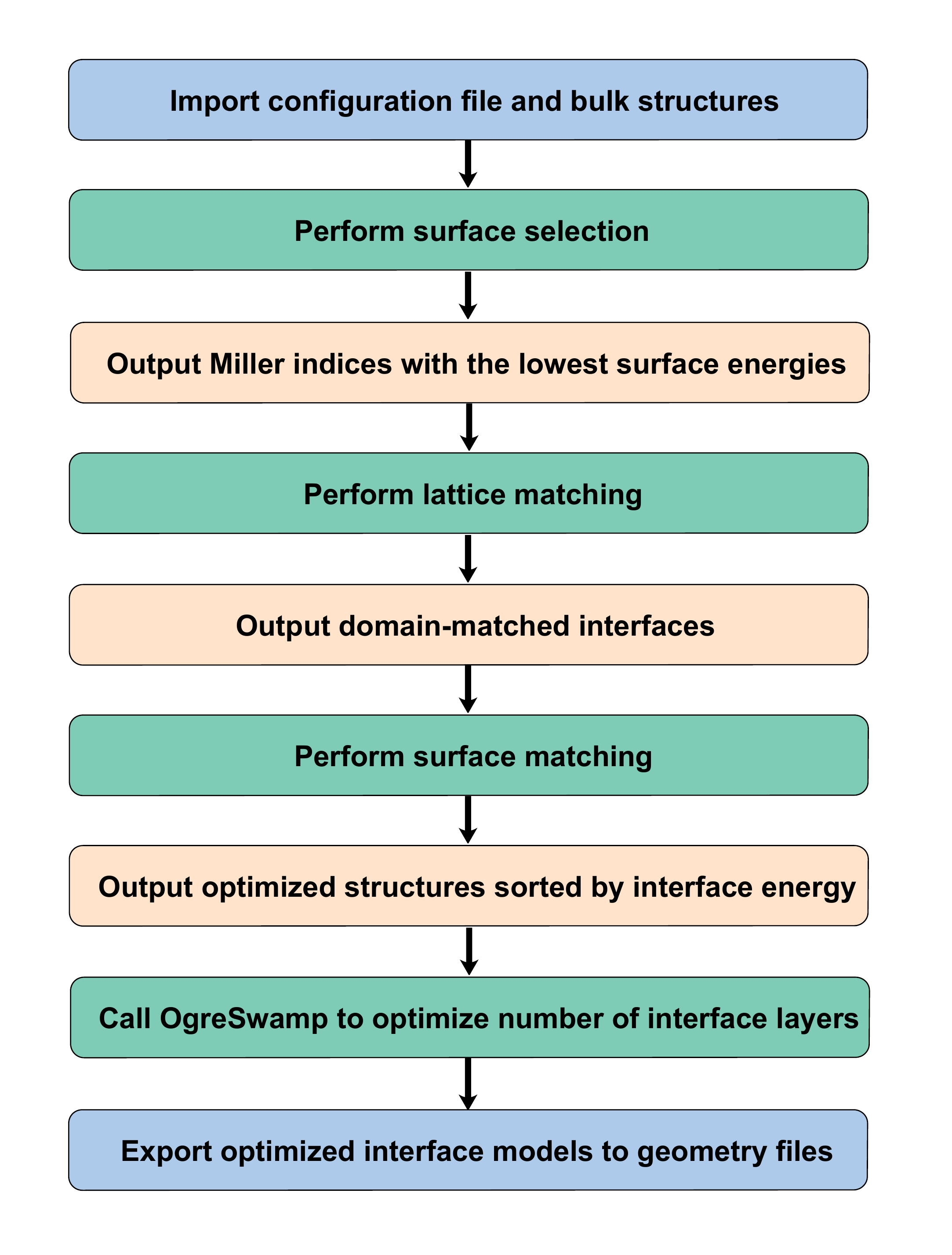}
\caption{workflow of interface structure prediction with Ogre. The blue boxes represent the input and output of the code. Code modules are represented by green boxes. The orange boxes represent module outputs that serve as inputs of the subsequent module. 
}
 \label{fig:lattice}
\end{figure}

As a preliminary step, Ogre may be used for substrate and surface selection. For organic epitaxy, substrates that form platelets are desirable.\cite{Dull2021} Ogre's morphology prediction capability \cite{yang2020ogre} could be used for substrate screening. If the substrate has already been chosen, as in the case of the TCNQ/TTF interface studied here, but the orientation is unknown, Ogre may be used to calculate the surface energy of all the symmetrically unique facets up to a selected Miller index.\cite{yang2020ogre} The lowest energy surfaces form the largest facets and are best suited to serve as the substrate for epitaxial growth. Lattice matching may then be performed for the chosen facets.

The lattice matching step for organic interfaces is similar to inorganic interfaces.\cite{Moayedpour2021} Ogre uses the algorithm of Zur and McGill\cite{Zur1984} to identify all domain-matched supercells of the substrate and film within the user-defined tolerances for interface area and lattice misfit. A Miller index scan can be performed if the substrate and/or film have multiple possible orientations. The maximum interface area, the misfit tolerance, and the Miller indices for the substrate and film are the input parameters for lattice matching. The lattice matching step generates a list of structures that are sorted by super-cell area misfit values. As explained above, owing to the weak nature of dispersion interactions in organic materials, robust epitaxial growth can occur at higher interface misfit values compared to inorganic interfaces.\cite{Sassella2008, Sassella2013} Therefore, a looser default tolerance of 5\%  has been adopted for the supercell area misfit, lattice vector length misfit, and angle misfit (the user may choose to alter the misfit tolerance). All the domain-matched interfaces within the misfit tolerance proceed to the surface matching step.

In the surface matching step, a search is performed to find the optimal distance in the $z$ direction and registry in the $xy$ plane between the substrate and film. Similar to the case of inorganic interfaces,\cite{Moayedpour2021} Bayesian optimization (BO) is used to find the most stable interface configuration by shifting the film in the $x$,$y$, and $z$ directions on top of the substrate. For inorganic interfaces the BO objective function is a geometric score function, based on the overlap and empty space between atomic spheres. Because this score function is not suitable for  molecules, organic interfaces require a different approach for fast evaluation of the relative stability of different interface configurations. Here, we use the ANI deep neural network interatomic potentials\cite{Smith2017} coupled with the Grimme D3 dispersion correction\cite{Grimme2010} as the BO objective function. To this end, Ogre uses the TorchANI\cite{Gao2020} package and the DFTD3 calculator class of the ASE package. Subsequently, the interface energy calculated with ANI+D3 is used for preliminary ranking of the optimized structures. The user can select a percentage of the generated interfaces to output. Finally, DFT may be used to accurately rank a small number of the best candidate interfaces and calculate their electronic properties. Ogre automates the construction of interface models with user-defined number of layers and vacuum space.

\subsection{\label{sec:surf_select}Surface Selection}

\begin{figure}[htbp]
\centering
\includegraphics[scale = 0.4]{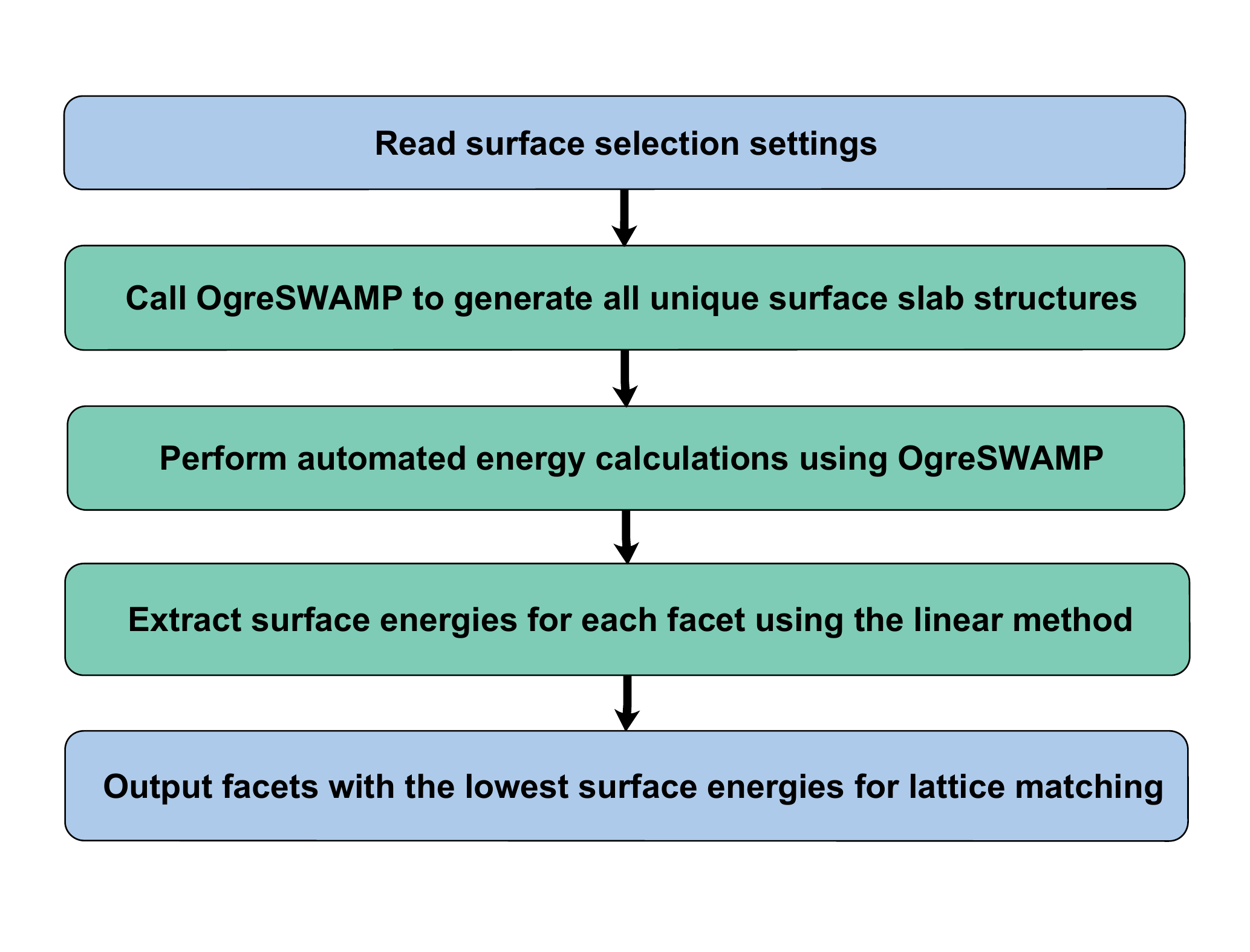}
\caption{The workflow of surface selection module in Ogre. The blue boxes represent the input and output of the surface selection step. Code modules are represented by green boxes.  }
 \label{fig:surface_selection_workflow}
\end{figure}

If the substrate orientation is unknown, as in the case of the TTF substrate used by Kattel \textit{et al.},\cite{Kattel2017} surface energy calculations may be performed by Ogre. We assume that the most stable facets are the most likely substrate orientations because they are the largest facets in the Wulff shape.\cite{yang2020ogre} The workflow of the surface selection step is shown in Figure \ref{fig:surface_selection_workflow}. In order to streamline the calculation of surface energies, Ogre is accompanied by the OgreSWAMP utilities\cite{yang2020ogre}. OgreSWAMP determines the Miller indices of all unique surfaces up to a user-defined maximum Miller index based on the space group symmetries of a given a molecular crystal structure. Automated DFT calculations are performed to obtain the total energy of surface slab models with an increasing number of layers. The surface energy is then calculated using the linear method\cite{Scholz2019,Sun2013}:
 \begin{equation}\label{eq:linear}
    E_{\text{Slab}} = NE_{\text{Bulk}} + 2A\gamma
\end{equation}{}
where $E_{slab}$ is the total energy of the surface slab, $N$, is the number of layers in the slab, and $A$ is the slab surface area. The energy of the bulk crystal, $E_{bulk}$, and the surface energy, $\gamma$, are extracted from the slope and intercept, respectively. 

\begin{figure}[htbp]
\centering
\includegraphics[scale = 0.27]{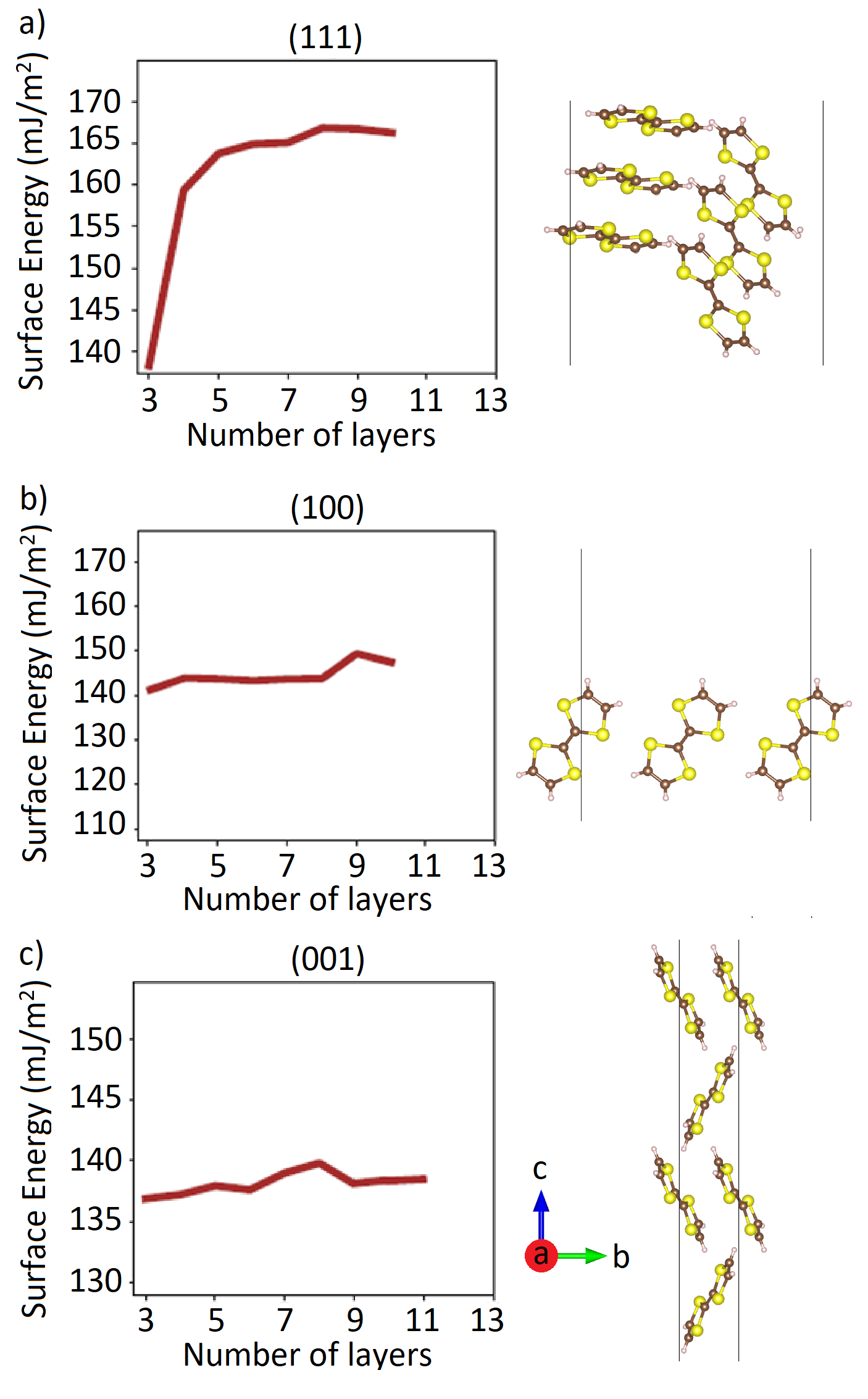}
\caption{Representative surface energy convergence plots obtained using PBE+TS for the (a) (111), (b) (100), and (c) (001) surfaces of TTF. The corresponding structures are also illustrated. C, S, and H atoms are colored in brown, yellow, and light pink, respectively. }
 \label{fig:surface energy}
\end{figure}

Figure \ref{fig:surface energy} shows the convergence of the surface energy with the number of layers for the (111), (001), and (100) facets of TTF. All other unique surfaces are shown in the SI. The stability of molecular crystal surfaces may be attributed to the chemical groups exposed on the surface.\cite{yang2020ogre} The (111) facet, shown in Figure \ref{fig:surface energy}a, cleaves through the strong interactions in stacking direction of the TTF molecules\cite{Antonijevic2019} and exposes the electron-rich $\pi$-system and the sulfur atoms of TTF on the surface. Therefore, this facet has a relatively high surface energy. In contrast, the (001) and (100) planes, shown in Figure \ref{fig:surface energy}b,c, cleave through a direction nearly perpendicular to the stacking direction of the TTF molecules, exposing hydrogen atoms on the surface. This produces more stable surfaces with lower surface energies. 
Table \ref{table:2} summarizes the surfaces energies obtained for all nine unique facets of TTF with a maximal Miller index of 1. The three facets with the lowest surface energies, (001), (100), and (011), were selected for the surface matching step.

\begin{table}[h!]
\centering
\begin{tabular}{|c | c |} 
 \hline
 Substrate (TTF) Miller index & Surface energy (\emph{mJ}/\emph{$m^2$})  \\ [0.5ex] 
 \hline\hline
 001 & 137.901  \\
100 & 144.843   \\
011 & 148.198 \\
$\bar{1}$01  & 150.802  \\
$\bar{1}$11 & 153.019  \\
010 & 153.754  \\
110 & 164.103  \\
111 & 164.117  \\
101 & 182.249  \\ [1ex] 
 \hline
\end{tabular}
\caption{Surface energy values for all symmetrically unique surface orientations of TTF obtained with OgreSWAMP module}
\label{table:2}
\end{table}

\subsection{\label{sec:lattice}Lattice Matching}
\begin{figure}[htbp]
\centering
\includegraphics[scale = 0.4]{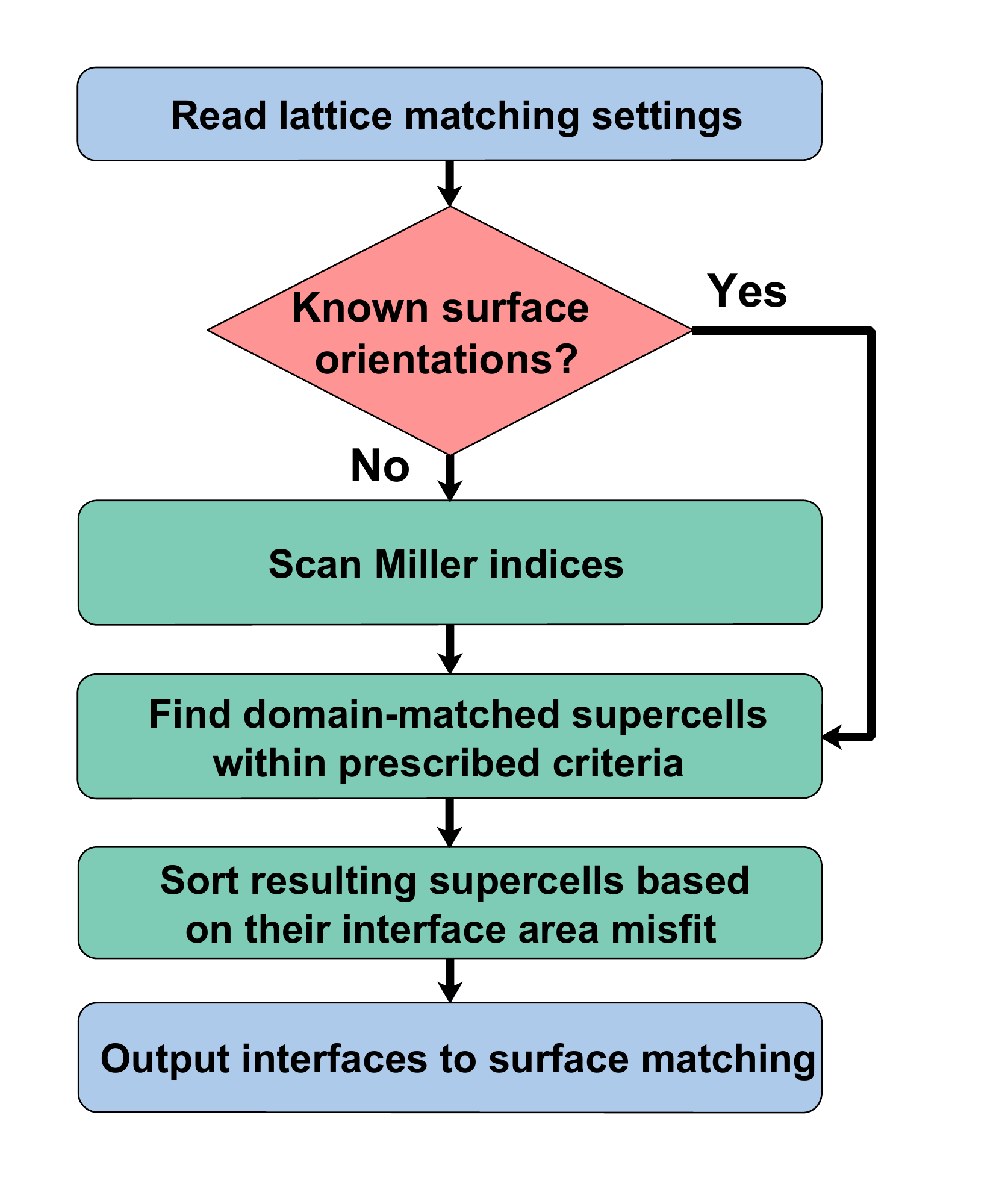}
\caption{The workflow of lattice matching module in Ogre. Different code modules are represented by green boxes. The red diamond indicates a decision whether a Miller index scan should be performed. 
}
 \label{fig:lattice}
\end{figure}

Figure \ref{fig:lattice} show the workflow of Ogre's lattice matching step. Lattice matching is controlled by the following input parameters: the substrate and film Miller indices, the maximum interface super-cell area, and the misfit tolerance for the (surface) lattice parameters, surface area, and angles. For organic interfaces, all misfit tolerances are set at 5\% by default, and the maximum super-cell area is set at 1000 {\AA}$^2$. ASE is used to obtain the basis vectors of the substrate and film surfaces by cleaving the bulk crystal structures along the specified Miller planes. In order to compare substrate and film two-dimensional lattices using a unique representation, the surface basis vectors are reduced to a pair of primitive basis vectors using Pymatgen. Next, all transformation matrices that would produce lattice matched super-cells within the user-defined tolerance of lattice vector length and angle misfit are constructed using the Pymatgen substrate analyzer module. Finally, all unique commensurate interfaces of a particular system are identified using a reduction scheme and the interface structure is generated. The user may choose to apply strain to either the substrate or the film, however in order to simulate an epitaxial growth experiment, the substrate's lattice parameter is fixed by default and the film layer is strained to conform to the substrate. 

In order to build a lattice matched interface, Ogre generates the matching substrate and film super-cells and aligns the atomic coordinates. The required parameters for creating the interface model, including the interfacial distance, the initial shift in the $xy$ plane, the number of layers, and the length of the vacuum region can be determined by the user. By default, the interfacial distance and vacuum region and are set to 2 {\AA} {}  and 40 {\AA}, respectively. The initial interface structure has no $xy$ shifts applied by default. The user may determine the substrate and film thickness by specifying the number of layers or a range of values to calculate. Ogre can detect all possible surface terminations for each Miller plane and automatically generate the corresponding interfaces. The models constructed by Ogre are subsequently used for the surface matching step.

\begin{figure}[htbp]
\centering
\includegraphics[scale = 0.38]{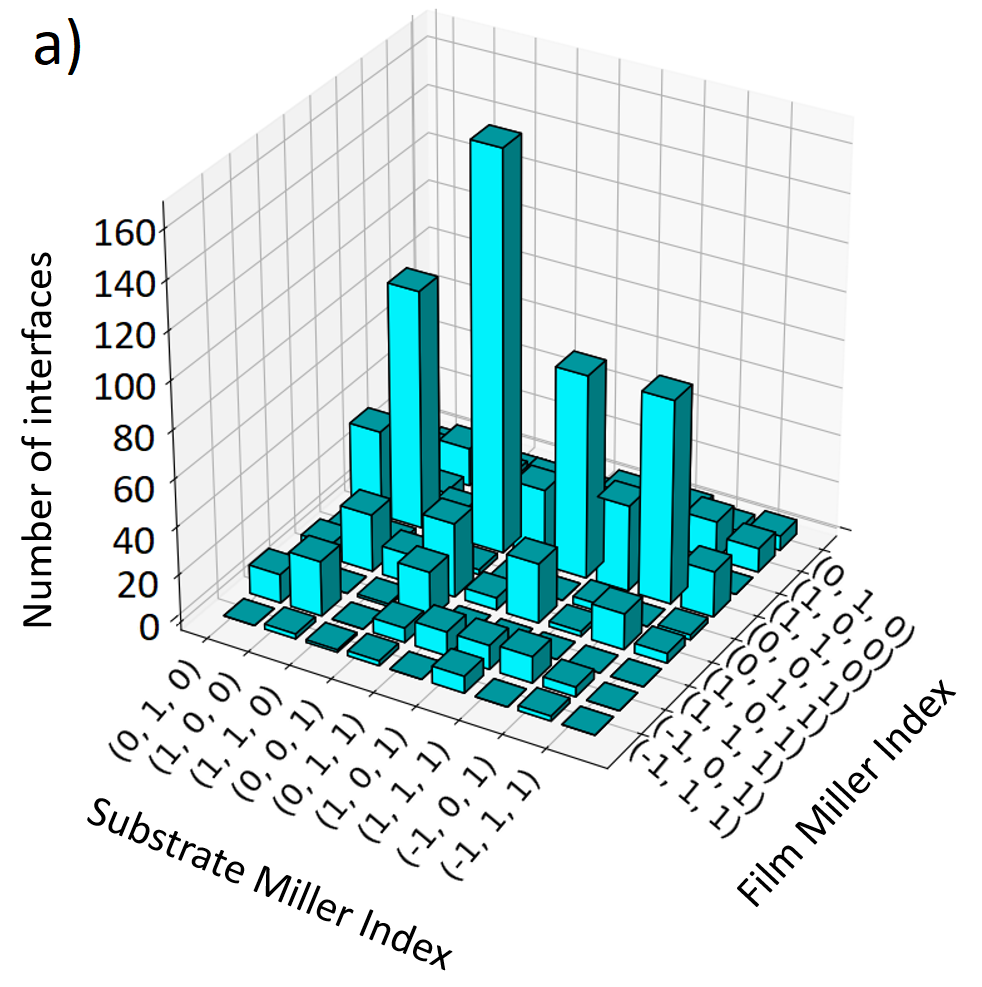}
\includegraphics[scale =0.55]{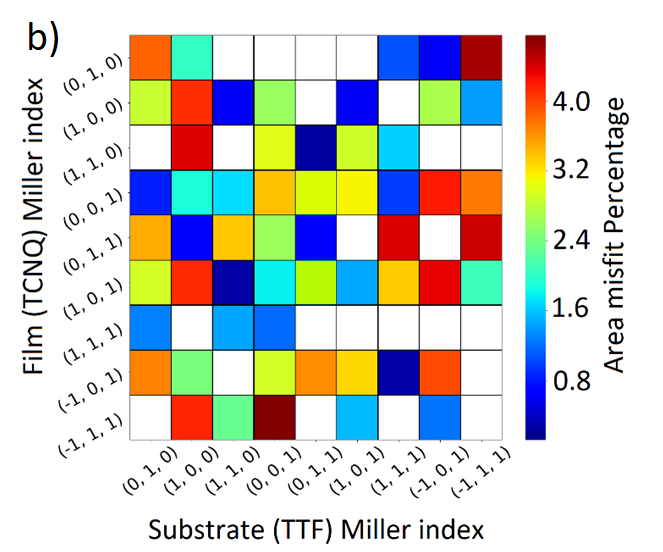}
\caption{Results of a Miller index scan for the TCNQ/TTF interface with a maximal Miller index of 1, maximum interface area of 1000 {\AA}$^2$ and area misfit tolerance of 5\%. (a) Histogram of the number of domain-matched interfaces generated for each combination of Miller indices. (b) Heat map plot of the lowest area misfit obtained for each combination of Miller indices. Miller index combinations for which no structures with an area misfit below 5\% were found are represented by white cells.} 
 \label{fig:Miller}
\end{figure}

The user may specify the substrate and film Miller indices. If multiple substrate orientations are feasible, and/or the film orientation is unknown Ogre may conduct a Miller index scan, where for each combination of substrate and film Miller indices lattice matching is performed. The input of the Miller index search module is the maximal single index. Ogre enumerates all possible symmetrically unique Miller indices as described in Ref. \cite{yang2020ogre}. For example, the results of a Miller index scan for a TCNQ/TTF interface with a maximal Miller index of 1, a misfit tolerance of 5\%, and a maximum area of 1000 {\AA}$^2$ are shown in Figure \ref{fig:Miller}. Panel (a) displays a histogram of the number of domain-matched interfaces generated for each combination of Miller indices. In total, 965  candidate  interfaces are generated with these settings. Additional candidate interfaces may be generated by selecting higher misfit tolerances or a larger maximum area, however, they are likely to be less stable. Panel (b) shows the minimum misfit percentage obtained for each interface orientation. Out of 81 possible Miller index combinations, 57 have a misfit under 5\%. To narrow down the number of candidates, only the three most stable facets of the TTF substrate, (100), (001), and (011), and only TCNQ facets that lead to an interface with a misfit lower than 2.5 \% are considered further. These leave 9 candidate interfaces that proceed to surface matching.

\subsection{\label{sec:surf_match}Surface Matching}

\begin{figure}[htbp]
\centering
\includegraphics[scale = 0.4]{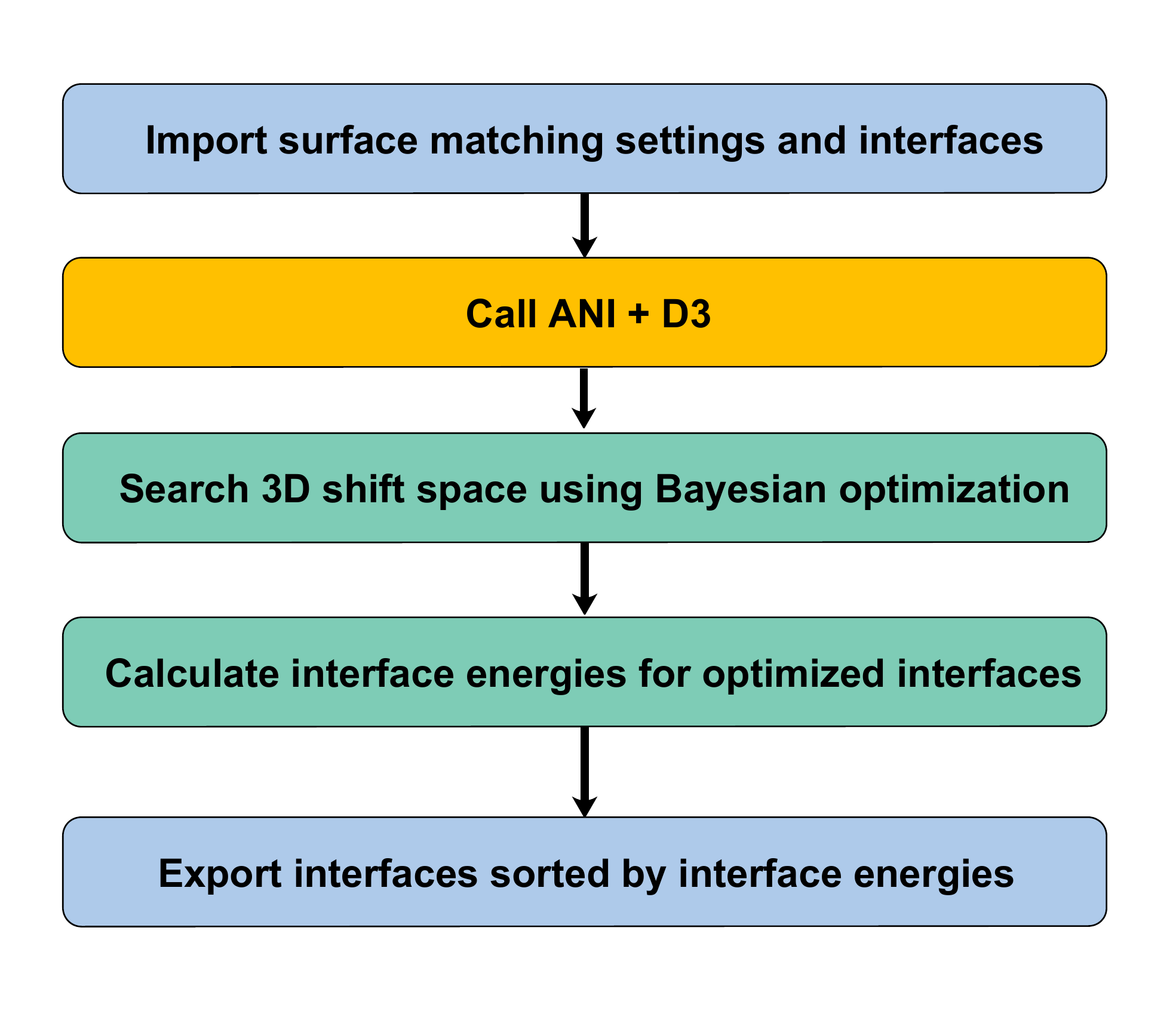}
\caption{The workflow of surface matching in Ogre. The blue boxes represent the inputs and outputs of the surface matching step. The yellow box represents the external codes that are called by Ogre for efficient calculation of total energies as the BO objective function. The green boxes represent different code modules.  }
 \label{fig:surf_workflow}
\end{figure}

Domain-matched interfaces proceed from the lattice matching step to the surface matching step, whose workflow is shown in Figure \ref{fig:surf_workflow}. The optimal position of the film above the substrate is determined using Bayesian optimization (BO) to efficiently search the 3D parameter space of the film position on top of the substrate in the $x$, $y$, and $z$ directions.  To evaluate the BO objective function, the user may choose DFT or ANI+D3. Ogre provides a grid search option that can generate potential energy surfaces or binding energy curves. Surface matching produces a list of structures ranked by the ANI+D3 interface energy. The final optimized interface structures are exported in the appropriate geometry file format (\textit{e.g.}, geometry.in for FHI-aims).

\subsubsection{\label{sec:level3}Performance of ANI+D3}

\begin{figure}[h]
\centering
\includegraphics[scale = 0.55]{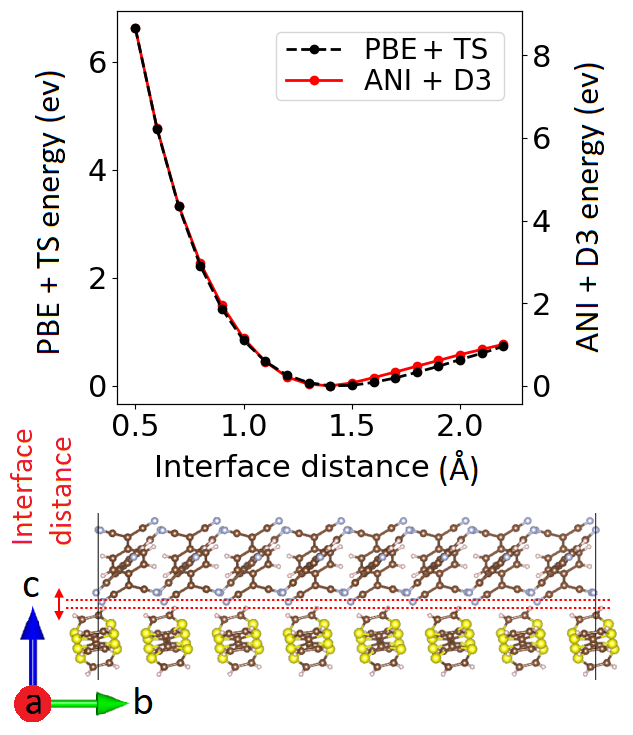}
\caption{Energy as a function of the interfacial distance, obtained with ANI+D3 (red) and PBE+TS (black) for the TCNQ(110)/TTF(011) interface. Both curves are referenced to their respective minima. An illustration of the interface is also shown. C, S, N, and H atoms are colored in brown, yellow, blue, and light pink, respectively. 
}
 \label{fig:surf_results_1}
\end{figure}

\begin{figure}[htbp]
\centering
\includegraphics[scale = 0.45]{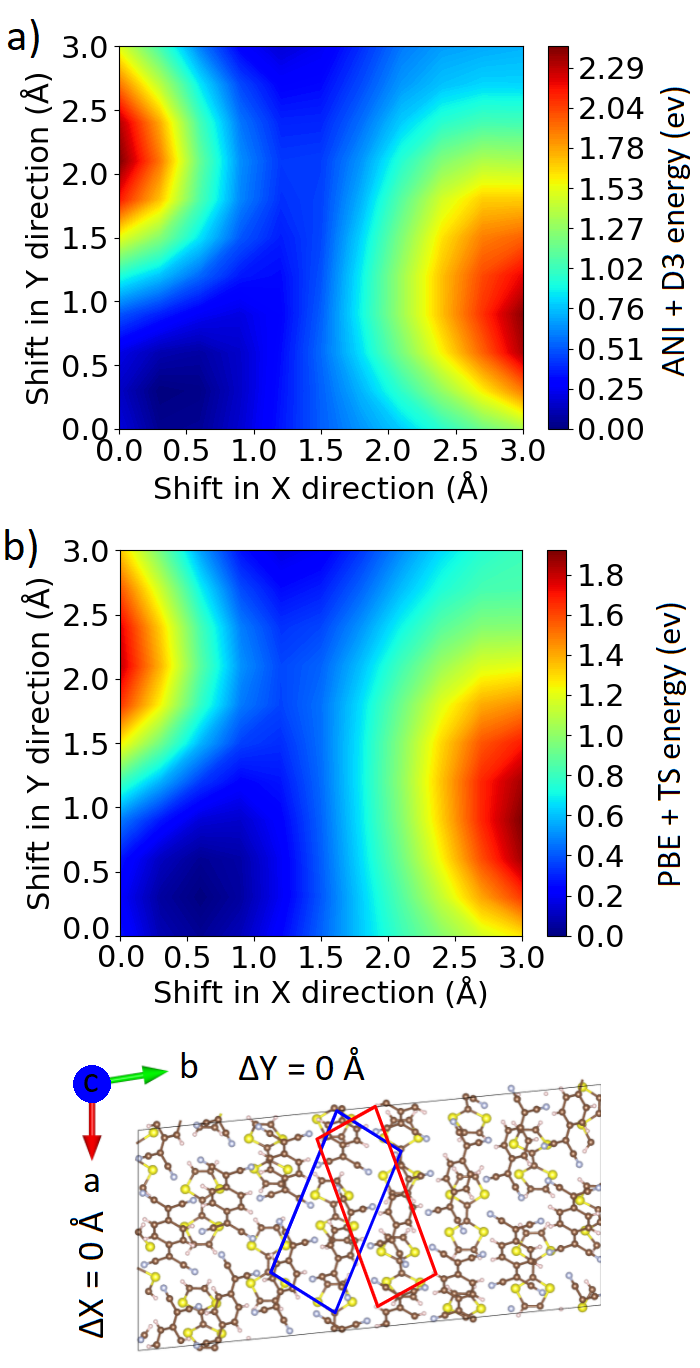}
\caption{Energy as a function of the registry in the $xy$ plane obtained with (a) ANI+D3 and (b) PBE+TS for the TCNQ(110)/TTF(011) interface at a fixed interfacial distance of 1.4 \AA. The ANI+D3 and PBE+TS energies are referenced to the respective minima. The illustration on the bottom shows the top view of the initial interface structure without any shifts applied in the $xy$ plane. C, S, N, and H atoms are colored in brown, yellow, blue, and light pink, respectively. The TTF substrate and TCNQ film unit cells are shown in blue and red, respectively. }
 \label{fig:surf_results_2}
\end{figure}

To validate the performance of ANI+D3, we compare its results to DFT, using PBE+TS, for a representative interface structure of TCNQ(110)/TTF(011), illustrated in Figure \ref{fig:surf_results_1}. This structure was selected because it has the lowest area mismatch of 0.23\% 
, as shown in Figure \ref{fig:Miller}. Figure \ref{fig:surf_results_1} shows a comparison of the potential energy curves as a function of the interfacial distance, obtained with ANI+D3 and PBE+TS. Both curves are referenced to their respective minima. Excellent agreement is demonstrated between ANI+D3 and PBE+TS for the position of the minimum and the shape of the curve. This is possibly because the interfacial distance is dominated by dispersion interactions, whose description by the TS and D3 pairwise methods is similar. Figure \ref{fig:surf_results_2} shows contour plots of the energy as a function of the position of the TCNQ film on top of the TTF substrate in the $xy$ plane at a fixed interface distance of 1.4 \AA, obtained with ANI+D3 compared with PBE+TS. ANI+D3 agrees well with the positions of the extrema and the features of the DFT potential energy surface. The in-plane registry depends mainly on the local electrostatic and exchange-correlation contributions because the dispersion contribution does not vary significantly at a fixed interfacial distance (a comparison of ANI to PBE without dispersion corrections is provided in the Fig. S3. 
Hence, the good agreement between ANI+D3 and PBE+TS may be attributed to the fact that ANI is trained to reproduce DFT data for local, intramolecular bonds.

\subsubsection{\label{sec:level3}Bayesian Optimization}

Bayesian optimization is a machine learning algorithm designed to find the extremum of a "black box" objective function with a minimal number of function queries.\cite{frazier2018tutorial} Based on the sampled points, a Gaussian process is typically used to construct a statistical model of the objective function, known as the prior.\cite{brochu2010tutorial} The following point to be sampled is then determined by an acquisition function. The prior is subsequently updated based on the newly learned information to create a new surrogate model, known as the posterior. The posterior serves as the new prior for the subsequent iteration and so on, until convergence is reached or a predetermined number of steps have been performed. 

 The BO objective function for surface matching is defined as the negative of the total energy of the interface, calculated using either ANI+D3 or DFT, such that the energy is minimized by maximizing the objective function. Ogre uses the upper confidence bound acquisition function\cite{Frazier2018, brochu2010tutorial, williams2006gaussian}:
\begin{equation}
     \vec{r}_{n+1} = \text{argmax}(\mu_{n}(\vec{r}) + \kappa \sigma_{n}(\vec{r}))
     \label{eq:UCB}
 \end{equation}
 where $\mu$ and $\sigma$ are the mean and standard deviation at each sampled point. The trade-off between exploitation and exploration is balanced by the hyperparameter $\kappa$. The higher the value of $\kappa$, the more exploration of regions with a high uncertainty is performed. A lower value of $\kappa$ accelerates convergence by favoring exploitation of promising regions, however if $\kappa$ is too low the BO might converge prematurely to a local extremum. 
 
\begin{figure}[htbp]
\centering
\includegraphics[scale = 0.2]{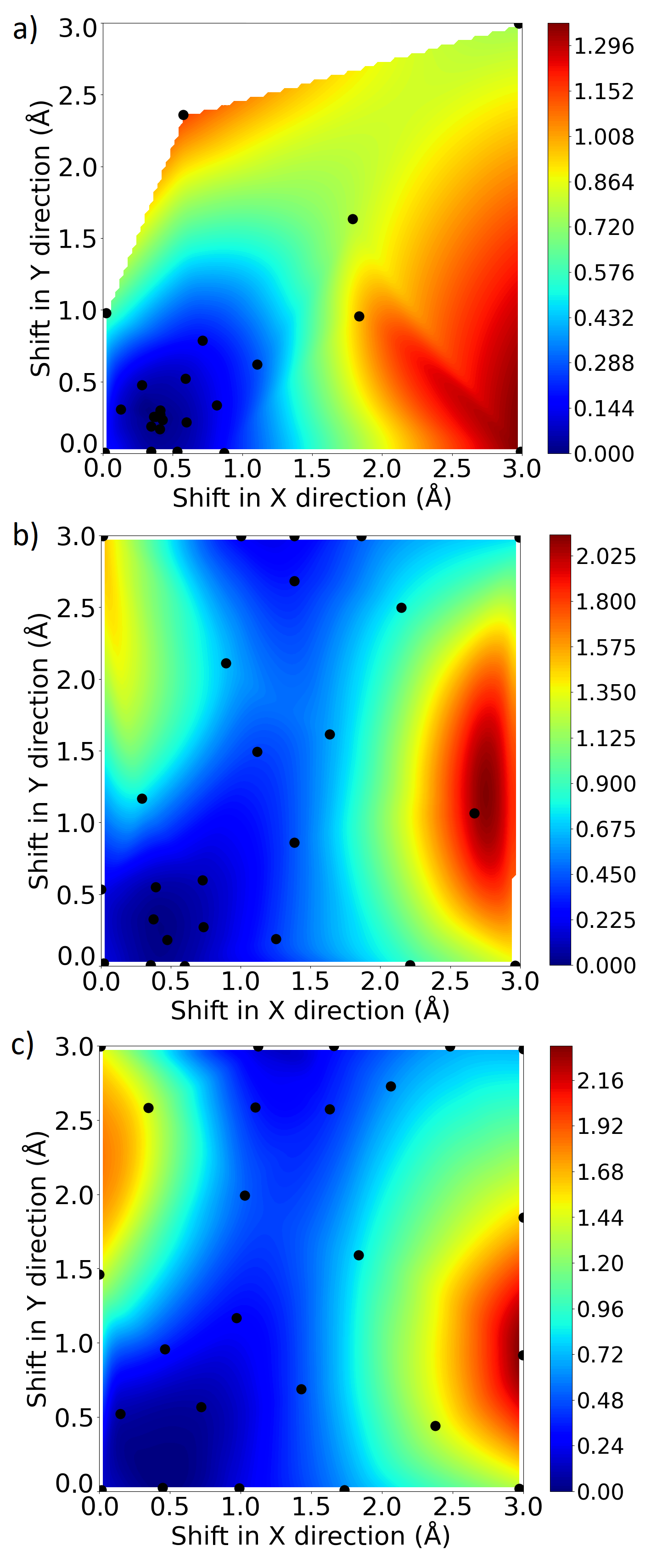}
\caption{BO predicted mean energy as a function of the registry in the $xy$ plane obtained with ANI+D3 using $\kappa$ values of (a) 1 (b) 5 (c) 10. all potential energy surfaces are referenced to their respective minima.}
 \label{fig:BO}
\end{figure}

 Figure \ref{fig:BO} illustrates the effect of the choice of $\kappa$ on the BO behavior for the two-dimensional surface of the TCNQ(110)/TTF(011) interface registry in the $xy$ plane at a fixed interfacial distance of 1.4 \AA. With $\kappa$=1 the BO algorithm converges to the global minimum very fast. Nearly all the points sampled are in the vicinity of the minimum and there is very little information on the remainder of the potential energy surface. The Gaussian process predicted mean bears little resemblance to the true ANI+D3 PES, shown in Figure \ref{fig:surf_results_2}a, and the local minimum at (1.25,3.0) is not found. With $\kappa$=5 the BO algorithm performs more exploration of the PES, while still converging relatively quickly to the global minimum. The Gaussian process predicted mean appears more similar to the true PES and the region of the local minimum is also sampled. With $\kappa$=10 significant exploration of the PES is performed, including the regions of the global and local maxima. As a result, the Gaussian process predicted mean appears very similar to the true PES. However, because computer time is wasted on exploring irrelevant regions of the PES, the convergence to the global minimum is slower. For the $\kappa$ values of 1 and 5 the BO started to exploit the global minimum region after 11 and 18 iterations, respectively. In comparison, for the $\kappa$ value of 10 it took over 25 iterations for the exploitation of the global minimum region to begin, although most of the sampled points were from the lower energy regions. Thus, $\kappa$=5 provides an optimal balance between exploration and exploitation.  
 
 The bayesian-optimization Python package\cite{Nogueira2014} is used by Ogre to perform BO. The bounds for shifts in the $x$, $y$,and $z$ directions  define the parameter space to be searched. These bounds have default values of  (0, $a$), (0, $b$), and ($d-1$ {\AA}, $d+1$ {\AA}), respectively, where $a$ and $b$ are the interface lattice parameters in the $xy$ plane and $d$ is the initial interface distance. The default values for the number of iterations, \emph{N}, and $\kappa$ are 100 and 5, respectively. The surface matching settings file enables the user to input different parameters than the default values. The most stable interface structure and any structures whose total energy is within a user-defined tolerance of the minimum are output by Ogre once the maximum number of BO iterations has been reached.
 
 BO is significantly more efficient than a grid search because it samples fewer points while maximizing the amount of information learned about the objective function. For example, for the TCNQ(110)/TTF(011) interface with a lattice parameter of 19.4 {\AA}, a grid search with a step size of 0.2 {\AA} and a 1 {\AA} range for the $z$ axis would require evaluating more than 47000 points. In comparison, the BO algorithm converges to the optimal interface configuration within 300 iterations. The computational cost of performing BO amounts to the number of sampled points multiplied by the cost of the objective function evaluation, which is the cost of an energy calculation with either ANI+D3 or DFT. ANI+D3 is faster than DFT by orders of magnitude. For example, single point energy evaluations  for the TCNQ(110)/TTF(011) interface took about 7 s using ANI+D3, compared to about 4000 s using DFT (PBE+TS).

\subsection{\label{sec:ranking}Ranking}

Every interface structure passed from the lattice matching step is surface matched, which could result in a significant number of candidate structures. Fast preliminary ranking is performed with ANI+D3 in order to down-select a smaller group of the most promising candidate structures for the final evaluation with DFT. The interface energy, $\sigma$, is defined as the energy required to eliminate two surfaces and form an interface: \cite{xiong2017first, christensen2002first, Moayedpour2021, Dardzinski2022}
\begin{equation}
\sigma = \gamma_{sub} + \gamma_{film} - W_{ad} 
\label{eq:interface_energy}
\end{equation}
where, $W_{ad}$ is the interface adhesive energy and $\gamma_{sub/film}$ are the surface energies of the substrate and film, calculated using Eq. \ref{eq:linear},\cite{yang2020ogre} The adhesive energy of an interface is defined as:\cite{li2016first,liu2004first,zhuo2018density,wang2020first}
\begin{equation}
W_{ad} = \frac{1}{A}(E_{sub} + E_{film} - E_{int})
\label{eq:adhesive_energy}
\end{equation}
where $E_{sub}$, $E_{film}$, and $E_{int}$ are total energies of the substrate surface slab, the film surface slab, and the interface, respectively. $E_{sub}$, $E_{film}$, and $E_{int}$ may be evaluated using either ANI+D3 or DFT. For inorganic interfaces, we have proposed a linear method for converging the interface energy as a function of the number of layers.\cite{Moayedpour2021, Dardzinski2022} For organic interfaces it is not feasible to calculate the interface energy for models with more than one layer of the substrate and one layer of the film, owing to the large system size. With only one layer of substrate and film, the smallest interface studied here, TCNQ(110)/TTF(011), contains 777 atoms and the largest interface studied here,  TCNQ(111)/TTF(001), contains 1655 atoms. In addition, the thickness of these organic interface models ranges from 93.58 \AA {} for TCNQ(110)/TTF(011) to 109.15 \AA {} for TCNQ(100)/TTF(001), which is roughly equivalent to an inorganic interface with 40 atomic layers. Owing to the weak nature of the dispersion interactions at organic interfaces and their large size, it is therefore a reasonable approximation to calculate the interface energy for models with only one layer of the substrate and one layer of the film. 

\begin{figure}[htbp]
\includegraphics[scale = 0.32]{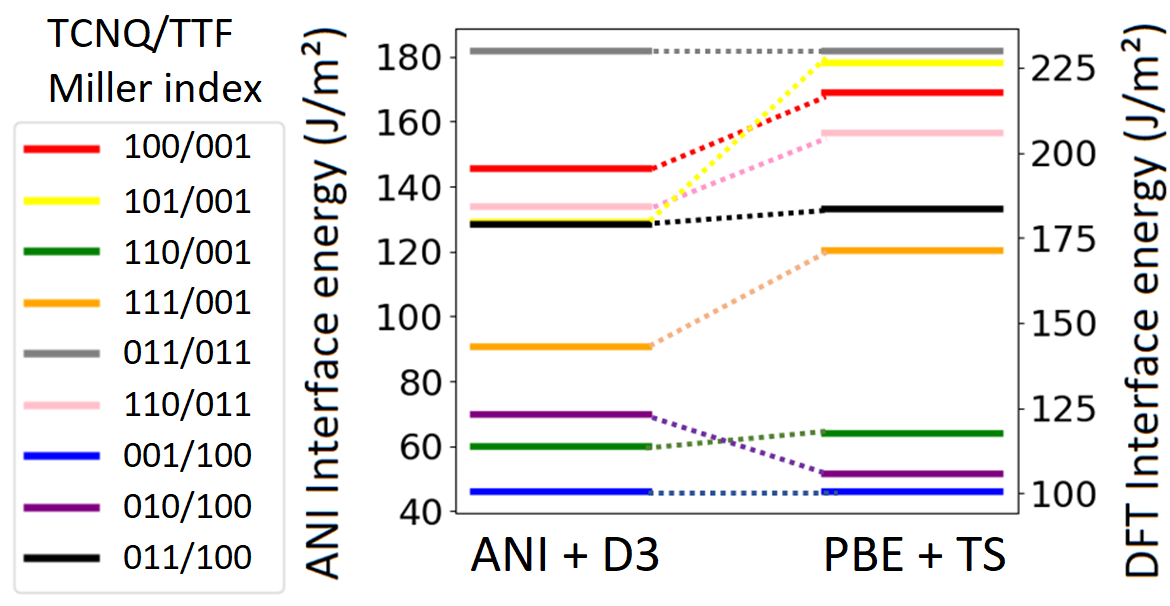}
\caption{ANI+D3 interface energies compared with DFT interface energies calculated using PBE+TS for the nine most stable interface structures of TCNQ/TTF.  }
 \label{fig:ranking}
\end{figure}

In figure \ref{fig:ranking} the interfaces energies obtained with ANI+D3 are compared to DFT interface energies obtained with PBE+TS for the TCNQ/TTF interfaces with the lowest substrate surface energies and interface area misfit. Because the ANI potentials have been trained on isolated molecules and have never "seen" an interface, we do not expect the interface energy values to be in close agreement with DFT values. Nevertheless, the ranking obtained with ANI+D3 is in reasonable agreement with the DFT ranking in the sense that the group of most stable structures is correctly identified. Hence we have confidence in the preliminary raking with ANI+D3 as a method of narrowing down the number of candidate structures to be calculated with DFT.

\subsection{\label{sec:Electronic}Electronic Structure}

We now proceed to study the electronic structure of the two most stable interface configurations of  TCNQ(001)/TTF(100) and TCNQ(010)/TTF(100). Previously, Beltr{\'{a}}n \textit{et al.} conducted DFT simulations of the TCNQ/TTF interface, whose structure was derived from a TTF-TCNQ co-crystal.\cite{Beltran2013} However, the structure of the co-crystal is significantly different than the bulk structure of either material. Therefore, such a model is not appropriate for the epitaxial growth of TCNQ on top of TTF by Kattel \textit{et al}.\cite{Kattel2017} Beltr{\'{a}}n \textit{et al.} used the local density approximation (LDA), with an empirical correction of the interface energy level alignment to compensate for the severe underestimation of band gaps by the LDA. Here, we use the HSE hybrid functional, which contains 25\% of exact (Fock) exchange. Atalla \textit{et al.} have shown for an isolated dimer of TTF and TCNQ molecules that a larger fraction of exact exchange is required to reproduce the correct energy level alignment.\cite{Atalla2016, Park2017} However, in extended systems the energy level alignment changes owing to band dispersion and polarization-induced gap narrowing.\cite{Wang2019} Hence, the HSE functional provides a correct description of the energy level alignment and the charge transfer at the interface of TTF and TCNQ, as we demonstrate below.

\begin{figure*}
    \centering
    \includegraphics[width=01\textwidth]{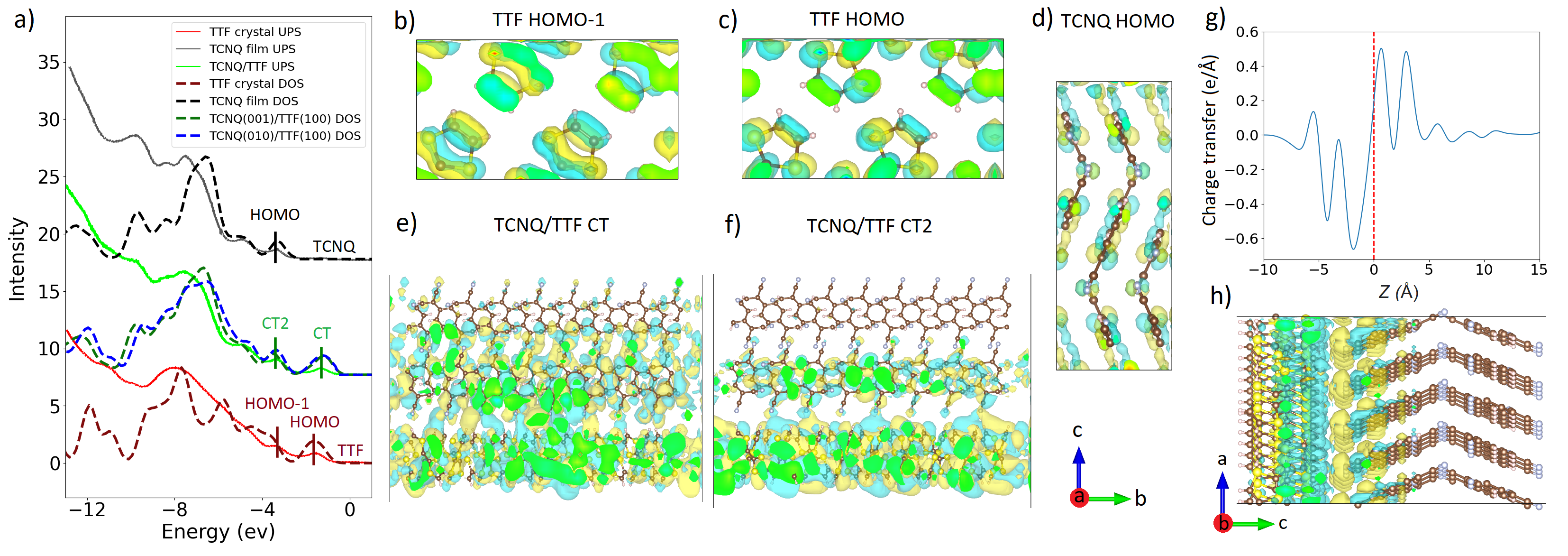}
    \caption{Electronic structure of the TCNQ/TTF interface: (a) Computed density of states of bulk TTF (red dashed line), bulk TCNQ (black dashed line), the TCNQ(001)/TTF(100) interface (green dashed line), and the TCNQ(010)/TTF(100) interface (blue dashed line) compared to UPS spectra of the TTF substrate (red solid line), a TCNQ thick film (gray solid line), and a 1 nm film of TCNQ on top of TTF (green solid line), reproduced from Ref. \cite{Kattel2017}.  Visualizations of (b) the HOMO and (c) HOMO-1 of TTF, (d) the HOMO of TCNQ,  and  orbitals corresponding to (e) the first and (f) the second DOS peaks of the  TCNQ(001)/TTF(100) interface, attributed to charge transfer (CT) states.  Charge transfer at the TCNQ(001)/TTF(100) interface: (g) Net charge transferred as a function of the distance from the interface for TTF(100)/TCNQ(001) interface, where the interface center is defined as $z = 0$. (h) Visualization of the charge density difference between the interface and the isolated substrate and film slabs as a function of position across the interface with an isosurface value of 0.004 \emph{e}/\AA$^3$. C, S, N, and H atoms are colored in brown, yellow, blue, and light pink. } 
    \label{fig:img1}
\end{figure*}

Kattle \textit{et al.} conducted UPS experiments for TCNQ films ranging in thickness from  0.3 nm to 10 nm, deposited on a single crystal of TTF.\cite{Kattel2017} In our model of the of TCNQ(001)/TTF(100) interface with a single layer of each material the thickness of the TCNQ film is 1.653 nm. Therefore, we compare our results to the UPS result of the 1 nm thick TCNQ film. The UPS results of bulk TTF and TCNQ are compared to calculations conducted for the respective crystals. Figure \ref{fig:img1}a shows the comparison of the computed density of states (DOS) with UPS experiments. The absolute energies of the Kohn-Sham orbitals from ground state DFT calculations using approximate exchange-correlation functionals do not strictly correspond to ionization energies.\cite{Kronik2014} Therefore, the computed DOS are shifted to align the first peak with the first UPS peak. In addition Gaussian broadening of 0.1 eV is applied to the DOS to simulate the resolution of the experiment. The comparison between theory and experiment is focused only on peak positions. We do not consider the secondary electron background, which leads to the increase of the spectral intensity towards lower binding energies,\cite{Kattel2017} vibrational effects, or intensity variations due to the scattering cross-section and penetration depth.\cite{UENO2008}  The positions of the peaks in the DOS of the TTF crystal, the TCNQ crystal, and the TCNQ/TTF interface are in a good agreement with the position of the peaks in the UPS spectra. The DOS of the TTF(100)/TCNQ(001) and the TCNQ(010)/TTF(100) interface structures are similar with the TTF(100)/TCNQ(001) DOS being in slightly better agreement with experiment. An experiment with a finer resolution may be required to unambiguously determine the interface structure. 

In the spectrum of the interface with the 1 nm thick TCNQ film, the peak observed around -1.5 eV was attributed by Kattle \textit{et al.} to a CT state, which is likely responsible for the high conductivity of the interface. Indeed, in our DFT simulations, the interface exhibits metallic behavior. Owing to the number of molecules included in the interface models, several orbitals contribute to each peak in the computed DOS. Within a 0.01 eV energy range of the peaks centers, 7 orbitals contribute to the first peak and 9 orbitals contribute to the second peak. The orbitals with energies closest to the peak centeres are depicted in  \ref{fig:img1} The first peak, labeled CT, originates from a state that is delocalized over the interface and is distinct from the HOMO of TTF, as shown in Figure \ref{fig:img1}. We attribute this interface state to ground state CT from TTF to TCNQ, rather than a CT excitation, in agreement with experiment. Kattle \textit{et al.} estimated the spatial extent of the CT state to be 1-2 nm or less, which is in agreement with our simulation. The second peak in the DOS of the interface, labeled CT2, arises from a delocalized state across the interface, comprising mixed contributions from the TTF HOMO-1 and the TCNQ HOMO, which are close in energy. The charge transfer at the TCNQ(001)/TTF(100) interface is given by the difference between the DFT charge density of the interface and the charge densities of isolated substrate and bulk slabs with the same geometry\cite{Yang2020}:
\begin{equation}
C_{net}(z) = C_{interface} - C_{substrate} - C_{film}
\label{eq:charge}
\end{equation}
where the charge is averaged over the $xy$ plane at each position along the $z$-axis. Figure \ref{fig:img1}f-g shows the charge transfer as a function of position across the interface. The net charge is negative at the interfacial layer of TTF(100) and positive at the TCNQ(001) layer, which confirms that the charge is transferred from the TTF substrate to the TCNQ film.

\section{\label{sec:level1}Conclusion}

We have presented a method for structure prediction of epitaxial organic interfaces by lattice and surface matching, implemented in the Ogre  open-source Python package. The lattice matching step identifies all potential domain-matched interface super-cells within user-defined tolerances for interface area and lattice mismatch. The most stable facets of the substrate may be pre-selected by calculating the surface energies.  If several substrate and/or film orientations are possible, a Miller index scan may be performed to find the combinations that result in the best candidate interface structures.  

In the surface matching step, Bayesian optimization (BO) is utilized  to determine the optimal configuration of each domain-matched interface in terms of the interfacial distance in the $z$ direction and the registry in the $xy$ plane. The BO objective function is based on the energy of the interface evaluated using the ANI deep neural network inter-atomic potentials with the Grimme D3 dispersion correction. ANI+D3 successfully reproduces the DFT results for the energy as a function of the interfacial distance and the features of the DFT potential energy surface in the $xy$ plane at a fraction of the computational cost. In addition, the preliminary ranking of interface structures with ANI+D3 is in reasonable agreement with the DFT ranking and the most stable interface structures are correctly identified. DFT simulations can be performed for a small number of the most promising candidate interface structures. Ogre streamlines the construction of surface and interface slab models. 

The application of Ogre has been demonstrated for an epitaxial interface of TCNQ on TTF, which is of interest for organic electronics. The electronic structure of the TCNQ/TTF interface had been probed by UPS, however its structure had not been experimentally characterized. The three facets of the TTF substrate with the lowest surface energies were found to be (001), (100), and (011). The stability of these surfaces may be attributed to the H-terminated edges of the TTF molecules being exposed on the surface, rather than the $\pi$ system in the plane of the rings. A Miller index scan was performed to identify the TCNQ film orientations that lead to interfaces with the lowest lattice mismatch. Nine interfaces with different Miller index combinations were selected for surface matching. In the final ranking, TCNQ(001)/TTF(100) was found to be the most stable interface configuration followed by  TCNQ(010)/TTF(100). Electronic structure calculations were performed for both of these interfaces and compared with UPS experiments. Excellent agreement was obtained between the computed DOS and UPS measurements in terms of the relative peak positions, although experiments with higher resolution would be required to unambiguously distinguish between the TCNQ(001)/TTF(100) and TCNQ(010)/TTF(100) interface configurations. For the most stable TCNQ(001)/TTF(100) interface configuration, we have also investigated the charge transfer. The first peak of the spectrum of the interface was found to be associated with an orbital delocalized over the interface, rather than states belonging to either the TTF or TCNQ. Ground state charge transfer from TTF to TCNQ was observed in agreement with experiment. 

In conclusion, we have shown that Ogre can contribute to advancements in the understanding of the structure and properties of epitaxial organic interfaces. Ogre can help interpret experiments by predicting the most likely interface configurations and correlating the structures with observed spectral features. Furthermore, Ogre may guide experimental growth efforts of epitaxial organic interfaces by identifying material combinations that are likely to form high-quality interfaces. In the future, Ogre may be integrated into automated materials discovery workflows to enable the computational design and discovery of epitaxial organic interfaces for high-performance organic electronic devices.

\begin{acknowledgments}
 We thank Wai-Lun Chan from the University of Kansas and Barry Rand from Princeton University for helpful discussions. Funding was provided by the U.S. Army Research Office (ARO) under grant W911NF1810148. O.I.  acknowledges  the Office of Naval Research (ONR) through support provided by the Energetic Materials Program (MURI grant no. N00014-21-1-2476). This research used resources of the National Energy Research Scientific Computing Center (NERSC), a DOE Office of Science User Facility supported by the Office of Science of the U.S. Department of Energy, under Contract DE-AC02-05CH11231. This work used the Extreme Science and Engineering Discovery Environment (XSEDE), which is supported by National Science Foundation grant number ACI-1548562. 
\end{acknowledgments}


\renewcommand\refname{References}
\bibliography{paper}

\begin{thebibliography}{127}%
\makeatletter
\providecommand \@ifxundefined [1]{%
 \@ifx{#1\undefined}
}%
\providecommand \@ifnum [1]{%
 \ifnum #1\expandafter \@firstoftwo
 \else \expandafter \@secondoftwo
 \fi
}%
\providecommand \@ifx [1]{%
 \ifx #1\expandafter \@firstoftwo
 \else \expandafter \@secondoftwo
 \fi
}%
\providecommand \natexlab [1]{#1}%
\providecommand \enquote  [1]{``#1''}%
\providecommand \bibnamefont  [1]{#1}%
\providecommand \bibfnamefont [1]{#1}%
\providecommand \citenamefont [1]{#1}%
\providecommand \href@noop [0]{\@secondoftwo}%
\providecommand \href [0]{\begingroup \@sanitize@url \@href}%
\providecommand \@href[1]{\@@startlink{#1}\@@href}%
\providecommand \@@href[1]{\endgroup#1\@@endlink}%
\providecommand \@sanitize@url [0]{\catcode `\\12\catcode `\$12\catcode
  `\&12\catcode `\#12\catcode `\^12\catcode `\_12\catcode `\%12\relax}%
\providecommand \@@startlink[1]{}%
\providecommand \@@endlink[0]{}%
\providecommand \url  [0]{\begingroup\@sanitize@url \@url }%
\providecommand \@url [1]{\endgroup\@href {#1}{\urlprefix }}%
\providecommand \urlprefix  [0]{URL }%
\providecommand \Eprint [0]{\href }%
\providecommand \doibase [0]{http://dx.doi.org/}%
\providecommand \selectlanguage [0]{\@gobble}%
\providecommand \bibinfo  [0]{\@secondoftwo}%
\providecommand \bibfield  [0]{\@secondoftwo}%
\providecommand \translation [1]{[#1]}%
\providecommand \BibitemOpen [0]{}%
\providecommand \bibitemStop [0]{}%
\providecommand \bibitemNoStop [0]{.\EOS\space}%
\providecommand \EOS [0]{\spacefactor3000\relax}%
\providecommand \BibitemShut  [1]{\csname bibitem#1\endcsname}%
\let\auto@bib@innerbib\@empty
\bibitem [{\citenamefont {Mazzio}\ and\ \citenamefont
  {Luscombe}(2014)}]{Mazzio2014}%
  \BibitemOpen
  \bibfield  {author} {\bibinfo {author} {\bibfnamefont {K.~A.}\ \bibnamefont
  {Mazzio}}\ and\ \bibinfo {author} {\bibfnamefont {C.~K.}\ \bibnamefont
  {Luscombe}},\ }\bibfield  {title} {\enquote {\bibinfo {title} {{The future of
  organic photovoltaics}},}\ }\href {\doibase 10.1039/C4CS00227J} {\bibfield
  {journal} {\bibinfo  {journal} {Chemical Society Reviews}\ }\textbf {\bibinfo
  {volume} {44}},\ \bibinfo {pages} {78--90} (\bibinfo {year}
  {2014})}\BibitemShut {NoStop}%
\bibitem [{\citenamefont {Kippelen}\ and\ \citenamefont
  {Brédas}(2009)}]{2009Kippelen}%
  \BibitemOpen
  \bibfield  {author} {\bibinfo {author} {\bibfnamefont {B.}~\bibnamefont
  {Kippelen}}\ and\ \bibinfo {author} {\bibfnamefont {J.-L.}\ \bibnamefont
  {Brédas}},\ }\bibfield  {title} {\enquote {\bibinfo {title} {Organic
  photovoltaics},}\ }\href {\doibase 10.1039/B812502N} {\bibfield  {journal}
  {\bibinfo  {journal} {Energy Environ. Sci.}\ }\textbf {\bibinfo {volume}
  {2}},\ \bibinfo {pages} {251--261} (\bibinfo {year} {2009})}\BibitemShut
  {NoStop}%
\bibitem [{\citenamefont {Reineke}\ \emph {et~al.}(2013)\citenamefont
  {Reineke}, \citenamefont {Thomschke}, \citenamefont {L\"ussem},\ and\
  \citenamefont {Leo}}]{Rieneke2013}%
  \BibitemOpen
  \bibfield  {author} {\bibinfo {author} {\bibfnamefont {S.}~\bibnamefont
  {Reineke}}, \bibinfo {author} {\bibfnamefont {M.}~\bibnamefont {Thomschke}},
  \bibinfo {author} {\bibfnamefont {B.}~\bibnamefont {L\"ussem}}, \ and\
  \bibinfo {author} {\bibfnamefont {K.}~\bibnamefont {Leo}},\ }\bibfield
  {title} {\enquote {\bibinfo {title} {White organic light-emitting diodes:
  Status and perspective},}\ }\href {\doibase 10.1103/RevModPhys.85.1245}
  {\bibfield  {journal} {\bibinfo  {journal} {Rev. Mod. Phys.}\ }\textbf
  {\bibinfo {volume} {85}},\ \bibinfo {pages} {1245--1293} (\bibinfo {year}
  {2013})}\BibitemShut {NoStop}%
\bibitem [{\citenamefont {Liu}\ \emph {et~al.}(2018)\citenamefont {Liu},
  \citenamefont {Li}, \citenamefont {Ren}, \citenamefont {Yan},\ and\
  \citenamefont {Bryce}}]{liu2018NatRevMater}%
  \BibitemOpen
  \bibfield  {author} {\bibinfo {author} {\bibfnamefont {Y.}~\bibnamefont
  {Liu}}, \bibinfo {author} {\bibfnamefont {C.}~\bibnamefont {Li}}, \bibinfo
  {author} {\bibfnamefont {Z.}~\bibnamefont {Ren}}, \bibinfo {author}
  {\bibfnamefont {S.}~\bibnamefont {Yan}}, \ and\ \bibinfo {author}
  {\bibfnamefont {M.~R.}\ \bibnamefont {Bryce}},\ }\bibfield  {title} {\enquote
  {\bibinfo {title} {All-organic thermally activated delayed fluorescence
  materials for organic light-emitting diodes},}\ }\href@noop {} {\bibfield
  {journal} {\bibinfo  {journal} {Nature Reviews Materials}\ }\textbf {\bibinfo
  {volume} {3}},\ \bibinfo {pages} {1--20} (\bibinfo {year}
  {2018})}\BibitemShut {NoStop}%
\bibitem [{\citenamefont {Brütting}, \citenamefont {Berleb},\ and\
  \citenamefont {Mückl}(2001)}]{Brutting2001}%
  \BibitemOpen
  \bibfield  {author} {\bibinfo {author} {\bibfnamefont {W.}~\bibnamefont
  {Brütting}}, \bibinfo {author} {\bibfnamefont {S.}~\bibnamefont {Berleb}}, \
  and\ \bibinfo {author} {\bibfnamefont {A.~G.}\ \bibnamefont {Mückl}},\
  }\bibfield  {title} {\enquote {\bibinfo {title} {Device physics of organic
  light-emitting diodes based on molecular materials},}\ }\href {\doibase
  https://doi.org/10.1016/S1566-1199(01)00009-X} {\bibfield  {journal}
  {\bibinfo  {journal} {Organic Electronics}\ }\textbf {\bibinfo {volume}
  {2}},\ \bibinfo {pages} {1--36} (\bibinfo {year} {2001})}\BibitemShut
  {NoStop}%
\bibitem [{\citenamefont {Armstrong}\ \emph {et~al.}(2009)\citenamefont
  {Armstrong}, \citenamefont {Wang}, \citenamefont {Alloway}, \citenamefont
  {Placencia}, \citenamefont {Ratcliff},\ and\ \citenamefont
  {Brumbach}}]{Armstrong2009}%
  \BibitemOpen
  \bibfield  {author} {\bibinfo {author} {\bibfnamefont {N.~R.}\ \bibnamefont
  {Armstrong}}, \bibinfo {author} {\bibfnamefont {W.}~\bibnamefont {Wang}},
  \bibinfo {author} {\bibfnamefont {D.~M.}\ \bibnamefont {Alloway}}, \bibinfo
  {author} {\bibfnamefont {D.}~\bibnamefont {Placencia}}, \bibinfo {author}
  {\bibfnamefont {E.}~\bibnamefont {Ratcliff}}, \ and\ \bibinfo {author}
  {\bibfnamefont {M.}~\bibnamefont {Brumbach}},\ }\bibfield  {title} {\enquote
  {\bibinfo {title} {Organic/organic heterojunctions: Organic light emitting
  diodes and organic photovoltaic devices},}\ }\href {\doibase
  https://doi.org/10.1002/marc.200900075} {\bibfield  {journal} {\bibinfo
  {journal} {Macromolecular Rapid Communications}\ }\textbf {\bibinfo {volume}
  {30}},\ \bibinfo {pages} {717--731} (\bibinfo {year} {2009})}\BibitemShut
  {NoStop}%
\bibitem [{\citenamefont {Mas-Torrent}\ and\ \citenamefont
  {Rovira}(2011)}]{Mas-Torrent2011}%
  \BibitemOpen
  \bibfield  {author} {\bibinfo {author} {\bibfnamefont {M.}~\bibnamefont
  {Mas-Torrent}}\ and\ \bibinfo {author} {\bibfnamefont {C.}~\bibnamefont
  {Rovira}},\ }\bibfield  {title} {\enquote {\bibinfo {title} {Role of
  molecular order and solid-state structure in organic field-effect
  transistors},}\ }\href {\doibase 10.1021/cr100142w} {\bibfield  {journal}
  {\bibinfo  {journal} {Chemical Reviews}\ }\textbf {\bibinfo {volume} {111}},\
  \bibinfo {pages} {4833--4856} (\bibinfo {year} {2011})}\BibitemShut {NoStop}%
\bibitem [{\citenamefont {Chen}\ \emph {et~al.}(2020)\citenamefont {Chen},
  \citenamefont {Zhang}, \citenamefont {Li}, \citenamefont {He},\ and\
  \citenamefont {Guo}}]{Chen2020ChemRev}%
  \BibitemOpen
  \bibfield  {author} {\bibinfo {author} {\bibfnamefont {H.}~\bibnamefont
  {Chen}}, \bibinfo {author} {\bibfnamefont {W.}~\bibnamefont {Zhang}},
  \bibinfo {author} {\bibfnamefont {M.}~\bibnamefont {Li}}, \bibinfo {author}
  {\bibfnamefont {G.}~\bibnamefont {He}}, \ and\ \bibinfo {author}
  {\bibfnamefont {X.}~\bibnamefont {Guo}},\ }\bibfield  {title} {\enquote
  {\bibinfo {title} {Interface engineering in organic field-effect transistors:
  Principles, applications, and perspectives},}\ }\href {\doibase
  10.1021/acs.chemrev.9b00532} {\bibfield  {journal} {\bibinfo  {journal}
  {Chemical Reviews}\ }\textbf {\bibinfo {volume} {120}},\ \bibinfo {pages}
  {2879--2949} (\bibinfo {year} {2020})}\BibitemShut {NoStop}%
\bibitem [{\citenamefont {Sirringhaus}(2014)}]{Sirringhaus2014}%
  \BibitemOpen
  \bibfield  {author} {\bibinfo {author} {\bibfnamefont {H.}~\bibnamefont
  {Sirringhaus}},\ }\bibfield  {title} {\enquote {\bibinfo {title} {25th
  anniversary article: Organic field-effect transistors: The path beyond
  amorphous silicon},}\ }\href {\doibase
  https://doi.org/10.1002/adma.201304346} {\bibfield  {journal} {\bibinfo
  {journal} {Advanced Materials}\ }\textbf {\bibinfo {volume} {26}},\ \bibinfo
  {pages} {1319--1335} (\bibinfo {year} {2014})}\BibitemShut {NoStop}%
\bibitem [{\citenamefont {Braga}\ and\ \citenamefont
  {Horowitz}(2009)}]{Braga2009}%
  \BibitemOpen
  \bibfield  {author} {\bibinfo {author} {\bibfnamefont {D.}~\bibnamefont
  {Braga}}\ and\ \bibinfo {author} {\bibfnamefont {G.}~\bibnamefont
  {Horowitz}},\ }\bibfield  {title} {\enquote {\bibinfo {title}
  {High-performance organic field-effect transistors},}\ }\href {\doibase
  https://doi.org/10.1002/adma.200802733} {\bibfield  {journal} {\bibinfo
  {journal} {Advanced Materials}\ }\textbf {\bibinfo {volume} {21}},\ \bibinfo
  {pages} {1473--1486} (\bibinfo {year} {2009})}\BibitemShut {NoStop}%
\bibitem [{\citenamefont {Wang}\ \emph {et~al.}(2022)\citenamefont {Wang},
  \citenamefont {Sawatzki}, \citenamefont {Darbandy}, \citenamefont {Talnack},
  \citenamefont {Vahland}, \citenamefont {Malfois}, \citenamefont {Kloes},
  \citenamefont {Mannsfeld}, \citenamefont {Kleemann},\ and\ \citenamefont
  {Leo}}]{wang2022organic}%
  \BibitemOpen
  \bibfield  {author} {\bibinfo {author} {\bibfnamefont {S.-J.}\ \bibnamefont
  {Wang}}, \bibinfo {author} {\bibfnamefont {M.}~\bibnamefont {Sawatzki}},
  \bibinfo {author} {\bibfnamefont {G.}~\bibnamefont {Darbandy}}, \bibinfo
  {author} {\bibfnamefont {F.}~\bibnamefont {Talnack}}, \bibinfo {author}
  {\bibfnamefont {J.}~\bibnamefont {Vahland}}, \bibinfo {author} {\bibfnamefont
  {M.}~\bibnamefont {Malfois}}, \bibinfo {author} {\bibfnamefont
  {A.}~\bibnamefont {Kloes}}, \bibinfo {author} {\bibfnamefont
  {S.}~\bibnamefont {Mannsfeld}}, \bibinfo {author} {\bibfnamefont
  {H.}~\bibnamefont {Kleemann}}, \ and\ \bibinfo {author} {\bibfnamefont
  {K.}~\bibnamefont {Leo}},\ }\bibfield  {title} {\enquote {\bibinfo {title}
  {Organic bipolar transistors},}\ }\href@noop {} {\bibfield  {journal}
  {\bibinfo  {journal} {Nature}\ }\textbf {\bibinfo {volume} {606}},\ \bibinfo
  {pages} {700--705} (\bibinfo {year} {2022})}\BibitemShut {NoStop}%
\bibitem [{\citenamefont {Khan}\ and\ \citenamefont {Rand}(2021)}]{Khan2021}%
  \BibitemOpen
  \bibfield  {author} {\bibinfo {author} {\bibfnamefont {S.-U.-Z.}\
  \bibnamefont {Khan}}\ and\ \bibinfo {author} {\bibfnamefont {B.~P.}\
  \bibnamefont {Rand}},\ }\bibfield  {title} {\enquote {\bibinfo {title}
  {Influence of disorder and state filling on charge-transfer-state absorption
  and emission spectra},}\ }\href {\doibase 10.1103/PhysRevApplied.16.044026}
  {\bibfield  {journal} {\bibinfo  {journal} {Phys. Rev. Applied}\ }\textbf
  {\bibinfo {volume} {16}},\ \bibinfo {pages} {044026} (\bibinfo {year}
  {2021})}\BibitemShut {NoStop}%
\bibitem [{\citenamefont {Khan}\ \emph {et~al.}(2019)\citenamefont {Khan},
  \citenamefont {Londi}, \citenamefont {Liu}, \citenamefont {Fusella},
  \citenamefont {D’Avino}, \citenamefont {Muccioli}, \citenamefont
  {Brigeman}, \citenamefont {Niesen}, \citenamefont {Yang}, \citenamefont
  {Olivier}, \citenamefont {Dull}, \citenamefont {Giebink}, \citenamefont
  {Beljonne},\ and\ \citenamefont {Rand}}]{Khan2019}%
  \BibitemOpen
  \bibfield  {author} {\bibinfo {author} {\bibfnamefont {S.-U.-Z.}\
  \bibnamefont {Khan}}, \bibinfo {author} {\bibfnamefont {G.}~\bibnamefont
  {Londi}}, \bibinfo {author} {\bibfnamefont {X.}~\bibnamefont {Liu}}, \bibinfo
  {author} {\bibfnamefont {M.~A.}\ \bibnamefont {Fusella}}, \bibinfo {author}
  {\bibfnamefont {G.}~\bibnamefont {D’Avino}}, \bibinfo {author}
  {\bibfnamefont {L.}~\bibnamefont {Muccioli}}, \bibinfo {author}
  {\bibfnamefont {A.~N.}\ \bibnamefont {Brigeman}}, \bibinfo {author}
  {\bibfnamefont {B.}~\bibnamefont {Niesen}}, \bibinfo {author} {\bibfnamefont
  {T.~C.-J.}\ \bibnamefont {Yang}}, \bibinfo {author} {\bibfnamefont
  {Y.}~\bibnamefont {Olivier}}, \bibinfo {author} {\bibfnamefont {J.~T.}\
  \bibnamefont {Dull}}, \bibinfo {author} {\bibfnamefont {N.~C.}\ \bibnamefont
  {Giebink}}, \bibinfo {author} {\bibfnamefont {D.}~\bibnamefont {Beljonne}}, \
  and\ \bibinfo {author} {\bibfnamefont {B.~P.}\ \bibnamefont {Rand}},\
  }\bibfield  {title} {\enquote {\bibinfo {title} {Multiple charge transfer
  states in donor–acceptor heterojunctions with large frontier orbital energy
  offsets},}\ }\href {\doibase 10.1021/acs.chemmater.9b01279} {\bibfield
  {journal} {\bibinfo  {journal} {Chemistry of Materials}\ }\textbf {\bibinfo
  {volume} {31}},\ \bibinfo {pages} {6808--6817} (\bibinfo {year}
  {2019})}\BibitemShut {NoStop}%
\bibitem [{\citenamefont {Choi}\ \emph {et~al.}(2020)\citenamefont {Choi},
  \citenamefont {Paterson}, \citenamefont {Fusella}, \citenamefont {Panidi},
  \citenamefont {Solomeshch}, \citenamefont {Tessler}, \citenamefont {Heeney},
  \citenamefont {Cho}, \citenamefont {Anthopoulos}, \citenamefont {Rand},\ and\
  \citenamefont {Podzorov}}]{Choi2020}%
  \BibitemOpen
  \bibfield  {author} {\bibinfo {author} {\bibfnamefont {H.~H.}\ \bibnamefont
  {Choi}}, \bibinfo {author} {\bibfnamefont {A.~F.}\ \bibnamefont {Paterson}},
  \bibinfo {author} {\bibfnamefont {M.~A.}\ \bibnamefont {Fusella}}, \bibinfo
  {author} {\bibfnamefont {J.}~\bibnamefont {Panidi}}, \bibinfo {author}
  {\bibfnamefont {O.}~\bibnamefont {Solomeshch}}, \bibinfo {author}
  {\bibfnamefont {N.}~\bibnamefont {Tessler}}, \bibinfo {author} {\bibfnamefont
  {M.}~\bibnamefont {Heeney}}, \bibinfo {author} {\bibfnamefont
  {K.}~\bibnamefont {Cho}}, \bibinfo {author} {\bibfnamefont {T.~D.}\
  \bibnamefont {Anthopoulos}}, \bibinfo {author} {\bibfnamefont {B.~P.}\
  \bibnamefont {Rand}}, \ and\ \bibinfo {author} {\bibfnamefont
  {V.}~\bibnamefont {Podzorov}},\ }\bibfield  {title} {\enquote {\bibinfo
  {title} {Hall effect in polycrystalline organic semiconductors: The effect of
  grain boundaries},}\ }\href {\doibase https://doi.org/10.1002/adfm.201903617}
  {\bibfield  {journal} {\bibinfo  {journal} {Advanced Functional Materials}\
  }\textbf {\bibinfo {volume} {30}},\ \bibinfo {pages} {1903617} (\bibinfo
  {year} {2020})}\BibitemShut {NoStop}%
\bibitem [{\citenamefont {Gershenson}, \citenamefont {Podzorov},\ and\
  \citenamefont {Morpurgo}(2006)}]{Gershenson2006}%
  \BibitemOpen
  \bibfield  {author} {\bibinfo {author} {\bibfnamefont {M.~E.}\ \bibnamefont
  {Gershenson}}, \bibinfo {author} {\bibfnamefont {V.}~\bibnamefont
  {Podzorov}}, \ and\ \bibinfo {author} {\bibfnamefont {A.~F.}\ \bibnamefont
  {Morpurgo}},\ }\bibfield  {title} {\enquote {\bibinfo {title} {Colloquium:
  Electronic transport in single-crystal organic transistors},}\ }\href
  {\doibase 10.1103/RevModPhys.78.973} {\bibfield  {journal} {\bibinfo
  {journal} {Rev. Mod. Phys.}\ }\textbf {\bibinfo {volume} {78}},\ \bibinfo
  {pages} {973--989} (\bibinfo {year} {2006})}\BibitemShut {NoStop}%
\bibitem [{\citenamefont {Coropceanu}\ \emph {et~al.}(2007)\citenamefont
  {Coropceanu}, \citenamefont {Cornil}, \citenamefont {da~Silva~Filho},
  \citenamefont {Olivier}, \citenamefont {Silbey},\ and\ \citenamefont
  {Brédas}}]{Coropceanu2007}%
  \BibitemOpen
  \bibfield  {author} {\bibinfo {author} {\bibfnamefont {V.}~\bibnamefont
  {Coropceanu}}, \bibinfo {author} {\bibfnamefont {J.}~\bibnamefont {Cornil}},
  \bibinfo {author} {\bibfnamefont {D.~A.}\ \bibnamefont {da~Silva~Filho}},
  \bibinfo {author} {\bibfnamefont {Y.}~\bibnamefont {Olivier}}, \bibinfo
  {author} {\bibfnamefont {R.}~\bibnamefont {Silbey}}, \ and\ \bibinfo {author}
  {\bibfnamefont {J.-L.}\ \bibnamefont {Brédas}},\ }\bibfield  {title}
  {\enquote {\bibinfo {title} {Charge transport in organic semiconductors},}\
  }\href {\doibase 10.1021/cr050140x} {\bibfield  {journal} {\bibinfo
  {journal} {Chemical Reviews}\ }\textbf {\bibinfo {volume} {107}},\ \bibinfo
  {pages} {926--952} (\bibinfo {year} {2007})}\BibitemShut {NoStop}%
\bibitem [{\citenamefont {Nakayama}, \citenamefont {Kera},\ and\ \citenamefont
  {Ueno}(2020)}]{Nakayama2020}%
  \BibitemOpen
  \bibfield  {author} {\bibinfo {author} {\bibfnamefont {Y.}~\bibnamefont
  {Nakayama}}, \bibinfo {author} {\bibfnamefont {S.}~\bibnamefont {Kera}}, \
  and\ \bibinfo {author} {\bibfnamefont {N.}~\bibnamefont {Ueno}},\ }\bibfield
  {title} {\enquote {\bibinfo {title} {Photoelectron spectroscopy on single
  crystals of organic semiconductors: experimental electronic band structure
  for optoelectronic properties},}\ }\href {\doibase 10.1039/D0TC00891E}
  {\bibfield  {journal} {\bibinfo  {journal} {J. Mater. Chem. C}\ }\textbf
  {\bibinfo {volume} {8}},\ \bibinfo {pages} {9090--9132} (\bibinfo {year}
  {2020})}\BibitemShut {NoStop}%
\bibitem [{\citenamefont {Podzorov}(2013)}]{Podzorov2013}%
  \BibitemOpen
  \bibfield  {author} {\bibinfo {author} {\bibfnamefont {V.}~\bibnamefont
  {Podzorov}},\ }\bibfield  {title} {\enquote {\bibinfo {title} {{Organic
  single crystals: Addressing the fundamentals of organic electronics}},}\
  }\href {\doibase 10.1557/MRS.2012.306} {\bibfield  {journal} {\bibinfo
  {journal} {MRS Bulletin}\ }\textbf {\bibinfo {volume} {38}},\ \bibinfo
  {pages} {15--24} (\bibinfo {year} {2013})}\BibitemShut {NoStop}%
\bibitem [{\citenamefont {Li}\ \emph {et~al.}(2010)\citenamefont {Li},
  \citenamefont {Hu}, \citenamefont {Liu},\ and\ \citenamefont {Zhu}}]{Li2010}%
  \BibitemOpen
  \bibfield  {author} {\bibinfo {author} {\bibfnamefont {R.}~\bibnamefont
  {Li}}, \bibinfo {author} {\bibfnamefont {W.}~\bibnamefont {Hu}}, \bibinfo
  {author} {\bibfnamefont {Y.}~\bibnamefont {Liu}}, \ and\ \bibinfo {author}
  {\bibfnamefont {D.}~\bibnamefont {Zhu}},\ }\bibfield  {title} {\enquote
  {\bibinfo {title} {{Micro- and nanocrystals of organic semiconductors}},}\
  }\href {\doibase
  10.1021/AR900228V/ASSET/IMAGES/LARGE/AR-2009-00228V_0014.JPEG} {\bibfield
  {journal} {\bibinfo  {journal} {Accounts of Chemical Research}\ }\textbf
  {\bibinfo {volume} {43}},\ \bibinfo {pages} {529--540} (\bibinfo {year}
  {2010})}\BibitemShut {NoStop}%
\bibitem [{\citenamefont {Reese}\ and\ \citenamefont {Bao}(2007)}]{Reese2007}%
  \BibitemOpen
  \bibfield  {author} {\bibinfo {author} {\bibfnamefont {C.}~\bibnamefont
  {Reese}}\ and\ \bibinfo {author} {\bibfnamefont {Z.}~\bibnamefont {Bao}},\
  }\bibfield  {title} {\enquote {\bibinfo {title} {{Organic single-crystal
  field-effect transistors}},}\ }\href {\doibase 10.1016/S1369-7021(07)70016-0}
  {\bibfield  {journal} {\bibinfo  {journal} {Materials Today}\ }\textbf
  {\bibinfo {volume} {10}},\ \bibinfo {pages} {20--27} (\bibinfo {year}
  {2007})}\BibitemShut {NoStop}%
\bibitem [{\citenamefont {Fusella}\ \emph {et~al.}(2018)\citenamefont
  {Fusella}, \citenamefont {Brigeman}, \citenamefont {Welborn}, \citenamefont
  {Purdum}, \citenamefont {Yan}, \citenamefont {Schaller}, \citenamefont {Lin},
  \citenamefont {Loo}, \citenamefont {Voorhis}, \citenamefont {Giebink},\ and\
  \citenamefont {Rand}}]{Fusella2018}%
  \BibitemOpen
  \bibfield  {author} {\bibinfo {author} {\bibfnamefont {M.~A.}\ \bibnamefont
  {Fusella}}, \bibinfo {author} {\bibfnamefont {A.~N.}\ \bibnamefont
  {Brigeman}}, \bibinfo {author} {\bibfnamefont {M.}~\bibnamefont {Welborn}},
  \bibinfo {author} {\bibfnamefont {G.~E.}\ \bibnamefont {Purdum}}, \bibinfo
  {author} {\bibfnamefont {Y.}~\bibnamefont {Yan}}, \bibinfo {author}
  {\bibfnamefont {R.~D.}\ \bibnamefont {Schaller}}, \bibinfo {author}
  {\bibfnamefont {Y.~H.~L.}\ \bibnamefont {Lin}}, \bibinfo {author}
  {\bibfnamefont {Y.~L.}\ \bibnamefont {Loo}}, \bibinfo {author} {\bibfnamefont
  {T.~V.}\ \bibnamefont {Voorhis}}, \bibinfo {author} {\bibfnamefont {N.~C.}\
  \bibnamefont {Giebink}}, \ and\ \bibinfo {author} {\bibfnamefont {B.~P.}\
  \bibnamefont {Rand}},\ }\bibfield  {title} {\enquote {\bibinfo {title}
  {{Band-like Charge Photogeneration at a Crystalline Organic Donor/Acceptor
  Interface}},}\ }\href {\doibase 10.1002/AENM.201701494} {\bibfield  {journal}
  {\bibinfo  {journal} {Advanced Energy Materials}\ }\textbf {\bibinfo {volume}
  {8}},\ \bibinfo {pages} {1701494} (\bibinfo {year} {2018})}\BibitemShut
  {NoStop}%
\bibitem [{\citenamefont {Raimondo}\ \emph {et~al.}(2011)\citenamefont
  {Raimondo}, \citenamefont {Moret}, \citenamefont {Campione}, \citenamefont
  {Borghesi},\ and\ \citenamefont {Sassella}}]{Raimondo2011}%
  \BibitemOpen
  \bibfield  {author} {\bibinfo {author} {\bibfnamefont {L.}~\bibnamefont
  {Raimondo}}, \bibinfo {author} {\bibfnamefont {M.}~\bibnamefont {Moret}},
  \bibinfo {author} {\bibfnamefont {M.}~\bibnamefont {Campione}}, \bibinfo
  {author} {\bibfnamefont {A.}~\bibnamefont {Borghesi}}, \ and\ \bibinfo
  {author} {\bibfnamefont {A.}~\bibnamefont {Sassella}},\ }\bibfield  {title}
  {\enquote {\bibinfo {title} {{Unique orientation of organic epitaxial thin
  films: The role of intermolecular interactions at the interface and surface
  symmetry}},}\ }\href {\doibase
  10.1021/JP111754R/SUPPL_FILE/JP111754R_SI_001.PDF} {\bibfield  {journal}
  {\bibinfo  {journal} {Journal of Physical Chemistry C}\ }\textbf {\bibinfo
  {volume} {115}},\ \bibinfo {pages} {5880--5885} (\bibinfo {year}
  {2011})}\BibitemShut {NoStop}%
\bibitem [{\citenamefont {Schreiber}(2004)}]{Schreiber2004}%
  \BibitemOpen
  \bibfield  {author} {\bibinfo {author} {\bibfnamefont {F.}~\bibnamefont
  {Schreiber}},\ }\bibfield  {title} {\enquote {\bibinfo {title} {Organic
  molecular beam deposition: Growth studies beyond the first monolayer},}\
  }\href {\doibase https://doi.org/10.1002/pssa.200404334} {\bibfield
  {journal} {\bibinfo  {journal} {Physica Status Solidi (a)}\ }\textbf
  {\bibinfo {volume} {201}},\ \bibinfo {pages} {1037--1054} (\bibinfo {year}
  {2004})}\BibitemShut {NoStop}%
\bibitem [{\citenamefont {Forrest}\ \emph {et~al.}(1994)\citenamefont
  {Forrest}, \citenamefont {Burrows}, \citenamefont {Haskal},\ and\
  \citenamefont {So}}]{Forrest1994}%
  \BibitemOpen
  \bibfield  {author} {\bibinfo {author} {\bibfnamefont {S.~R.}\ \bibnamefont
  {Forrest}}, \bibinfo {author} {\bibfnamefont {P.~E.}\ \bibnamefont
  {Burrows}}, \bibinfo {author} {\bibfnamefont {E.~I.}\ \bibnamefont {Haskal}},
  \ and\ \bibinfo {author} {\bibfnamefont {F.~F.}\ \bibnamefont {So}},\
  }\bibfield  {title} {\enquote {\bibinfo {title} {{Ultrahigh-vacuum
  quasiepitaxial growth of model van der Waals thin films. II. Experiment}},}\
  }\href {\doibase 10.1103/PhysRevB.49.11309} {\bibfield  {journal} {\bibinfo
  {journal} {Physical Review B}\ }\textbf {\bibinfo {volume} {49}},\ \bibinfo
  {pages} {11309} (\bibinfo {year} {1994})}\BibitemShut {NoStop}%
\bibitem [{\citenamefont {Moret}\ \emph {et~al.}(2011)\citenamefont {Moret},
  \citenamefont {Borghesi}, \citenamefont {Campione}, \citenamefont
  {Fumagalli}, \citenamefont {Raimondo},\ and\ \citenamefont
  {Sassella}}]{Moret2011}%
  \BibitemOpen
  \bibfield  {author} {\bibinfo {author} {\bibfnamefont {M.}~\bibnamefont
  {Moret}}, \bibinfo {author} {\bibfnamefont {A.}~\bibnamefont {Borghesi}},
  \bibinfo {author} {\bibfnamefont {M.}~\bibnamefont {Campione}}, \bibinfo
  {author} {\bibfnamefont {E.}~\bibnamefont {Fumagalli}}, \bibinfo {author}
  {\bibfnamefont {L.}~\bibnamefont {Raimondo}}, \ and\ \bibinfo {author}
  {\bibfnamefont {A.}~\bibnamefont {Sassella}},\ }\bibfield  {title} {\enquote
  {\bibinfo {title} {{Organic-organic heteroepitaxy: facts, concepts and
  perspectives}},}\ }\href {\doibase 10.1002/CRAT.201000581} {\bibfield
  {journal} {\bibinfo  {journal} {Crystal Research and Technology}\ }\textbf
  {\bibinfo {volume} {46}},\ \bibinfo {pages} {827--832} (\bibinfo {year}
  {2011})}\BibitemShut {NoStop}%
\bibitem [{\citenamefont {Sassella}(2013)}]{Sassella2013}%
  \BibitemOpen
  \bibfield  {author} {\bibinfo {author} {\bibfnamefont {A.}~\bibnamefont
  {Sassella}},\ }\bibfield  {title} {\enquote {\bibinfo {title} {{Organic
  epitaxial layers on organic substrates}},}\ }\href {\doibase
  10.1002/CRAT.201200709} {\bibfield  {journal} {\bibinfo  {journal} {Crystal
  Research and Technology}\ }\textbf {\bibinfo {volume} {48}},\ \bibinfo
  {pages} {840--848} (\bibinfo {year} {2013})}\BibitemShut {NoStop}%
\bibitem [{\citenamefont {Koma}(1995)}]{Koma1995}%
  \BibitemOpen
  \bibfield  {author} {\bibinfo {author} {\bibfnamefont {A.}~\bibnamefont
  {Koma}},\ }\bibfield  {title} {\enquote {\bibinfo {title} {{Molecular beam
  epitaxial growth of organic thin films}},}\ }\href {\doibase
  10.1016/0960-8974(95)00009-V} {\bibfield  {journal} {\bibinfo  {journal}
  {Progress in Crystal Growth and Characterization of Materials}\ }\textbf
  {\bibinfo {volume} {30}},\ \bibinfo {pages} {129--152} (\bibinfo {year}
  {1995})}\BibitemShut {NoStop}%
\bibitem [{\citenamefont {Koma}(1999)}]{Koma1999}%
  \BibitemOpen
  \bibfield  {author} {\bibinfo {author} {\bibfnamefont {A.}~\bibnamefont
  {Koma}},\ }\bibfield  {title} {\enquote {\bibinfo {title} {{Van der Waals
  epitaxy for highly lattice-mismatched systems}},}\ }\href {\doibase
  10.1016/S0022-0248(98)01329-3} {\bibfield  {journal} {\bibinfo  {journal}
  {Journal of Crystal Growth}\ }\textbf {\bibinfo {volume} {201-202}},\
  \bibinfo {pages} {236--241} (\bibinfo {year} {1999})}\BibitemShut {NoStop}%
\bibitem [{\citenamefont {Fenter}\ \emph {et~al.}(1997)\citenamefont {Fenter},
  \citenamefont {Schreiber}, \citenamefont {Zhou}, \citenamefont
  {Eisenberger},\ and\ \citenamefont {Forrest}}]{Fenter1997}%
  \BibitemOpen
  \bibfield  {author} {\bibinfo {author} {\bibfnamefont {P.}~\bibnamefont
  {Fenter}}, \bibinfo {author} {\bibfnamefont {F.}~\bibnamefont {Schreiber}},
  \bibinfo {author} {\bibfnamefont {L.}~\bibnamefont {Zhou}}, \bibinfo {author}
  {\bibfnamefont {P.}~\bibnamefont {Eisenberger}}, \ and\ \bibinfo {author}
  {\bibfnamefont {S.}~\bibnamefont {Forrest}},\ }\bibfield  {title} {\enquote
  {\bibinfo {title} {{In situ studies of morphology, strain, and growth modes
  of a molecular organic thin film}},}\ }\href {\doibase
  10.1103/PhysRevB.56.3046} {\bibfield  {journal} {\bibinfo  {journal}
  {Physical Review B}\ }\textbf {\bibinfo {volume} {56}},\ \bibinfo {pages}
  {3046--3053} (\bibinfo {year} {1997})}\BibitemShut {NoStop}%
\bibitem [{\citenamefont {Forrest}\ and\ \citenamefont
  {Zhang}(1994)}]{Forrest1994a}%
  \BibitemOpen
  \bibfield  {author} {\bibinfo {author} {\bibfnamefont {S.~R.}\ \bibnamefont
  {Forrest}}\ and\ \bibinfo {author} {\bibfnamefont {Y.}~\bibnamefont
  {Zhang}},\ }\bibfield  {title} {\enquote {\bibinfo {title} {{Ultrahigh-vacuum
  quasiepitaxial growth of model van der Waals thin films. I. Theory}},}\
  }\href {\doibase 10.1103/PhysRevB.49.11297} {\bibfield  {journal} {\bibinfo
  {journal} {Physical Review B}\ }\textbf {\bibinfo {volume} {49}},\ \bibinfo
  {pages} {11297} (\bibinfo {year} {1994})}\BibitemShut {NoStop}%
\bibitem [{\citenamefont {Hooks}, \citenamefont {Fritz},\ and\ \citenamefont
  {Ward}(2001)}]{Hooks2001}%
  \BibitemOpen
  \bibfield  {author} {\bibinfo {author} {\bibfnamefont {D.~E.}\ \bibnamefont
  {Hooks}}, \bibinfo {author} {\bibfnamefont {T.}~\bibnamefont {Fritz}}, \ and\
  \bibinfo {author} {\bibfnamefont {M.~D.}\ \bibnamefont {Ward}},\ }\bibfield
  {title} {\enquote {\bibinfo {title} {Epitaxy and molecular organization on
  solid substrates},}\ }\href {\doibase
  https://doi.org/10.1002/1521-4095(200102)13:4<227::AID-ADMA227>3.0.CO;2-P}
  {\bibfield  {journal} {\bibinfo  {journal} {Advanced Materials}\ }\textbf
  {\bibinfo {volume} {13}},\ \bibinfo {pages} {227--241} (\bibinfo {year}
  {2001})}\BibitemShut {NoStop}%
\bibitem [{\citenamefont {Yang}\ and\ \citenamefont
  {Yan}(2009)}]{Yang2009ChemSocRev}%
  \BibitemOpen
  \bibfield  {author} {\bibinfo {author} {\bibfnamefont {J.}~\bibnamefont
  {Yang}}\ and\ \bibinfo {author} {\bibfnamefont {D.}~\bibnamefont {Yan}},\
  }\bibfield  {title} {\enquote {\bibinfo {title} {Weak epitaxy growth of
  organic semiconductor thin films},}\ }\href {\doibase 10.1039/B815723P}
  {\bibfield  {journal} {\bibinfo  {journal} {Chem. Soc. Rev.}\ }\textbf
  {\bibinfo {volume} {38}},\ \bibinfo {pages} {2634--2645} (\bibinfo {year}
  {2009})}\BibitemShut {NoStop}%
\bibitem [{\citenamefont {Yang}, \citenamefont {Yan},\ and\ \citenamefont
  {Jones}(2015)}]{Yang2015}%
  \BibitemOpen
  \bibfield  {author} {\bibinfo {author} {\bibfnamefont {J.}~\bibnamefont
  {Yang}}, \bibinfo {author} {\bibfnamefont {D.}~\bibnamefont {Yan}}, \ and\
  \bibinfo {author} {\bibfnamefont {T.~S.}\ \bibnamefont {Jones}},\ }\bibfield
  {title} {\enquote {\bibinfo {title} {{Molecular Template Growth and Its
  Applications in Organic Electronics and Optoelectronics}},}\ }\href {\doibase
  10.1021/ACS.CHEMREV.5B00142/ASSET/IMAGES/ACS.CHEMREV.5B00142.SOCIAL.JPEG_V03}
  {\bibfield  {journal} {\bibinfo  {journal} {Chemical Reviews}\ }\textbf
  {\bibinfo {volume} {115}},\ \bibinfo {pages} {5570--5603} (\bibinfo {year}
  {2015})}\BibitemShut {NoStop}%
\bibitem [{\citenamefont {Forrest}(1997)}]{Forrest1997}%
  \BibitemOpen
  \bibfield  {author} {\bibinfo {author} {\bibfnamefont {S.~R.}\ \bibnamefont
  {Forrest}},\ }\bibfield  {title} {\enquote {\bibinfo {title} {{Ultrathin
  organic films grown by organic molecular beam deposition and related
  techniques}},}\ }\href {\doibase
  10.1021/CR941014O/ASSET/IMAGES/LARGE/CR941014OF00036.JPEG} {\bibfield
  {journal} {\bibinfo  {journal} {Chemical Reviews}\ }\textbf {\bibinfo
  {volume} {97}},\ \bibinfo {pages} {1793--1896} (\bibinfo {year}
  {1997})}\BibitemShut {NoStop}%
\bibitem [{\citenamefont {Campione}\ \emph {et~al.}(2009)\citenamefont
  {Campione}, \citenamefont {Raimondo}, \citenamefont {Moret}, \citenamefont
  {Campiglio}, \citenamefont {Fumagalli},\ and\ \citenamefont
  {Sassella}}]{Campione2009}%
  \BibitemOpen
  \bibfield  {author} {\bibinfo {author} {\bibfnamefont {M.}~\bibnamefont
  {Campione}}, \bibinfo {author} {\bibfnamefont {L.}~\bibnamefont {Raimondo}},
  \bibinfo {author} {\bibfnamefont {M.}~\bibnamefont {Moret}}, \bibinfo
  {author} {\bibfnamefont {P.}~\bibnamefont {Campiglio}}, \bibinfo {author}
  {\bibfnamefont {E.}~\bibnamefont {Fumagalli}}, \ and\ \bibinfo {author}
  {\bibfnamefont {A.}~\bibnamefont {Sassella}},\ }\bibfield  {title} {\enquote
  {\bibinfo {title} {{Organic-organic heteroepitaxy of semiconductor crystals:
  $\alpha$-quaterthiophene on rubrene}},}\ }\href {\doibase
  10.1021/CM901463U/SUPPL_FILE/CM901463U_SI_001.PDF} {\bibfield  {journal}
  {\bibinfo  {journal} {Chemistry of Materials}\ }\textbf {\bibinfo {volume}
  {21}},\ \bibinfo {pages} {4859--4867} (\bibinfo {year} {2009})}\BibitemShut
  {NoStop}%
\bibitem [{\citenamefont {Al-Mahboob}\ \emph {et~al.}(2009)\citenamefont
  {Al-Mahboob}, \citenamefont {Sadowski}, \citenamefont {Fujikawa},\ and\
  \citenamefont {Sakurai}}]{Al-Mahboob2009}%
  \BibitemOpen
  \bibfield  {author} {\bibinfo {author} {\bibfnamefont {A.}~\bibnamefont
  {Al-Mahboob}}, \bibinfo {author} {\bibfnamefont {J.~T.}\ \bibnamefont
  {Sadowski}}, \bibinfo {author} {\bibfnamefont {Y.}~\bibnamefont {Fujikawa}},
  \ and\ \bibinfo {author} {\bibfnamefont {T.}~\bibnamefont {Sakurai}},\
  }\bibfield  {title} {\enquote {\bibinfo {title} {{The growth mechanism of
  pentacene-fullerene heteroepitaxial films}},}\ }\href {\doibase
  10.1016/J.SUSC.2009.02.032} {\bibfield  {journal} {\bibinfo  {journal}
  {Surface Science}\ }\textbf {\bibinfo {volume} {603}},\ \bibinfo {pages}
  {L53--L56} (\bibinfo {year} {2009})}\BibitemShut {NoStop}%
\bibitem [{\citenamefont {Nakayama}\ \emph {et~al.}(2016)\citenamefont
  {Nakayama}, \citenamefont {Mizuno}, \citenamefont {Hosokai}, \citenamefont
  {Koganezawa}, \citenamefont {Tsuruta}, \citenamefont {Hinderhofer},
  \citenamefont {Gerlach}, \citenamefont {Broch}, \citenamefont {Belova},
  \citenamefont {Frank}, \citenamefont {Yamamoto}, \citenamefont
  {Niederhausen}, \citenamefont {Glowatzki}, \citenamefont {Rabe},
  \citenamefont {Koch}, \citenamefont {Ishii}, \citenamefont {Schreiber},\ and\
  \citenamefont {Ueno}}]{Nakayama2016}%
  \BibitemOpen
  \bibfield  {author} {\bibinfo {author} {\bibfnamefont {Y.}~\bibnamefont
  {Nakayama}}, \bibinfo {author} {\bibfnamefont {Y.}~\bibnamefont {Mizuno}},
  \bibinfo {author} {\bibfnamefont {T.}~\bibnamefont {Hosokai}}, \bibinfo
  {author} {\bibfnamefont {T.}~\bibnamefont {Koganezawa}}, \bibinfo {author}
  {\bibfnamefont {R.}~\bibnamefont {Tsuruta}}, \bibinfo {author} {\bibfnamefont
  {A.}~\bibnamefont {Hinderhofer}}, \bibinfo {author} {\bibfnamefont
  {A.}~\bibnamefont {Gerlach}}, \bibinfo {author} {\bibfnamefont
  {K.}~\bibnamefont {Broch}}, \bibinfo {author} {\bibfnamefont
  {V.}~\bibnamefont {Belova}}, \bibinfo {author} {\bibfnamefont
  {H.}~\bibnamefont {Frank}}, \bibinfo {author} {\bibfnamefont
  {M.}~\bibnamefont {Yamamoto}}, \bibinfo {author} {\bibfnamefont
  {J.}~\bibnamefont {Niederhausen}}, \bibinfo {author} {\bibfnamefont
  {H.}~\bibnamefont {Glowatzki}}, \bibinfo {author} {\bibfnamefont {J.~P.}\
  \bibnamefont {Rabe}}, \bibinfo {author} {\bibfnamefont {N.}~\bibnamefont
  {Koch}}, \bibinfo {author} {\bibfnamefont {H.}~\bibnamefont {Ishii}},
  \bibinfo {author} {\bibfnamefont {F.}~\bibnamefont {Schreiber}}, \ and\
  \bibinfo {author} {\bibfnamefont {N.}~\bibnamefont {Ueno}},\ }\bibfield
  {title} {\enquote {\bibinfo {title} {Epitaxial growth of an organic p-n
  heterojunction: C60 on single-crystal pentacene},}\ }\href
  {https://pubs.acs.org/doi/full/10.1021/acsami.6b02744} {\bibfield  {journal}
  {\bibinfo  {journal} {ACS Applied Materials and Interfaces}\ }\textbf
  {\bibinfo {volume} {8}},\ \bibinfo {pages} {13499--13505} (\bibinfo {year}
  {2016})}\BibitemShut {NoStop}%
\bibitem [{\citenamefont {Nakayama}\ \emph {et~al.}(2018)\citenamefont
  {Nakayama}, \citenamefont {Tsuruta}, \citenamefont {Hinderhofer},
  \citenamefont {Mizuno}, \citenamefont {Broch}, \citenamefont {Gerlach},
  \citenamefont {Koganezawa}, \citenamefont {Hosokai},\ and\ \citenamefont
  {Schreiber}}]{Nakayama2018}%
  \BibitemOpen
  \bibfield  {author} {\bibinfo {author} {\bibfnamefont {Y.}~\bibnamefont
  {Nakayama}}, \bibinfo {author} {\bibfnamefont {R.}~\bibnamefont {Tsuruta}},
  \bibinfo {author} {\bibfnamefont {A.}~\bibnamefont {Hinderhofer}}, \bibinfo
  {author} {\bibfnamefont {Y.}~\bibnamefont {Mizuno}}, \bibinfo {author}
  {\bibfnamefont {K.}~\bibnamefont {Broch}}, \bibinfo {author} {\bibfnamefont
  {A.}~\bibnamefont {Gerlach}}, \bibinfo {author} {\bibfnamefont
  {T.}~\bibnamefont {Koganezawa}}, \bibinfo {author} {\bibfnamefont
  {T.}~\bibnamefont {Hosokai}}, \ and\ \bibinfo {author} {\bibfnamefont
  {F.}~\bibnamefont {Schreiber}},\ }\bibfield  {title} {\enquote {\bibinfo
  {title} {Temperature dependent epitaxial growth of {C60} overlayers on single
  crystal pentacene},}\ }\href@noop {} {\bibfield  {journal} {\bibinfo
  {journal} {Advanced Materials Interfaces}\ }\textbf {\bibinfo {volume} {5}},\
  \bibinfo {pages} {1800084} (\bibinfo {year} {2018})}\BibitemShut {NoStop}%
\bibitem [{\citenamefont {Conrad}\ \emph {et~al.}(2009)\citenamefont {Conrad},
  \citenamefont {Tosado}, \citenamefont {Dutton}, \citenamefont {Dougherty},
  \citenamefont {Jin}, \citenamefont {Bonnen}, \citenamefont {Schuldenfrei},
  \citenamefont {Cullen}, \citenamefont {Williams}, \citenamefont
  {Reutt-Robey},\ and\ \citenamefont {Robey}}]{Conrad2009}%
  \BibitemOpen
  \bibfield  {author} {\bibinfo {author} {\bibfnamefont {B.~R.}\ \bibnamefont
  {Conrad}}, \bibinfo {author} {\bibfnamefont {J.}~\bibnamefont {Tosado}},
  \bibinfo {author} {\bibfnamefont {G.}~\bibnamefont {Dutton}}, \bibinfo
  {author} {\bibfnamefont {D.~B.}\ \bibnamefont {Dougherty}}, \bibinfo {author}
  {\bibfnamefont {W.}~\bibnamefont {Jin}}, \bibinfo {author} {\bibfnamefont
  {T.}~\bibnamefont {Bonnen}}, \bibinfo {author} {\bibfnamefont
  {A.}~\bibnamefont {Schuldenfrei}}, \bibinfo {author} {\bibfnamefont {W.~G.}\
  \bibnamefont {Cullen}}, \bibinfo {author} {\bibfnamefont {E.~D.}\
  \bibnamefont {Williams}}, \bibinfo {author} {\bibfnamefont {J.~E.}\
  \bibnamefont {Reutt-Robey}}, \ and\ \bibinfo {author} {\bibfnamefont {S.~W.}\
  \bibnamefont {Robey}},\ }\bibfield  {title} {\enquote {\bibinfo {title} {{C60
  cluster formation at interfaces with pentacene thin-film phases}},}\
  }\href@noop {} {\bibfield  {journal} {\bibinfo  {journal} {Applied Physics
  Letters}\ }\textbf {\bibinfo {volume} {95}},\ \bibinfo {pages} {213302}
  (\bibinfo {year} {2009})}\BibitemShut {NoStop}%
\bibitem [{\citenamefont {Huttner}, \citenamefont {Breuer},\ and\ \citenamefont
  {Witte}(2019)}]{Huttner2019}%
  \BibitemOpen
  \bibfield  {author} {\bibinfo {author} {\bibfnamefont {A.}~\bibnamefont
  {Huttner}}, \bibinfo {author} {\bibfnamefont {T.}~\bibnamefont {Breuer}}, \
  and\ \bibinfo {author} {\bibfnamefont {G.}~\bibnamefont {Witte}},\ }\bibfield
   {title} {\enquote {\bibinfo {title} {Controlling interface morphology and
  layer crystallinity in organic heterostructures: Microscopic view on c60
  island formation on pentacene buffer layers},}\ }\href@noop {} {\bibfield
  {journal} {\bibinfo  {journal} {ACS Applied Materials and Interfaces}\
  }\textbf {\bibinfo {volume} {11}},\ \bibinfo {pages} {35177--35184} (\bibinfo
  {year} {2019})}\BibitemShut {NoStop}%
\bibitem [{\citenamefont {Iwasawa}\ \emph {et~al.}(2020)\citenamefont
  {Iwasawa}, \citenamefont {Tsuruta}, \citenamefont {Nakayama}, \citenamefont
  {Sasaki}, \citenamefont {Hosokai}, \citenamefont {Lee}, \citenamefont
  {Fukumoto},\ and\ \citenamefont {Yamada}}]{Iwasawa2020}%
  \BibitemOpen
  \bibfield  {author} {\bibinfo {author} {\bibfnamefont {M.}~\bibnamefont
  {Iwasawa}}, \bibinfo {author} {\bibfnamefont {R.}~\bibnamefont {Tsuruta}},
  \bibinfo {author} {\bibfnamefont {Y.}~\bibnamefont {Nakayama}}, \bibinfo
  {author} {\bibfnamefont {M.}~\bibnamefont {Sasaki}}, \bibinfo {author}
  {\bibfnamefont {T.}~\bibnamefont {Hosokai}}, \bibinfo {author} {\bibfnamefont
  {S.}~\bibnamefont {Lee}}, \bibinfo {author} {\bibfnamefont {K.}~\bibnamefont
  {Fukumoto}}, \ and\ \bibinfo {author} {\bibfnamefont {Y.}~\bibnamefont
  {Yamada}},\ }\bibfield  {title} {\enquote {\bibinfo {title} {Exciton
  dissociation and electron transfer at a well-defined organic interface of an
  epitaxial {C60} layer on a pentacene single crystal},}\ }\href {\doibase
  10.1021/acs.jpcc.0c02796} {\bibfield  {journal} {\bibinfo  {journal} {The
  Journal of Physical Chemistry C}\ }\textbf {\bibinfo {volume} {124}},\
  \bibinfo {pages} {13572--13579} (\bibinfo {year} {2020})}\BibitemShut
  {NoStop}%
\bibitem [{\citenamefont {Mitsuta}\ \emph {et~al.}(2017)\citenamefont
  {Mitsuta}, \citenamefont {Miyadera}, \citenamefont {Ohashi}, \citenamefont
  {Zhou}, \citenamefont {Taima}, \citenamefont {Koganezawa}, \citenamefont
  {Yoshida},\ and\ \citenamefont {Tamura}}]{Mitsuta2017}%
  \BibitemOpen
  \bibfield  {author} {\bibinfo {author} {\bibfnamefont {H.}~\bibnamefont
  {Mitsuta}}, \bibinfo {author} {\bibfnamefont {T.}~\bibnamefont {Miyadera}},
  \bibinfo {author} {\bibfnamefont {N.}~\bibnamefont {Ohashi}}, \bibinfo
  {author} {\bibfnamefont {Y.}~\bibnamefont {Zhou}}, \bibinfo {author}
  {\bibfnamefont {T.}~\bibnamefont {Taima}}, \bibinfo {author} {\bibfnamefont
  {T.}~\bibnamefont {Koganezawa}}, \bibinfo {author} {\bibfnamefont
  {Y.}~\bibnamefont {Yoshida}}, \ and\ \bibinfo {author} {\bibfnamefont
  {M.}~\bibnamefont {Tamura}},\ }\bibfield  {title} {\enquote {\bibinfo {title}
  {Epitaxial growth of {C60} on rubrene single crystals for a highly ordered
  organic donor/acceptor interface},}\ }\href {\doibase
  10.1021/acs.cgd.7b00467} {\bibfield  {journal} {\bibinfo  {journal} {Crystal
  Growth \& Design}\ }\textbf {\bibinfo {volume} {17}},\ \bibinfo {pages}
  {4622--4627} (\bibinfo {year} {2017})}\BibitemShut {NoStop}%
\bibitem [{\citenamefont {Nakayama}\ \emph {et~al.}(2019)\citenamefont
  {Nakayama}, \citenamefont {Tsuruta}, \citenamefont {Moriya}, \citenamefont
  {Hikasa}, \citenamefont {Meissner}, \citenamefont {Yamaguchi}, \citenamefont
  {Mizuno}, \citenamefont {Suzuki}, \citenamefont {Koganezawa}, \citenamefont
  {Hosokai}, \citenamefont {Ueba},\ and\ \citenamefont {Kera}}]{Nakayama2019}%
  \BibitemOpen
  \bibfield  {author} {\bibinfo {author} {\bibfnamefont {Y.}~\bibnamefont
  {Nakayama}}, \bibinfo {author} {\bibfnamefont {R.}~\bibnamefont {Tsuruta}},
  \bibinfo {author} {\bibfnamefont {N.}~\bibnamefont {Moriya}}, \bibinfo
  {author} {\bibfnamefont {M.}~\bibnamefont {Hikasa}}, \bibinfo {author}
  {\bibfnamefont {M.}~\bibnamefont {Meissner}}, \bibinfo {author}
  {\bibfnamefont {T.}~\bibnamefont {Yamaguchi}}, \bibinfo {author}
  {\bibfnamefont {Y.}~\bibnamefont {Mizuno}}, \bibinfo {author} {\bibfnamefont
  {T.}~\bibnamefont {Suzuki}}, \bibinfo {author} {\bibfnamefont
  {T.}~\bibnamefont {Koganezawa}}, \bibinfo {author} {\bibfnamefont
  {T.}~\bibnamefont {Hosokai}}, \bibinfo {author} {\bibfnamefont
  {T.}~\bibnamefont {Ueba}}, \ and\ \bibinfo {author} {\bibfnamefont
  {S.}~\bibnamefont {Kera}},\ }\bibfield  {title} {\enquote {\bibinfo {title}
  {Widely dispersed intermolecular valence bands of epitaxially grown
  perfluoropentacene on pentacene single crystals},}\ }\href {\doibase
  10.1021/acs.jpclett.8b03866} {\bibfield  {journal} {\bibinfo  {journal} {The
  Journal of Physical Chemistry Letters}\ }\textbf {\bibinfo {volume} {10}},\
  \bibinfo {pages} {1312--1318} (\bibinfo {year} {2019})}\BibitemShut {NoStop}%
\bibitem [{\citenamefont {Hinderhofer}\ \emph {et~al.}(2011)\citenamefont
  {Hinderhofer}, \citenamefont {Hosokai}, \citenamefont {Frank}, \citenamefont
  {Nov{\'{a}}k}, \citenamefont {Gerlach},\ and\ \citenamefont
  {Schreiber}}]{Hinderhofer2011}%
  \BibitemOpen
  \bibfield  {author} {\bibinfo {author} {\bibfnamefont {A.}~\bibnamefont
  {Hinderhofer}}, \bibinfo {author} {\bibfnamefont {T.}~\bibnamefont
  {Hosokai}}, \bibinfo {author} {\bibfnamefont {C.}~\bibnamefont {Frank}},
  \bibinfo {author} {\bibfnamefont {J.}~\bibnamefont {Nov{\'{a}}k}}, \bibinfo
  {author} {\bibfnamefont {A.}~\bibnamefont {Gerlach}}, \ and\ \bibinfo
  {author} {\bibfnamefont {F.}~\bibnamefont {Schreiber}},\ }\bibfield  {title}
  {\enquote {\bibinfo {title} {{Templating effect for organic heterostructure
  film growth: Perfluoropentacene on diindenoperylene}},}\ }\href {\doibase
  10.1021/JP203003M/ASSET/IMAGES/MEDIUM/JP-2011-03003M_0004.GIF} {\bibfield
  {journal} {\bibinfo  {journal} {Journal of Physical Chemistry C}\ }\textbf
  {\bibinfo {volume} {115}},\ \bibinfo {pages} {16155--16160} (\bibinfo {year}
  {2011})}\BibitemShut {NoStop}%
\bibitem [{\citenamefont {Barrena}\ \emph {et~al.}(2006)\citenamefont
  {Barrena}, \citenamefont {{De Oteyza}}, \citenamefont {Sellner},
  \citenamefont {Dosch}, \citenamefont {Oss{o}},\ and\ \citenamefont
  {Struth}}]{Barrena2006}%
  \BibitemOpen
  \bibfield  {author} {\bibinfo {author} {\bibfnamefont {E.}~\bibnamefont
  {Barrena}}, \bibinfo {author} {\bibfnamefont {D.~G.}\ \bibnamefont {{De
  Oteyza}}}, \bibinfo {author} {\bibfnamefont {S.}~\bibnamefont {Sellner}},
  \bibinfo {author} {\bibfnamefont {H.}~\bibnamefont {Dosch}}, \bibinfo
  {author} {\bibfnamefont {J.~O.}\ \bibnamefont {Oss{o}}}, \ and\ \bibinfo
  {author} {\bibfnamefont {B.}~\bibnamefont {Struth}},\ }\bibfield  {title}
  {\enquote {\bibinfo {title} {{In situ study of the growth of nanodots in
  organic heteroepitaxy}},}\ }\href {\doibase 10.1103/PHYSREVLETT.97.076102}
  {\bibfield  {journal} {\bibinfo  {journal} {Physical Review Letters}\
  }\textbf {\bibinfo {volume} {97}},\ \bibinfo {pages} {076102} (\bibinfo
  {year} {2006})}\BibitemShut {NoStop}%
\bibitem [{\citenamefont {Takahashi}\ \emph {et~al.}(2021)\citenamefont
  {Takahashi}, \citenamefont {Izawa}, \citenamefont {Ohtsuka}, \citenamefont
  {Izumiseki}, \citenamefont {Tsuruta}, \citenamefont {Takeuchi}, \citenamefont
  {Gunjo}, \citenamefont {Nakanishi}, \citenamefont {Mase}, \citenamefont
  {Koganezawa}, \citenamefont {Momiyama}, \citenamefont {Hiramoto},\ and\
  \citenamefont {Nakayama}}]{Takahashi2021}%
  \BibitemOpen
  \bibfield  {author} {\bibinfo {author} {\bibfnamefont {K.}~\bibnamefont
  {Takahashi}}, \bibinfo {author} {\bibfnamefont {S.}~\bibnamefont {Izawa}},
  \bibinfo {author} {\bibfnamefont {N.}~\bibnamefont {Ohtsuka}}, \bibinfo
  {author} {\bibfnamefont {A.}~\bibnamefont {Izumiseki}}, \bibinfo {author}
  {\bibfnamefont {R.}~\bibnamefont {Tsuruta}}, \bibinfo {author} {\bibfnamefont
  {R.}~\bibnamefont {Takeuchi}}, \bibinfo {author} {\bibfnamefont
  {Y.}~\bibnamefont {Gunjo}}, \bibinfo {author} {\bibfnamefont
  {Y.}~\bibnamefont {Nakanishi}}, \bibinfo {author} {\bibfnamefont
  {K.}~\bibnamefont {Mase}}, \bibinfo {author} {\bibfnamefont {T.}~\bibnamefont
  {Koganezawa}}, \bibinfo {author} {\bibfnamefont {N.}~\bibnamefont
  {Momiyama}}, \bibinfo {author} {\bibfnamefont {M.}~\bibnamefont {Hiramoto}},
  \ and\ \bibinfo {author} {\bibfnamefont {Y.}~\bibnamefont {Nakayama}},\
  }\bibfield  {title} {\enquote {\bibinfo {title} {Quasi-homoepitaxial junction
  of organic semiconductors: A structurally seamless but electronically abrupt
  interface between rubrene and bis(trifluoromethyl)dimethylrubrene},}\ }\href
  {\doibase 10.1021/acs.jpclett.1c03094} {\bibfield  {journal} {\bibinfo
  {journal} {The Journal of Physical Chemistry Letters}\ }\textbf {\bibinfo
  {volume} {12}},\ \bibinfo {pages} {11430--11437} (\bibinfo {year}
  {2021})}\BibitemShut {NoStop}%
\bibitem [{\citenamefont {Gunjo}\ \emph {et~al.}(2021)\citenamefont {Gunjo},
  \citenamefont {Kamebuchi}, \citenamefont {Tsuruta}, \citenamefont {Iwashita},
  \citenamefont {Takahashi}, \citenamefont {Takeuchi}, \citenamefont {Kanai},
  \citenamefont {Koganezawa}, \citenamefont {Mase}, \citenamefont {Tadokoro},\
  and\ \citenamefont {Nakayama}}]{Gunjo2021}%
  \BibitemOpen
  \bibfield  {author} {\bibinfo {author} {\bibfnamefont {Y.}~\bibnamefont
  {Gunjo}}, \bibinfo {author} {\bibfnamefont {H.}~\bibnamefont {Kamebuchi}},
  \bibinfo {author} {\bibfnamefont {R.}~\bibnamefont {Tsuruta}}, \bibinfo
  {author} {\bibfnamefont {M.}~\bibnamefont {Iwashita}}, \bibinfo {author}
  {\bibfnamefont {K.}~\bibnamefont {Takahashi}}, \bibinfo {author}
  {\bibfnamefont {R.}~\bibnamefont {Takeuchi}}, \bibinfo {author}
  {\bibfnamefont {K.}~\bibnamefont {Kanai}}, \bibinfo {author} {\bibfnamefont
  {T.}~\bibnamefont {Koganezawa}}, \bibinfo {author} {\bibfnamefont
  {K.}~\bibnamefont {Mase}}, \bibinfo {author} {\bibfnamefont {M.}~\bibnamefont
  {Tadokoro}}, \ and\ \bibinfo {author} {\bibfnamefont {Y.}~\bibnamefont
  {Nakayama}},\ }\bibfield  {title} {\enquote {\bibinfo {title} {Interface
  structures and electronic states of epitaxial tetraazanaphthacene on
  single-crystal pentacene},}\ }\href {\doibase 10.3390/ma14051088} {\bibfield
  {journal} {\bibinfo  {journal} {Materials}\ }\textbf {\bibinfo {volume}
  {14}},\ \bibinfo {pages} {1088} (\bibinfo {year} {2021})}\BibitemShut
  {NoStop}%
\bibitem [{\citenamefont {Kattel}\ \emph {et~al.}(2017)\citenamefont {Kattel},
  \citenamefont {Wang}, \citenamefont {Kafle},\ and\ \citenamefont
  {Chan}}]{Kattel2017}%
  \BibitemOpen
  \bibfield  {author} {\bibinfo {author} {\bibfnamefont {B.}~\bibnamefont
  {Kattel}}, \bibinfo {author} {\bibfnamefont {T.}~\bibnamefont {Wang}},
  \bibinfo {author} {\bibfnamefont {T.~R.}\ \bibnamefont {Kafle}}, \ and\
  \bibinfo {author} {\bibfnamefont {W.~L.}\ \bibnamefont {Chan}},\ }\bibfield
  {title} {\enquote {\bibinfo {title} {{The thickness of the two-dimensional
  charge transfer state at the TTF-TCNQ interface}},}\ }\href {\doibase
  10.1016/J.ORGEL.2017.06.018} {\bibfield  {journal} {\bibinfo  {journal}
  {Organic Electronics}\ }\textbf {\bibinfo {volume} {48}},\ \bibinfo {pages}
  {371--376} (\bibinfo {year} {2017})}\BibitemShut {NoStop}%
\bibitem [{\citenamefont {D\"{o}ring}\ \emph {et~al.}(2019)\citenamefont
  {D\"{o}ring}, \citenamefont {Rosemann}, \citenamefont {Huttner},
  \citenamefont {Breuer}, \citenamefont {Witte},\ and\ \citenamefont
  {Chatterjee}}]{Doring2019}%
  \BibitemOpen
  \bibfield  {author} {\bibinfo {author} {\bibfnamefont {R.~C.}\ \bibnamefont
  {D\"{o}ring}}, \bibinfo {author} {\bibfnamefont {N.~W.}\ \bibnamefont
  {Rosemann}}, \bibinfo {author} {\bibfnamefont {A.}~\bibnamefont {Huttner}},
  \bibinfo {author} {\bibfnamefont {T.}~\bibnamefont {Breuer}}, \bibinfo
  {author} {\bibfnamefont {G.}~\bibnamefont {Witte}}, \ and\ \bibinfo {author}
  {\bibfnamefont {S.}~\bibnamefont {Chatterjee}},\ }\bibfield  {title}
  {\enquote {\bibinfo {title} {Charge-transfer processes and carrier dynamics
  at the pentacene-{C60} interface},}\ }\href {\doibase
  10.1088/1361-648x/aafea7} {\bibfield  {journal} {\bibinfo  {journal} {Journal
  of Physics: Condensed Matter}\ }\textbf {\bibinfo {volume} {31}},\ \bibinfo
  {pages} {134001} (\bibinfo {year} {2019})}\BibitemShut {NoStop}%
\bibitem [{\citenamefont {Dardzinski}\ \emph {et~al.}(2022)\citenamefont
  {Dardzinski}, \citenamefont {Yu}, \citenamefont {Moayedpour},\ and\
  \citenamefont {Marom}}]{Dardzinski2022}%
  \BibitemOpen
  \bibfield  {author} {\bibinfo {author} {\bibfnamefont {D.}~\bibnamefont
  {Dardzinski}}, \bibinfo {author} {\bibfnamefont {M.}~\bibnamefont {Yu}},
  \bibinfo {author} {\bibfnamefont {S.}~\bibnamefont {Moayedpour}}, \ and\
  \bibinfo {author} {\bibfnamefont {N.}~\bibnamefont {Marom}},\ }\bibfield
  {title} {\enquote {\bibinfo {title} {Best practices for first-principles
  simulations of epitaxial inorganic interfaces},}\ }\href {\doibase
  10.1088/1361-648x/ac577b} {\bibfield  {journal} {\bibinfo  {journal} {Journal
  of Physics: Condensed Matter}\ }\textbf {\bibinfo {volume} {34}},\ \bibinfo
  {pages} {233002} (\bibinfo {year} {2022})}\BibitemShut {NoStop}%
\bibitem [{\citenamefont {Moayedpour}\ \emph {et~al.}(2021)\citenamefont
  {Moayedpour}, \citenamefont {Dardzinski}, \citenamefont {Yang}, \citenamefont
  {Hwang},\ and\ \citenamefont {Marom}}]{Moayedpour2021}%
  \BibitemOpen
  \bibfield  {author} {\bibinfo {author} {\bibfnamefont {S.}~\bibnamefont
  {Moayedpour}}, \bibinfo {author} {\bibfnamefont {D.}~\bibnamefont
  {Dardzinski}}, \bibinfo {author} {\bibfnamefont {S.}~\bibnamefont {Yang}},
  \bibinfo {author} {\bibfnamefont {A.}~\bibnamefont {Hwang}}, \ and\ \bibinfo
  {author} {\bibfnamefont {N.}~\bibnamefont {Marom}},\ }\bibfield  {title}
  {\enquote {\bibinfo {title} {{Structure prediction of epitaxial inorganic
  interfaces by lattice and surface matching with Ogre}},}\ }\href {\doibase
  10.1063/5.0051343} {\bibfield  {journal} {\bibinfo  {journal} {The Journal of
  Chemical Physics}\ }\textbf {\bibinfo {volume} {155}},\ \bibinfo {pages}
  {034111} (\bibinfo {year} {2021})}\BibitemShut {NoStop}%
\bibitem [{\citenamefont {Mathew}\ \emph {et~al.}(2016)\citenamefont {Mathew},
  \citenamefont {Singh}, \citenamefont {Gabriel}, \citenamefont {Choudhary},
  \citenamefont {Sinnott}, \citenamefont {Davydov}, \citenamefont {Tavazza},\
  and\ \citenamefont {Hennig}}]{Mathew2016}%
  \BibitemOpen
  \bibfield  {author} {\bibinfo {author} {\bibfnamefont {K.}~\bibnamefont
  {Mathew}}, \bibinfo {author} {\bibfnamefont {A.~K.}\ \bibnamefont {Singh}},
  \bibinfo {author} {\bibfnamefont {J.~J.}\ \bibnamefont {Gabriel}}, \bibinfo
  {author} {\bibfnamefont {K.}~\bibnamefont {Choudhary}}, \bibinfo {author}
  {\bibfnamefont {S.~B.}\ \bibnamefont {Sinnott}}, \bibinfo {author}
  {\bibfnamefont {A.~V.}\ \bibnamefont {Davydov}}, \bibinfo {author}
  {\bibfnamefont {F.}~\bibnamefont {Tavazza}}, \ and\ \bibinfo {author}
  {\bibfnamefont {R.~G.}\ \bibnamefont {Hennig}},\ }\bibfield  {title}
  {\enquote {\bibinfo {title} {{MPInterfaces: A Materials Project based Python
  tool for high-throughput computational screening of interfacial systems}},}\
  }\href {\doibase 10.1016/j.commatsci.2016.05.020} {\bibfield  {journal}
  {\bibinfo  {journal} {Computational Materials Science}\ }\textbf {\bibinfo
  {volume} {122}},\ \bibinfo {pages} {183--190} (\bibinfo {year}
  {2016})}\BibitemShut {NoStop}%
\bibitem [{\citenamefont {Ding}\ \emph {et~al.}(2016)\citenamefont {Ding},
  \citenamefont {Dwaraknath}, \citenamefont {Garten}, \citenamefont {Ndione},
  \citenamefont {Ginley},\ and\ \citenamefont {Persson}}]{Ding2016}%
  \BibitemOpen
  \bibfield  {author} {\bibinfo {author} {\bibfnamefont {H.}~\bibnamefont
  {Ding}}, \bibinfo {author} {\bibfnamefont {S.~S.}\ \bibnamefont
  {Dwaraknath}}, \bibinfo {author} {\bibfnamefont {L.}~\bibnamefont {Garten}},
  \bibinfo {author} {\bibfnamefont {P.}~\bibnamefont {Ndione}}, \bibinfo
  {author} {\bibfnamefont {D.}~\bibnamefont {Ginley}}, \ and\ \bibinfo {author}
  {\bibfnamefont {K.~A.}\ \bibnamefont {Persson}},\ }\bibfield  {title}
  {\enquote {\bibinfo {title} {{Computational Approach for Epitaxial Polymorph
  Stabilization through Substrate Selection}},}\ }\href {\doibase
  10.1021/ACSAMI.6B01630/ASSET/IMAGES/ACSAMI.6B01630.SOCIAL.JPEG_V03}
  {\bibfield  {journal} {\bibinfo  {journal} {ACS Applied Materials and
  Interfaces}\ }\textbf {\bibinfo {volume} {8}},\ \bibinfo {pages}
  {13086--13093} (\bibinfo {year} {2016})}\BibitemShut {NoStop}%
\bibitem [{\citenamefont {Raclariu}\ \emph {et~al.}(2015)\citenamefont
  {Raclariu}, \citenamefont {Deshpande}, \citenamefont {Bruggemann},
  \citenamefont {Zhuge}, \citenamefont {Yu}, \citenamefont {Ratsch},\ and\
  \citenamefont {Shankar}}]{Raclariu2015}%
  \BibitemOpen
  \bibfield  {author} {\bibinfo {author} {\bibfnamefont {A.~M.}\ \bibnamefont
  {Raclariu}}, \bibinfo {author} {\bibfnamefont {S.}~\bibnamefont {Deshpande}},
  \bibinfo {author} {\bibfnamefont {J.}~\bibnamefont {Bruggemann}}, \bibinfo
  {author} {\bibfnamefont {W.}~\bibnamefont {Zhuge}}, \bibinfo {author}
  {\bibfnamefont {T.~H.}\ \bibnamefont {Yu}}, \bibinfo {author} {\bibfnamefont
  {C.}~\bibnamefont {Ratsch}}, \ and\ \bibinfo {author} {\bibfnamefont
  {S.}~\bibnamefont {Shankar}},\ }\bibfield  {title} {\enquote {\bibinfo
  {title} {{A fast method for predicting the formation of crystal interfaces
  and heterocrystals}},}\ }\href {\doibase 10.1016/j.commatsci.2015.05.023}
  {\bibfield  {journal} {\bibinfo  {journal} {Computational Materials Science}\
  }\textbf {\bibinfo {volume} {108}},\ \bibinfo {pages} {88--93} (\bibinfo
  {year} {2015})}\BibitemShut {NoStop}%
\bibitem [{\citenamefont {Sun}\ and\ \citenamefont {Ceder}(2013)}]{Sun2013}%
  \BibitemOpen
  \bibfield  {author} {\bibinfo {author} {\bibfnamefont {W.}~\bibnamefont
  {Sun}}\ and\ \bibinfo {author} {\bibfnamefont {G.}~\bibnamefont {Ceder}},\
  }\bibfield  {title} {\enquote {\bibinfo {title} {{Efficient creation and
  convergence of surface slabs}},}\ }\href {\doibase
  10.1016/j.susc.2013.05.016} {\bibfield  {journal} {\bibinfo  {journal}
  {Surface Science}\ }\textbf {\bibinfo {volume} {617}},\ \bibinfo {pages}
  {53--59} (\bibinfo {year} {2013})}\BibitemShut {NoStop}%
\bibitem [{\citenamefont {Gao}\ \emph {et~al.}(2019)\citenamefont {Gao},
  \citenamefont {Gao}, \citenamefont {Lu}, \citenamefont {Lv}, \citenamefont
  {Wang},\ and\ \citenamefont {Ma}}]{Gao2019}%
  \BibitemOpen
  \bibfield  {author} {\bibinfo {author} {\bibfnamefont {B.}~\bibnamefont
  {Gao}}, \bibinfo {author} {\bibfnamefont {P.}~\bibnamefont {Gao}}, \bibinfo
  {author} {\bibfnamefont {S.}~\bibnamefont {Lu}}, \bibinfo {author}
  {\bibfnamefont {J.}~\bibnamefont {Lv}}, \bibinfo {author} {\bibfnamefont
  {Y.}~\bibnamefont {Wang}}, \ and\ \bibinfo {author} {\bibfnamefont
  {Y.}~\bibnamefont {Ma}},\ }\bibfield  {title} {\enquote {\bibinfo {title}
  {{Interface structure prediction via CALYPSO method}},}\ }\href {\doibase
  10.1016/J.SCIB.2019.02.009} {\bibfield  {journal} {\bibinfo  {journal}
  {Science Bulletin}\ }\textbf {\bibinfo {volume} {64}},\ \bibinfo {pages}
  {301--309} (\bibinfo {year} {2019})}\BibitemShut {NoStop}%
\bibitem [{\citenamefont {Mannsfeld}\ and\ \citenamefont
  {Fritz}(2011)}]{Mannsfeld2011}%
  \BibitemOpen
  \bibfield  {author} {\bibinfo {author} {\bibfnamefont {S.~C.}\ \bibnamefont
  {Mannsfeld}}\ and\ \bibinfo {author} {\bibfnamefont {T.}~\bibnamefont
  {Fritz}},\ }\bibfield  {title} {\enquote {\bibinfo {title} {Advanced
  modelling of epitaxial ordering of organic layers on crystalline surfaces},}\
  }\href {\doibase 10.1142/S0217984906011189} {\bibfield  {journal} {\bibinfo
  {journal} {Modern Physics Letters B}\ }\textbf {\bibinfo {volume} {20}},\
  \bibinfo {pages} {585--605} (\bibinfo {year} {2011})}\BibitemShut {NoStop}%
\bibitem [{\citenamefont {Packwood}\ and\ \citenamefont
  {Hitosugi}(2017)}]{Packwood2017}%
  \BibitemOpen
  \bibfield  {author} {\bibinfo {author} {\bibfnamefont {D.~M.}\ \bibnamefont
  {Packwood}}\ and\ \bibinfo {author} {\bibfnamefont {T.}~\bibnamefont
  {Hitosugi}},\ }\bibfield  {title} {\enquote {\bibinfo {title} {{Rapid
  prediction of molecule arrangements on metal surfaces via Bayesian
  optimization}},}\ }\href {\doibase 10.7567/APEX.10.065502/XML} {\bibfield
  {journal} {\bibinfo  {journal} {Applied Physics Express}\ }\textbf {\bibinfo
  {volume} {10}},\ \bibinfo {pages} {065502} (\bibinfo {year}
  {2017})}\BibitemShut {NoStop}%
\bibitem [{\citenamefont {Packwood}, \citenamefont {Han},\ and\ \citenamefont
  {Hitosugi}(2017)}]{Packwood2017a}%
  \BibitemOpen
  \bibfield  {author} {\bibinfo {author} {\bibfnamefont {D.~M.}\ \bibnamefont
  {Packwood}}, \bibinfo {author} {\bibfnamefont {P.}~\bibnamefont {Han}}, \
  and\ \bibinfo {author} {\bibfnamefont {T.}~\bibnamefont {Hitosugi}},\
  }\bibfield  {title} {\enquote {\bibinfo {title} {{Chemical and entropic
  control on the molecular self-assembly process}},}\ }\href {\doibase
  10.1038/ncomms14463} {\bibfield  {journal} {\bibinfo  {journal} {Nature
  Communications}\ }\textbf {\bibinfo {volume} {8}},\ \bibinfo {pages} {1--8}
  (\bibinfo {year} {2017})}\BibitemShut {NoStop}%
\bibitem [{\citenamefont {Obersteiner}\ \emph {et~al.}(2017)\citenamefont
  {Obersteiner}, \citenamefont {Scherbela}, \citenamefont {H{\"{o}}rmann},
  \citenamefont {Wegner},\ and\ \citenamefont {Hofmann}}]{Obersteiner2017}%
  \BibitemOpen
  \bibfield  {author} {\bibinfo {author} {\bibfnamefont {V.}~\bibnamefont
  {Obersteiner}}, \bibinfo {author} {\bibfnamefont {M.}~\bibnamefont
  {Scherbela}}, \bibinfo {author} {\bibfnamefont {L.}~\bibnamefont
  {H{\"{o}}rmann}}, \bibinfo {author} {\bibfnamefont {D.}~\bibnamefont
  {Wegner}}, \ and\ \bibinfo {author} {\bibfnamefont {O.~T.}\ \bibnamefont
  {Hofmann}},\ }\bibfield  {title} {\enquote {\bibinfo {title} {{Structure
  Prediction for Surface-Induced Phases of Organic Monolayers: Overcoming the
  Combinatorial Bottleneck}},}\ }\href {\doibase
  10.1021/ACS.NANOLETT.7B01637/SUPPL_FILE/NL7B01637_SI_001.PDF} {\bibfield
  {journal} {\bibinfo  {journal} {Nano Letters}\ }\textbf {\bibinfo {volume}
  {17}},\ \bibinfo {pages} {4453--4460} (\bibinfo {year} {2017})}\BibitemShut
  {NoStop}%
\bibitem [{\citenamefont {Scherbela}\ \emph {et~al.}(2018)\citenamefont
  {Scherbela}, \citenamefont {H{\"{o}}rmann}, \citenamefont {Jeindl},
  \citenamefont {Obersteiner},\ and\ \citenamefont {Hofmann}}]{Scherbela2018}%
  \BibitemOpen
  \bibfield  {author} {\bibinfo {author} {\bibfnamefont {M.}~\bibnamefont
  {Scherbela}}, \bibinfo {author} {\bibfnamefont {L.}~\bibnamefont
  {H{\"{o}}rmann}}, \bibinfo {author} {\bibfnamefont {A.}~\bibnamefont
  {Jeindl}}, \bibinfo {author} {\bibfnamefont {V.}~\bibnamefont {Obersteiner}},
  \ and\ \bibinfo {author} {\bibfnamefont {O.~T.}\ \bibnamefont {Hofmann}},\
  }\bibfield  {title} {\enquote {\bibinfo {title} {{Charting the energy
  landscape of metal/organic interfaces via machine learning}},}\ }\href
  {\doibase 10.1103/PHYSREVMATERIALS.2.043803/FIGURES/8/MEDIUM} {\bibfield
  {journal} {\bibinfo  {journal} {Physical Review Materials}\ }\textbf
  {\bibinfo {volume} {2}},\ \bibinfo {pages} {043803} (\bibinfo {year}
  {2018})}\BibitemShut {NoStop}%
\bibitem [{\citenamefont {Han}, \citenamefont {Shen},\ and\ \citenamefont
  {Yi}(2015)}]{Han2015}%
  \BibitemOpen
  \bibfield  {author} {\bibinfo {author} {\bibfnamefont {G.}~\bibnamefont
  {Han}}, \bibinfo {author} {\bibfnamefont {X.}~\bibnamefont {Shen}}, \ and\
  \bibinfo {author} {\bibfnamefont {Y.}~\bibnamefont {Yi}},\ }\bibfield
  {title} {\enquote {\bibinfo {title} {Deposition growth and morphologies of
  {C60} on {DTDCTB} surfaces: An atomistic insight into the integrated impact
  of surface stability, landscape, and molecular orientation},}\ }\href
  {\doibase https://doi.org/10.1002/admi.201500329} {\bibfield  {journal}
  {\bibinfo  {journal} {Advanced Materials Interfaces}\ }\textbf {\bibinfo
  {volume} {2}},\ \bibinfo {pages} {1500329} (\bibinfo {year}
  {2015})}\BibitemShut {NoStop}%
\bibitem [{\citenamefont {Fu}\ \emph {et~al.}(2014)\citenamefont {Fu},
  \citenamefont {da~Silva~Filho}, \citenamefont {Sini}, \citenamefont {Asiri},
  \citenamefont {Aziz}, \citenamefont {Risko},\ and\ \citenamefont
  {Brédas}}]{Fu2014}%
  \BibitemOpen
  \bibfield  {author} {\bibinfo {author} {\bibfnamefont {Y.-T.}\ \bibnamefont
  {Fu}}, \bibinfo {author} {\bibfnamefont {D.~A.}\ \bibnamefont
  {da~Silva~Filho}}, \bibinfo {author} {\bibfnamefont {G.}~\bibnamefont
  {Sini}}, \bibinfo {author} {\bibfnamefont {A.~M.}\ \bibnamefont {Asiri}},
  \bibinfo {author} {\bibfnamefont {S.~G.}\ \bibnamefont {Aziz}}, \bibinfo
  {author} {\bibfnamefont {C.}~\bibnamefont {Risko}}, \ and\ \bibinfo {author}
  {\bibfnamefont {J.-L.}\ \bibnamefont {Brédas}},\ }\bibfield  {title}
  {\enquote {\bibinfo {title} {Structure and disorder in squaraine–{C60}
  organic solar cells: A theoretical description of molecular packing and
  electronic coupling at the donor–acceptor interface},}\ }\href {\doibase
  https://doi.org/10.1002/adfm.201303941} {\bibfield  {journal} {\bibinfo
  {journal} {Advanced Functional Materials}\ }\textbf {\bibinfo {volume}
  {24}},\ \bibinfo {pages} {3790--3798} (\bibinfo {year} {2014})}\BibitemShut
  {NoStop}%
\bibitem [{\citenamefont {Fu}, \citenamefont {Risko},\ and\ \citenamefont
  {Brédas}(2013)}]{Fu2013}%
  \BibitemOpen
  \bibfield  {author} {\bibinfo {author} {\bibfnamefont {Y.-T.}\ \bibnamefont
  {Fu}}, \bibinfo {author} {\bibfnamefont {C.}~\bibnamefont {Risko}}, \ and\
  \bibinfo {author} {\bibfnamefont {J.-L.}\ \bibnamefont {Brédas}},\
  }\bibfield  {title} {\enquote {\bibinfo {title} {Intermixing at the
  pentacene-fullerene bilayer interface: A molecular dynamics study},}\ }\href
  {\doibase https://doi.org/10.1002/adma.201203412} {\bibfield  {journal}
  {\bibinfo  {journal} {Advanced Materials}\ }\textbf {\bibinfo {volume}
  {25}},\ \bibinfo {pages} {878--882} (\bibinfo {year} {2013})}\BibitemShut
  {NoStop}%
\bibitem [{\citenamefont {Yoo}\ \emph {et~al.}(2019)\citenamefont {Yoo},
  \citenamefont {Song}, \citenamefont {Youn}, \citenamefont {Jeon},
  \citenamefont {Cho},\ and\ \citenamefont {Han}}]{Yoo2019}%
  \BibitemOpen
  \bibfield  {author} {\bibinfo {author} {\bibfnamefont {D.}~\bibnamefont
  {Yoo}}, \bibinfo {author} {\bibfnamefont {H.}~\bibnamefont {Song}}, \bibinfo
  {author} {\bibfnamefont {Y.}~\bibnamefont {Youn}}, \bibinfo {author}
  {\bibfnamefont {S.~H.}\ \bibnamefont {Jeon}}, \bibinfo {author}
  {\bibfnamefont {Y.}~\bibnamefont {Cho}}, \ and\ \bibinfo {author}
  {\bibfnamefont {S.}~\bibnamefont {Han}},\ }\bibfield  {title} {\enquote
  {\bibinfo {title} {A molecular dynamics study on the interface morphology of
  vapor-deposited amorphous organic thin films},}\ }\href {\doibase
  10.1039/C8CP05294H} {\bibfield  {journal} {\bibinfo  {journal} {Phys. Chem.
  Chem. Phys.}\ }\textbf {\bibinfo {volume} {21}},\ \bibinfo {pages}
  {1484--1490} (\bibinfo {year} {2019})}\BibitemShut {NoStop}%
\bibitem [{\citenamefont {Yang}\ \emph {et~al.}(2020)\citenamefont {Yang},
  \citenamefont {Bier}, \citenamefont {Wen}, \citenamefont {Zhan},
  \citenamefont {Moayedpour},\ and\ \citenamefont {Marom}}]{yang2020ogre}%
  \BibitemOpen
  \bibfield  {author} {\bibinfo {author} {\bibfnamefont {S.}~\bibnamefont
  {Yang}}, \bibinfo {author} {\bibfnamefont {I.}~\bibnamefont {Bier}}, \bibinfo
  {author} {\bibfnamefont {W.}~\bibnamefont {Wen}}, \bibinfo {author}
  {\bibfnamefont {J.}~\bibnamefont {Zhan}}, \bibinfo {author} {\bibfnamefont
  {S.}~\bibnamefont {Moayedpour}}, \ and\ \bibinfo {author} {\bibfnamefont
  {N.}~\bibnamefont {Marom}},\ }\bibfield  {title} {\enquote {\bibinfo {title}
  {Ogre: A python package for molecular crystal surface generation with
  applications to surface energy and crystal habit prediction},}\ }\href@noop
  {} {\bibfield  {journal} {\bibinfo  {journal} {The Journal of Chemical
  Physics}\ }\textbf {\bibinfo {volume} {152}},\ \bibinfo {pages} {244122}
  (\bibinfo {year} {2020})}\BibitemShut {NoStop}%
\bibitem [{\citenamefont {Zur}\ and\ \citenamefont {McGill}(1998)}]{Zur1998}%
  \BibitemOpen
  \bibfield  {author} {\bibinfo {author} {\bibfnamefont {A.}~\bibnamefont
  {Zur}}\ and\ \bibinfo {author} {\bibfnamefont {T.~C.}\ \bibnamefont
  {McGill}},\ }\bibfield  {title} {\enquote {\bibinfo {title} {{Lattice match:
  An application to heteroepitaxy}},}\ }\href {\doibase 10.1063/1.333084}
  {\bibfield  {journal} {\bibinfo  {journal} {Journal of Applied Physics}\
  }\textbf {\bibinfo {volume} {55}},\ \bibinfo {pages} {378} (\bibinfo {year}
  {1998})}\BibitemShut {NoStop}%
\bibitem [{\citenamefont {Spackman}\ and\ \citenamefont
  {Jayatilaka}(2009)}]{Spackman2009}%
  \BibitemOpen
  \bibfield  {author} {\bibinfo {author} {\bibfnamefont {M.~A.}\ \bibnamefont
  {Spackman}}\ and\ \bibinfo {author} {\bibfnamefont {D.}~\bibnamefont
  {Jayatilaka}},\ }\bibfield  {title} {\enquote {\bibinfo {title} {{Hirshfeld
  surface analysis}},}\ }\href {\doibase 10.1039/B818330A} {\bibfield
  {journal} {\bibinfo  {journal} {CrystEngComm}\ }\textbf {\bibinfo {volume}
  {11}},\ \bibinfo {pages} {19--32} (\bibinfo {year} {2009})}\BibitemShut
  {NoStop}%
\bibitem [{\citenamefont {Smith}, \citenamefont {Isayev},\ and\ \citenamefont
  {Roitberg}(2017)}]{Smith2017}%
  \BibitemOpen
  \bibfield  {author} {\bibinfo {author} {\bibfnamefont {J.~S.}\ \bibnamefont
  {Smith}}, \bibinfo {author} {\bibfnamefont {O.}~\bibnamefont {Isayev}}, \
  and\ \bibinfo {author} {\bibfnamefont {A.~E.}\ \bibnamefont {Roitberg}},\
  }\bibfield  {title} {\enquote {\bibinfo {title} {{ANI-1: an extensible neural
  network potential with DFT accuracy at force field computational cost}},}\
  }\href {\doibase 10.1039/C6SC05720A} {\bibfield  {journal} {\bibinfo
  {journal} {Chemical Science}\ }\textbf {\bibinfo {volume} {8}},\ \bibinfo
  {pages} {3192--3203} (\bibinfo {year} {2017})}\BibitemShut {NoStop}%
\bibitem [{\citenamefont {Smith}\ \emph {et~al.}(2020)\citenamefont {Smith},
  \citenamefont {Zubatyuk}, \citenamefont {Nebgen}, \citenamefont {Lubbers},
  \citenamefont {Barros}, \citenamefont {Roitberg}, \citenamefont {Isayev},\
  and\ \citenamefont {Tretiak}}]{Smith2020}%
  \BibitemOpen
  \bibfield  {author} {\bibinfo {author} {\bibfnamefont {J.~S.}\ \bibnamefont
  {Smith}}, \bibinfo {author} {\bibfnamefont {R.}~\bibnamefont {Zubatyuk}},
  \bibinfo {author} {\bibfnamefont {B.}~\bibnamefont {Nebgen}}, \bibinfo
  {author} {\bibfnamefont {N.}~\bibnamefont {Lubbers}}, \bibinfo {author}
  {\bibfnamefont {K.}~\bibnamefont {Barros}}, \bibinfo {author} {\bibfnamefont
  {A.~E.}\ \bibnamefont {Roitberg}}, \bibinfo {author} {\bibfnamefont
  {O.}~\bibnamefont {Isayev}}, \ and\ \bibinfo {author} {\bibfnamefont
  {S.}~\bibnamefont {Tretiak}},\ }\bibfield  {title} {\enquote {\bibinfo
  {title} {{The ANI-1ccx and ANI-1x data sets, coupled-cluster and density
  functional theory properties for molecules}},}\ }\href {\doibase
  10.1038/s41597-020-0473-z} {\bibfield  {journal} {\bibinfo  {journal}
  {Scientific Data}\ }\textbf {\bibinfo {volume} {7}},\ \bibinfo {pages}
  {1--10} (\bibinfo {year} {2020})}\BibitemShut {NoStop}%
\bibitem [{\citenamefont {Smith}\ \emph {et~al.}(2018)\citenamefont {Smith},
  \citenamefont {Nebgen}, \citenamefont {Lubbers}, \citenamefont {Isayev},\
  and\ \citenamefont {Roitberg}}]{Smith2018}%
  \BibitemOpen
  \bibfield  {author} {\bibinfo {author} {\bibfnamefont {J.~S.}\ \bibnamefont
  {Smith}}, \bibinfo {author} {\bibfnamefont {B.}~\bibnamefont {Nebgen}},
  \bibinfo {author} {\bibfnamefont {N.}~\bibnamefont {Lubbers}}, \bibinfo
  {author} {\bibfnamefont {O.}~\bibnamefont {Isayev}}, \ and\ \bibinfo {author}
  {\bibfnamefont {A.~E.}\ \bibnamefont {Roitberg}},\ }\bibfield  {title}
  {\enquote {\bibinfo {title} {{Less is more: Sampling chemical space with
  active learning}},}\ }\href {\doibase 10.1063/1.5023802} {\bibfield
  {journal} {\bibinfo  {journal} {The Journal of Chemical Physics}\ }\textbf
  {\bibinfo {volume} {148}},\ \bibinfo {pages} {241733} (\bibinfo {year}
  {2018})}\BibitemShut {NoStop}%
\bibitem [{\citenamefont {Smith}\ \emph {et~al.}(2016)\citenamefont {Smith},
  \citenamefont {Burns}, \citenamefont {Patkowski},\ and\ \citenamefont
  {Sherrill}}]{Smith2016}%
  \BibitemOpen
  \bibfield  {author} {\bibinfo {author} {\bibfnamefont {D.~G.}\ \bibnamefont
  {Smith}}, \bibinfo {author} {\bibfnamefont {L.~A.}\ \bibnamefont {Burns}},
  \bibinfo {author} {\bibfnamefont {K.}~\bibnamefont {Patkowski}}, \ and\
  \bibinfo {author} {\bibfnamefont {C.~D.}\ \bibnamefont {Sherrill}},\
  }\bibfield  {title} {\enquote {\bibinfo {title} {{Revised Damping Parameters
  for the D3 Dispersion Correction to Density Functional Theory}},}\ }\href
  {\doibase 10.1021/ACS.JPCLETT.6B00780/SUPPL_FILE/JZ6B00780_SI_002.ZIP}
  {\bibfield  {journal} {\bibinfo  {journal} {Journal of Physical Chemistry
  Letters}\ }\textbf {\bibinfo {volume} {7}},\ \bibinfo {pages} {2197--2203}
  (\bibinfo {year} {2016})}\BibitemShut {NoStop}%
\bibitem [{\citenamefont {Grimme}, \citenamefont {Ehrlich},\ and\ \citenamefont
  {Goerigk}(2011)}]{Grimme2011}%
  \BibitemOpen
  \bibfield  {author} {\bibinfo {author} {\bibfnamefont {S.}~\bibnamefont
  {Grimme}}, \bibinfo {author} {\bibfnamefont {S.}~\bibnamefont {Ehrlich}}, \
  and\ \bibinfo {author} {\bibfnamefont {L.}~\bibnamefont {Goerigk}},\
  }\bibfield  {title} {\enquote {\bibinfo {title} {{Effect of the damping
  function in dispersion corrected density functional theory}},}\ }\href
  {\doibase 10.1002/JCC.21759} {\bibfield  {journal} {\bibinfo  {journal}
  {Journal of Computational Chemistry}\ }\textbf {\bibinfo {volume} {32}},\
  \bibinfo {pages} {1456--1465} (\bibinfo {year} {2011})}\BibitemShut {NoStop}%
\bibitem [{\citenamefont {Ehrlich}\ \emph {et~al.}(2011)\citenamefont
  {Ehrlich}, \citenamefont {Moellmann}, \citenamefont {Reckien}, \citenamefont
  {Bredow},\ and\ \citenamefont {Grimme}}]{Ehrlich2011}%
  \BibitemOpen
  \bibfield  {author} {\bibinfo {author} {\bibfnamefont {S.}~\bibnamefont
  {Ehrlich}}, \bibinfo {author} {\bibfnamefont {J.}~\bibnamefont {Moellmann}},
  \bibinfo {author} {\bibfnamefont {W.}~\bibnamefont {Reckien}}, \bibinfo
  {author} {\bibfnamefont {T.}~\bibnamefont {Bredow}}, \ and\ \bibinfo {author}
  {\bibfnamefont {S.}~\bibnamefont {Grimme}},\ }\bibfield  {title} {\enquote
  {\bibinfo {title} {{System-Dependent Dispersion Coefficients for the DFT-D3
  Treatment of Adsorption Processes on Ionic Surfaces}},}\ }\href {\doibase
  10.1002/CPHC.201100521} {\bibfield  {journal} {\bibinfo  {journal}
  {ChemPhysChem}\ }\textbf {\bibinfo {volume} {12}},\ \bibinfo {pages}
  {3414--3420} (\bibinfo {year} {2011})}\BibitemShut {NoStop}%
\bibitem [{\citenamefont {Murdey}\ and\ \citenamefont
  {Salaneck}(2005)}]{Murdey2005}%
  \BibitemOpen
  \bibfield  {author} {\bibinfo {author} {\bibfnamefont {R.~J.}\ \bibnamefont
  {Murdey}}\ and\ \bibinfo {author} {\bibfnamefont {W.~R.}\ \bibnamefont
  {Salaneck}},\ }\bibfield  {title} {\enquote {\bibinfo {title} {Charge
  injection barrier heights across multilayer organic thin films},}\ }\href
  {\doibase 10.1143/jjap.44.3751} {\bibfield  {journal} {\bibinfo  {journal}
  {Japanese Journal of Applied Physics}\ }\textbf {\bibinfo {volume} {44}},\
  \bibinfo {pages} {3751--3756} (\bibinfo {year} {2005})}\BibitemShut {NoStop}%
\bibitem [{\citenamefont {Braun}\ \emph {et~al.}(2010)\citenamefont {Braun},
  \citenamefont {Liu}, \citenamefont {Salaneck},\ and\ \citenamefont
  {Fahlman}}]{BRAUN2010}%
  \BibitemOpen
  \bibfield  {author} {\bibinfo {author} {\bibfnamefont {S.}~\bibnamefont
  {Braun}}, \bibinfo {author} {\bibfnamefont {X.}~\bibnamefont {Liu}}, \bibinfo
  {author} {\bibfnamefont {W.}~\bibnamefont {Salaneck}}, \ and\ \bibinfo
  {author} {\bibfnamefont {M.}~\bibnamefont {Fahlman}},\ }\bibfield  {title}
  {\enquote {\bibinfo {title} {Fermi level equilibrium at donor–acceptor
  interfaces in multi-layered thin film stack of ttf and tcnq},}\ }\href
  {\doibase https://doi.org/10.1016/j.orgel.2009.10.018} {\bibfield  {journal}
  {\bibinfo  {journal} {Organic Electronics}\ }\textbf {\bibinfo {volume}
  {11}},\ \bibinfo {pages} {212--217} (\bibinfo {year} {2010})}\BibitemShut
  {NoStop}%
\bibitem [{\citenamefont {Beltr{\'{a}}n}\ \emph {et~al.}(2013)\citenamefont
  {Beltr{\'{a}}n}, \citenamefont {Flores}, \citenamefont {Mart{\'{i}}nez},\
  and\ \citenamefont {Ortega}}]{Beltran2013}%
  \BibitemOpen
  \bibfield  {author} {\bibinfo {author} {\bibfnamefont {J.~I.}\ \bibnamefont
  {Beltr{\'{a}}n}}, \bibinfo {author} {\bibfnamefont {F.}~\bibnamefont
  {Flores}}, \bibinfo {author} {\bibfnamefont {J.~I.}\ \bibnamefont
  {Mart{\'{i}}nez}}, \ and\ \bibinfo {author} {\bibfnamefont {J.}~\bibnamefont
  {Ortega}},\ }\bibfield  {title} {\enquote {\bibinfo {title} {{Energy level
  alignment in organic-organic heterojunctions: The TTF/TCNQ interface}},}\
  }\href {\doibase 10.1021/JP306079T/ASSET/IMAGES/JP306079T.SOCIAL.JPEG_V03}
  {\bibfield  {journal} {\bibinfo  {journal} {Journal of Physical Chemistry C}\
  }\textbf {\bibinfo {volume} {117}},\ \bibinfo {pages} {3888--3894} (\bibinfo
  {year} {2013})}\BibitemShut {NoStop}%
\bibitem [{\citenamefont {Atalla}\ \emph {et~al.}(2016)\citenamefont {Atalla},
  \citenamefont {Zhang}, \citenamefont {Hofmann}, \citenamefont {Ren},
  \citenamefont {Rinke},\ and\ \citenamefont {Scheffler}}]{Atalla2016}%
  \BibitemOpen
  \bibfield  {author} {\bibinfo {author} {\bibfnamefont {V.}~\bibnamefont
  {Atalla}}, \bibinfo {author} {\bibfnamefont {I.~Y.}\ \bibnamefont {Zhang}},
  \bibinfo {author} {\bibfnamefont {O.~T.}\ \bibnamefont {Hofmann}}, \bibinfo
  {author} {\bibfnamefont {X.}~\bibnamefont {Ren}}, \bibinfo {author}
  {\bibfnamefont {P.}~\bibnamefont {Rinke}}, \ and\ \bibinfo {author}
  {\bibfnamefont {M.}~\bibnamefont {Scheffler}},\ }\bibfield  {title} {\enquote
  {\bibinfo {title} {{Enforcing the linear behavior of the total energy with
  hybrid functionals: Implications for charge transfer, interaction energies,
  and the random-phase approximation}},}\ }\href {\doibase
  10.1103/PHYSREVB.94.035140/FIGURES/13/MEDIUM} {\bibfield  {journal} {\bibinfo
   {journal} {Physical Review B}\ }\textbf {\bibinfo {volume} {94}},\ \bibinfo
  {pages} {035140} (\bibinfo {year} {2016})}\BibitemShut {NoStop}%
\bibitem [{\citenamefont {Ando}\ \emph {et~al.}(2008)\citenamefont {Ando},
  \citenamefont {Fowler}, \citenamefont {Stern}, \citenamefont {Kistenmacher},
  \citenamefont {Phillips}, \citenamefont {Cowan}, \citenamefont {Endo},
  \citenamefont {Iye},\ and\ \citenamefont {Brinkman}}]{Ando2008}%
  \BibitemOpen
  \bibfield  {author} {\bibinfo {author} {\bibfnamefont {T.}~\bibnamefont
  {Ando}}, \bibinfo {author} {\bibfnamefont {A.~B.}\ \bibnamefont {Fowler}},
  \bibinfo {author} {\bibfnamefont {F.}~\bibnamefont {Stern}}, \bibinfo
  {author} {\bibfnamefont {T.~J.}\ \bibnamefont {Kistenmacher}}, \bibinfo
  {author} {\bibfnamefont {T.~E.}\ \bibnamefont {Phillips}}, \bibinfo {author}
  {\bibfnamefont {D.~O.}\ \bibnamefont {Cowan}}, \bibinfo {author}
  {\bibfnamefont {A.}~\bibnamefont {Endo}}, \bibinfo {author} {\bibfnamefont
  {Y.}~\bibnamefont {Iye}}, \ and\ \bibinfo {author} {\bibfnamefont
  {A.}~\bibnamefont {Brinkman}},\ }\bibfield  {title} {\enquote {\bibinfo
  {title} {{When TTF met TCNQ}},}\ }\href {\doibase 10.1038/nmat2211}
  {\bibfield  {journal} {\bibinfo  {journal} {Nature Materials}\ }\textbf
  {\bibinfo {volume} {7}},\ \bibinfo {pages} {520--521} (\bibinfo {year}
  {2008})}\BibitemShut {NoStop}%
\bibitem [{\citenamefont {Cohen}\ \emph {et~al.}(1974)\citenamefont {Cohen},
  \citenamefont {Coleman}, \citenamefont {Garito},\ and\ \citenamefont
  {Heeger}}]{Cohen1974}%
  \BibitemOpen
  \bibfield  {author} {\bibinfo {author} {\bibfnamefont {M.~J.}\ \bibnamefont
  {Cohen}}, \bibinfo {author} {\bibfnamefont {L.~B.}\ \bibnamefont {Coleman}},
  \bibinfo {author} {\bibfnamefont {A.~F.}\ \bibnamefont {Garito}}, \ and\
  \bibinfo {author} {\bibfnamefont {A.~J.}\ \bibnamefont {Heeger}},\ }\bibfield
   {title} {\enquote {\bibinfo {title} {{Electrical conductivity of
  tetrathiofulvalinium tetracyanoquinodimethan (TTF) (TCNQ)}},}\ }\href
  {\doibase 10.1103/PhysRevB.10.1298} {\bibfield  {journal} {\bibinfo
  {journal} {Physical Review B}\ }\textbf {\bibinfo {volume} {10}},\ \bibinfo
  {pages} {1298} (\bibinfo {year} {1974})}\BibitemShut {NoStop}%
\bibitem [{\citenamefont {Ferraris}\ \emph {et~al.}(2002)\citenamefont
  {Ferraris}, \citenamefont {Cowan}, \citenamefont {Walatka},\ and\
  \citenamefont {Perlstein}}]{Ferraris2002}%
  \BibitemOpen
  \bibfield  {author} {\bibinfo {author} {\bibfnamefont {J.}~\bibnamefont
  {Ferraris}}, \bibinfo {author} {\bibfnamefont {D.~O.}\ \bibnamefont {Cowan}},
  \bibinfo {author} {\bibfnamefont {V.}~\bibnamefont {Walatka}}, \ and\
  \bibinfo {author} {\bibfnamefont {J.~H.}\ \bibnamefont {Perlstein}},\
  }\bibfield  {title} {\enquote {\bibinfo {title} {{Electron transfer in a new
  highly conducting donor-acceptor complex}},}\ }\href {\doibase
  10.1021/JA00784A066} {\bibfield  {journal} {\bibinfo  {journal} {Journal of
  the American Chemical Society}\ }\textbf {\bibinfo {volume} {95}},\ \bibinfo
  {pages} {948--949} (\bibinfo {year} {2002})}\BibitemShut {NoStop}%
\bibitem [{\citenamefont {Alves}, \citenamefont {Pinto},\ and\ \citenamefont
  {Ma{\c{c}}{\^{O}}as}(2013)}]{Alves2013}%
  \BibitemOpen
  \bibfield  {author} {\bibinfo {author} {\bibfnamefont {H.}~\bibnamefont
  {Alves}}, \bibinfo {author} {\bibfnamefont {R.~M.}\ \bibnamefont {Pinto}}, \
  and\ \bibinfo {author} {\bibfnamefont {E.~S.}\ \bibnamefont
  {Ma{\c{c}}{\^{O}}as}},\ }\bibfield  {title} {\enquote {\bibinfo {title}
  {{Photoconductive response in organic charge transfer interfaces with high
  quantum efficiency}},}\ }\href {\doibase 10.1038/NCOMMS2890} {\bibfield
  {journal} {\bibinfo  {journal} {Nature Communications}\ }\textbf {\bibinfo
  {volume} {4}},\ \bibinfo {pages} {1842} (\bibinfo {year} {2013})}\BibitemShut
  {NoStop}%
\bibitem [{\citenamefont {Krupskaya}\ \emph {et~al.}(2016)\citenamefont
  {Krupskaya}, \citenamefont {R{\"{u}}ckerl}, \citenamefont {Knupfer},
  \citenamefont {Morpurgo}, \citenamefont {Krupskaya}, \citenamefont
  {Morpurgo}, \citenamefont {R{\"{u}}ckerl},\ and\ \citenamefont
  {Knupfer}}]{Krupskaya2016}%
  \BibitemOpen
  \bibfield  {author} {\bibinfo {author} {\bibfnamefont {Y.}~\bibnamefont
  {Krupskaya}}, \bibinfo {author} {\bibfnamefont {F.}~\bibnamefont
  {R{\"{u}}ckerl}}, \bibinfo {author} {\bibfnamefont {M.}~\bibnamefont
  {Knupfer}}, \bibinfo {author} {\bibfnamefont {A.~F.}\ \bibnamefont
  {Morpurgo}}, \bibinfo {author} {\bibfnamefont {Y.}~\bibnamefont {Krupskaya}},
  \bibinfo {author} {\bibfnamefont {A.~F.}\ \bibnamefont {Morpurgo}}, \bibinfo
  {author} {\bibfnamefont {F.}~\bibnamefont {R{\"{u}}ckerl}}, \ and\ \bibinfo
  {author} {\bibfnamefont {M.}~\bibnamefont {Knupfer}},\ }\bibfield  {title}
  {\enquote {\bibinfo {title} {{Charge Transfer, Band-Like Transport, and
  Magnetic Ions at F16CoPc/Rubrene Interfaces}},}\ }\href {\doibase
  10.1002/ADMI.201500863} {\bibfield  {journal} {\bibinfo  {journal} {Advanced
  Materials Interfaces}\ }\textbf {\bibinfo {volume} {3}},\ \bibinfo {pages}
  {1500863} (\bibinfo {year} {2016})}\BibitemShut {NoStop}%
\bibitem [{\citenamefont {Krupskaya}, \citenamefont {Lezama},\ and\
  \citenamefont {Morpurgo}(2016)}]{Krupskaya2016a}%
  \BibitemOpen
  \bibfield  {author} {\bibinfo {author} {\bibfnamefont {Y.}~\bibnamefont
  {Krupskaya}}, \bibinfo {author} {\bibfnamefont {I.~G.}\ \bibnamefont
  {Lezama}}, \ and\ \bibinfo {author} {\bibfnamefont {A.~F.}\ \bibnamefont
  {Morpurgo}},\ }\bibfield  {title} {\enquote {\bibinfo {title} {{Tuning the
  Charge Transfer in Fx-TCNQ/Rubrene Single-Crystal Interfaces}},}\ }\href
  {\doibase 10.1002/ADFM.201502082} {\bibfield  {journal} {\bibinfo  {journal}
  {Advanced Functional Materials}\ }\textbf {\bibinfo {volume} {26}},\ \bibinfo
  {pages} {2334--2340} (\bibinfo {year} {2016})}\BibitemShut {NoStop}%
\bibitem [{\citenamefont {Takahashi}\ \emph {et~al.}(2012)\citenamefont
  {Takahashi}, \citenamefont {Hayakawa}, \citenamefont {Naito},\ and\
  \citenamefont {Inabe}}]{Takahashi2012}%
  \BibitemOpen
  \bibfield  {author} {\bibinfo {author} {\bibfnamefont {Y.}~\bibnamefont
  {Takahashi}}, \bibinfo {author} {\bibfnamefont {K.}~\bibnamefont {Hayakawa}},
  \bibinfo {author} {\bibfnamefont {T.}~\bibnamefont {Naito}}, \ and\ \bibinfo
  {author} {\bibfnamefont {T.}~\bibnamefont {Inabe}},\ }\bibfield  {title}
  {\enquote {\bibinfo {title} {{What happens at the interface between TTF and
  TCNQ crystals (TTF = tetrathiafulvalene and TCNQ =
  7,7,8,8-tetracyanoquinodimethane)?}}}\ }\href {\doibase
  10.1021/JP2074368/SUPPL_FILE/JP2074368_SI_002.PDF} {\bibfield  {journal}
  {\bibinfo  {journal} {Journal of Physical Chemistry C}\ }\textbf {\bibinfo
  {volume} {116}},\ \bibinfo {pages} {700--703} (\bibinfo {year}
  {2012})}\BibitemShut {NoStop}%
\bibitem [{\citenamefont {Mathis}\ \emph {et~al.}(2012)\citenamefont {Mathis},
  \citenamefont {Mattenberger}, \citenamefont {Moll},\ and\ \citenamefont
  {Batlogg}}]{Mathis2012}%
  \BibitemOpen
  \bibfield  {author} {\bibinfo {author} {\bibfnamefont {T.}~\bibnamefont
  {Mathis}}, \bibinfo {author} {\bibfnamefont {K.}~\bibnamefont
  {Mattenberger}}, \bibinfo {author} {\bibfnamefont {P.}~\bibnamefont {Moll}},
  \ and\ \bibinfo {author} {\bibfnamefont {B.}~\bibnamefont {Batlogg}},\
  }\bibfield  {title} {\enquote {\bibinfo {title} {{Tetrathiofulvalene and
  tetracyanoquinodimethane crystals: Conducting surface versus interface}},}\
  }\href {\doibase 10.1063/1.4731244} {\bibfield  {journal} {\bibinfo
  {journal} {Applied Physics Letters}\ }\textbf {\bibinfo {volume} {101}},\
  \bibinfo {pages} {023302} (\bibinfo {year} {2012})}\BibitemShut {NoStop}%
\bibitem [{\citenamefont {Lezama}\ \emph {et~al.}(2012)\citenamefont {Lezama},
  \citenamefont {Nakano}, \citenamefont {Minder}, \citenamefont {Chen},
  \citenamefont {{Di Girolamo}}, \citenamefont {Facchetti},\ and\ \citenamefont
  {Morpurgo}}]{Lezama2012}%
  \BibitemOpen
  \bibfield  {author} {\bibinfo {author} {\bibfnamefont {I.~G.}\ \bibnamefont
  {Lezama}}, \bibinfo {author} {\bibfnamefont {M.}~\bibnamefont {Nakano}},
  \bibinfo {author} {\bibfnamefont {N.~A.}\ \bibnamefont {Minder}}, \bibinfo
  {author} {\bibfnamefont {Z.}~\bibnamefont {Chen}}, \bibinfo {author}
  {\bibfnamefont {F.~V.}\ \bibnamefont {{Di Girolamo}}}, \bibinfo {author}
  {\bibfnamefont {A.}~\bibnamefont {Facchetti}}, \ and\ \bibinfo {author}
  {\bibfnamefont {A.~F.}\ \bibnamefont {Morpurgo}},\ }\bibfield  {title}
  {\enquote {\bibinfo {title} {{Single-crystal organic charge-transfer
  interfaces probed using Schottky-gated heterostructures}},}\ }\href {\doibase
  10.1038/nmat3383} {\bibfield  {journal} {\bibinfo  {journal} {Nature
  Materials}\ }\textbf {\bibinfo {volume} {11}},\ \bibinfo {pages} {788--794}
  (\bibinfo {year} {2012})}\BibitemShut {NoStop}%
\bibitem [{\citenamefont {Calhoun}\ \emph {et~al.}(2007)\citenamefont
  {Calhoun}, \citenamefont {Sanchez}, \citenamefont {Olaya}, \citenamefont
  {Gershenson},\ and\ \citenamefont {Podzorov}}]{Calhoun2007}%
  \BibitemOpen
  \bibfield  {author} {\bibinfo {author} {\bibfnamefont {M.~F.}\ \bibnamefont
  {Calhoun}}, \bibinfo {author} {\bibfnamefont {J.}~\bibnamefont {Sanchez}},
  \bibinfo {author} {\bibfnamefont {D.}~\bibnamefont {Olaya}}, \bibinfo
  {author} {\bibfnamefont {M.~E.}\ \bibnamefont {Gershenson}}, \ and\ \bibinfo
  {author} {\bibfnamefont {V.}~\bibnamefont {Podzorov}},\ }\bibfield  {title}
  {\enquote {\bibinfo {title} {{Electronic functionalization of the surface
  of organic semiconductors with self-assembled monolayers}},}\ }\href
  {\doibase 10.1038/nmat2059} {\bibfield  {journal} {\bibinfo  {journal}
  {Nature Materials}\ }\textbf {\bibinfo {volume} {7}},\ \bibinfo {pages}
  {84--89} (\bibinfo {year} {2007})}\BibitemShut {NoStop}%
\bibitem [{\citenamefont {Alves}\ \emph {et~al.}(2008)\citenamefont {Alves},
  \citenamefont {Molinari}, \citenamefont {Xie},\ and\ \citenamefont
  {Morpurgo}}]{Alves2008}%
  \BibitemOpen
  \bibfield  {author} {\bibinfo {author} {\bibfnamefont {H.}~\bibnamefont
  {Alves}}, \bibinfo {author} {\bibfnamefont {A.~S.}\ \bibnamefont {Molinari}},
  \bibinfo {author} {\bibfnamefont {H.}~\bibnamefont {Xie}}, \ and\ \bibinfo
  {author} {\bibfnamefont {A.~F.}\ \bibnamefont {Morpurgo}},\ }\bibfield
  {title} {\enquote {\bibinfo {title} {{Metallic conduction at organic
  charge-transfer interfaces}},}\ }\href {\doibase 10.1038/nmat2205} {\bibfield
   {journal} {\bibinfo  {journal} {Nature Materials}\ }\textbf {\bibinfo
  {volume} {7}},\ \bibinfo {pages} {574--580} (\bibinfo {year}
  {2008})}\BibitemShut {NoStop}%
\bibitem [{\citenamefont {Blum}\ \emph {et~al.}(2009)\citenamefont {Blum},
  \citenamefont {Gehrke}, \citenamefont {Hanke}, \citenamefont {Havu},
  \citenamefont {Havu}, \citenamefont {Ren}, \citenamefont {Reuter},\ and\
  \citenamefont {Scheffler}}]{Blum2009}%
  \BibitemOpen
  \bibfield  {author} {\bibinfo {author} {\bibfnamefont {V.}~\bibnamefont
  {Blum}}, \bibinfo {author} {\bibfnamefont {R.}~\bibnamefont {Gehrke}},
  \bibinfo {author} {\bibfnamefont {F.}~\bibnamefont {Hanke}}, \bibinfo
  {author} {\bibfnamefont {P.}~\bibnamefont {Havu}}, \bibinfo {author}
  {\bibfnamefont {V.}~\bibnamefont {Havu}}, \bibinfo {author} {\bibfnamefont
  {X.}~\bibnamefont {Ren}}, \bibinfo {author} {\bibfnamefont {K.}~\bibnamefont
  {Reuter}}, \ and\ \bibinfo {author} {\bibfnamefont {M.}~\bibnamefont
  {Scheffler}},\ }\bibfield  {title} {\enquote {\bibinfo {title} {{Ab initio
  molecular simulations with numeric atom-centered orbitals}},}\ }\href
  {\doibase 10.1016/j.cpc.2009.06.022} {\bibfield  {journal} {\bibinfo
  {journal} {Computer Physics Communications}\ }\textbf {\bibinfo {volume}
  {180}},\ \bibinfo {pages} {2175--2196} (\bibinfo {year} {2009})}\BibitemShut
  {NoStop}%
\bibitem [{\citenamefont {Perdew}, \citenamefont {Burke},\ and\ \citenamefont
  {Ernzerhof}(1996)}]{PhysRevLett.77.3865}%
  \BibitemOpen
  \bibfield  {author} {\bibinfo {author} {\bibfnamefont {J.~P.}\ \bibnamefont
  {Perdew}}, \bibinfo {author} {\bibfnamefont {K.}~\bibnamefont {Burke}}, \
  and\ \bibinfo {author} {\bibfnamefont {M.}~\bibnamefont {Ernzerhof}},\
  }\bibfield  {title} {\enquote {\bibinfo {title} {Generalized gradient
  approximation made simple},}\ }\href {\doibase 10.1103/PhysRevLett.77.3865}
  {\bibfield  {journal} {\bibinfo  {journal} {Phys. Rev. Lett.}\ }\textbf
  {\bibinfo {volume} {77}},\ \bibinfo {pages} {3865--3868} (\bibinfo {year}
  {1996})}\BibitemShut {NoStop}%
\bibitem [{\citenamefont {Perdew}, \citenamefont {Burke},\ and\ \citenamefont
  {Ernzerhof}(1997)}]{PhysRevLett.78.1396}%
  \BibitemOpen
  \bibfield  {author} {\bibinfo {author} {\bibfnamefont {J.~P.}\ \bibnamefont
  {Perdew}}, \bibinfo {author} {\bibfnamefont {K.}~\bibnamefont {Burke}}, \
  and\ \bibinfo {author} {\bibfnamefont {M.}~\bibnamefont {Ernzerhof}},\
  }\bibfield  {title} {\enquote {\bibinfo {title} {Generalized gradient
  approximation made simple [phys. rev. lett. 77, 3865 (1996)]},}\ }\href
  {\doibase 10.1103/PhysRevLett.78.1396} {\bibfield  {journal} {\bibinfo
  {journal} {Phys. Rev. Lett.}\ }\textbf {\bibinfo {volume} {78}},\ \bibinfo
  {pages} {1396--1396} (\bibinfo {year} {1997})}\BibitemShut {NoStop}%
\bibitem [{\citenamefont {Tkatchenko}\ and\ \citenamefont
  {Scheffler}(2009)}]{tkatchenko2009accurate}%
  \BibitemOpen
  \bibfield  {author} {\bibinfo {author} {\bibfnamefont {A.}~\bibnamefont
  {Tkatchenko}}\ and\ \bibinfo {author} {\bibfnamefont {M.}~\bibnamefont
  {Scheffler}},\ }\bibfield  {title} {\enquote {\bibinfo {title} {Accurate
  molecular van der waals interactions from ground-state electron density and
  free-atom reference data},}\ }\href@noop {} {\bibfield  {journal} {\bibinfo
  {journal} {Physical review letters}\ }\textbf {\bibinfo {volume} {102}},\
  \bibinfo {pages} {073005} (\bibinfo {year} {2009})}\BibitemShut {NoStop}%
\bibitem [{\citenamefont {Batsanov}(2006)}]{Batsanov2006}%
  \BibitemOpen
  \bibfield  {author} {\bibinfo {author} {\bibfnamefont {A.~S.}\ \bibnamefont
  {Batsanov}},\ }\bibfield  {title} {\enquote {\bibinfo {title}
  {{Tetrathiafulvalene revisited}},}\ }\href {\doibase
  10.1107/S0108270106022554/GA3012I_90SUP4.HKL} {\bibfield  {journal} {\bibinfo
   {journal} {Acta Crystallographica Section C: Crystal Structure
  Communications}\ }\textbf {\bibinfo {volume} {62}},\ \bibinfo {pages}
  {o501--o504} (\bibinfo {year} {2006})}\BibitemShut {NoStop}%
\bibitem [{\citenamefont {Krupskaya}\ \emph {et~al.}(2015)\citenamefont
  {Krupskaya}, \citenamefont {Gibertini}, \citenamefont {Marzari},
  \citenamefont {Morpurgo}, \citenamefont {Krupskaya}, \citenamefont
  {Morpurgo}, \citenamefont {Gibertini},\ and\ \citenamefont
  {Marzari}}]{Krupskaya2015}%
  \BibitemOpen
  \bibfield  {author} {\bibinfo {author} {\bibfnamefont {Y.}~\bibnamefont
  {Krupskaya}}, \bibinfo {author} {\bibfnamefont {M.}~\bibnamefont
  {Gibertini}}, \bibinfo {author} {\bibfnamefont {N.}~\bibnamefont {Marzari}},
  \bibinfo {author} {\bibfnamefont {A.~F.}\ \bibnamefont {Morpurgo}}, \bibinfo
  {author} {\bibfnamefont {Y.}~\bibnamefont {Krupskaya}}, \bibinfo {author}
  {\bibfnamefont {A.~F.}\ \bibnamefont {Morpurgo}}, \bibinfo {author}
  {\bibfnamefont {M.}~\bibnamefont {Gibertini}}, \ and\ \bibinfo {author}
  {\bibfnamefont {N.}~\bibnamefont {Marzari}},\ }\bibfield  {title} {\enquote
  {\bibinfo {title} {{Band-Like Electron Transport with Record-High Mobility in
  the TCNQ Family}},}\ }\href {\doibase 10.1002/ADMA.201405699} {\bibfield
  {journal} {\bibinfo  {journal} {Advanced Materials}\ }\textbf {\bibinfo
  {volume} {27}},\ \bibinfo {pages} {2453--2458} (\bibinfo {year}
  {2015})}\BibitemShut {NoStop}%
\bibitem [{\citenamefont {Heyd}, \citenamefont {Scuseria},\ and\ \citenamefont
  {Ernzerhof}(2003)}]{heyd2003hybrid}%
  \BibitemOpen
  \bibfield  {author} {\bibinfo {author} {\bibfnamefont {J.}~\bibnamefont
  {Heyd}}, \bibinfo {author} {\bibfnamefont {G.~E.}\ \bibnamefont {Scuseria}},
  \ and\ \bibinfo {author} {\bibfnamefont {M.}~\bibnamefont {Ernzerhof}},\
  }\bibfield  {title} {\enquote {\bibinfo {title} {Hybrid functionals based on
  a screened coulomb potential},}\ }\href@noop {} {\bibfield  {journal}
  {\bibinfo  {journal} {The Journal of chemical physics}\ }\textbf {\bibinfo
  {volume} {118}},\ \bibinfo {pages} {8207--8215} (\bibinfo {year}
  {2003})}\BibitemShut {NoStop}%
\bibitem [{\citenamefont {Heyd}, \citenamefont {Scuseria},\ and\ \citenamefont
  {Ernzerhof}(2006)}]{ge2006erratum}%
  \BibitemOpen
  \bibfield  {author} {\bibinfo {author} {\bibfnamefont {J.}~\bibnamefont
  {Heyd}}, \bibinfo {author} {\bibfnamefont {G.~E.}\ \bibnamefont {Scuseria}},
  \ and\ \bibinfo {author} {\bibfnamefont {M.}~\bibnamefont {Ernzerhof}},\
  }\bibfield  {title} {\enquote {\bibinfo {title} {Erratum:“hybrid
  functionals based on a screened coulomb potential”{[J. Chem. Phys. 118,
  8207 (2003)]}},}\ }\href@noop {} {\bibfield  {journal} {\bibinfo  {journal}
  {J. Chem. Phys}\ }\textbf {\bibinfo {volume} {124}},\ \bibinfo {pages}
  {219906} (\bibinfo {year} {2006})}\BibitemShut {NoStop}%
\bibitem [{\citenamefont {Devereux}\ \emph {et~al.}(2020)\citenamefont
  {Devereux}, \citenamefont {Smith}, \citenamefont {Davis}, \citenamefont
  {Barros}, \citenamefont {Zubatyuk}, \citenamefont {Isayev},\ and\
  \citenamefont {Roitberg}}]{Devereux2020}%
  \BibitemOpen
  \bibfield  {author} {\bibinfo {author} {\bibfnamefont {C.}~\bibnamefont
  {Devereux}}, \bibinfo {author} {\bibfnamefont {J.~S.}\ \bibnamefont {Smith}},
  \bibinfo {author} {\bibfnamefont {K.~K.}\ \bibnamefont {Davis}}, \bibinfo
  {author} {\bibfnamefont {K.}~\bibnamefont {Barros}}, \bibinfo {author}
  {\bibfnamefont {R.}~\bibnamefont {Zubatyuk}}, \bibinfo {author}
  {\bibfnamefont {O.}~\bibnamefont {Isayev}}, \ and\ \bibinfo {author}
  {\bibfnamefont {A.~E.}\ \bibnamefont {Roitberg}},\ }\bibfield  {title}
  {\enquote {\bibinfo {title} {{Extending the Applicability of the ANI Deep
  Learning Molecular Potential to Sulfur and Halogens}},}\ }\href {\doibase
  10.1021/ACS.JCTC.0C00121/SUPPL_FILE/CT0C00121_SI_001.PDF} {\bibfield
  {journal} {\bibinfo  {journal} {Journal of Chemical Theory and Computation}\
  }\textbf {\bibinfo {volume} {16}},\ \bibinfo {pages} {4192--4202} (\bibinfo
  {year} {2020})}\BibitemShut {NoStop}%
\bibitem [{\citenamefont {Grimme}\ \emph {et~al.}(2010)\citenamefont {Grimme},
  \citenamefont {Antony}, \citenamefont {Ehrlich},\ and\ \citenamefont
  {Krieg}}]{Grimme2010}%
  \BibitemOpen
  \bibfield  {author} {\bibinfo {author} {\bibfnamefont {S.}~\bibnamefont
  {Grimme}}, \bibinfo {author} {\bibfnamefont {J.}~\bibnamefont {Antony}},
  \bibinfo {author} {\bibfnamefont {S.}~\bibnamefont {Ehrlich}}, \ and\
  \bibinfo {author} {\bibfnamefont {H.}~\bibnamefont {Krieg}},\ }\bibfield
  {title} {\enquote {\bibinfo {title} {{A consistent and accurate ab initio
  parametrization of density functional dispersion correction (DFT-D) for the
  94 elements H-Pu}},}\ }\href {\doibase 10.1063/1.3382344} {\bibfield
  {journal} {\bibinfo  {journal} {The Journal of Chemical Physics}\ }\textbf
  {\bibinfo {volume} {132}},\ \bibinfo {pages} {154104} (\bibinfo {year}
  {2010})}\BibitemShut {NoStop}%
\bibitem [{\citenamefont {Gao}\ \emph {et~al.}(2020)\citenamefont {Gao},
  \citenamefont {Ramezanghorbani}, \citenamefont {Isayev}, \citenamefont
  {Smith},\ and\ \citenamefont {Roitberg}}]{Gao2020}%
  \BibitemOpen
  \bibfield  {author} {\bibinfo {author} {\bibfnamefont {X.}~\bibnamefont
  {Gao}}, \bibinfo {author} {\bibfnamefont {F.}~\bibnamefont
  {Ramezanghorbani}}, \bibinfo {author} {\bibfnamefont {O.}~\bibnamefont
  {Isayev}}, \bibinfo {author} {\bibfnamefont {J.~S.}\ \bibnamefont {Smith}}, \
  and\ \bibinfo {author} {\bibfnamefont {A.~E.}\ \bibnamefont {Roitberg}},\
  }\bibfield  {title} {\enquote {\bibinfo {title} {{TorchANI: A Free and Open
  Source PyTorch-Based Deep Learning Implementation of the ANI Neural Network
  Potentials}},}\ }\href {\doibase
  10.1021/ACS.JCIM.0C00451/ASSET/IMAGES/MEDIUM/CI0C00451_0009.GIF} {\bibfield
  {journal} {\bibinfo  {journal} {Journal of Chemical Information and
  Modeling}\ }\textbf {\bibinfo {volume} {60}},\ \bibinfo {pages} {3408--3415}
  (\bibinfo {year} {2020})}\BibitemShut {NoStop}%
\bibitem [{\citenamefont {Ong}\ \emph {et~al.}(2013)\citenamefont {Ong},
  \citenamefont {Richards}, \citenamefont {Jain}, \citenamefont {Hautier},
  \citenamefont {Kocher}, \citenamefont {Cholia}, \citenamefont {Gunter},
  \citenamefont {Chevrier}, \citenamefont {Persson},\ and\ \citenamefont
  {Ceder}}]{ong2013python}%
  \BibitemOpen
  \bibfield  {author} {\bibinfo {author} {\bibfnamefont {S.~P.}\ \bibnamefont
  {Ong}}, \bibinfo {author} {\bibfnamefont {W.~D.}\ \bibnamefont {Richards}},
  \bibinfo {author} {\bibfnamefont {A.}~\bibnamefont {Jain}}, \bibinfo {author}
  {\bibfnamefont {G.}~\bibnamefont {Hautier}}, \bibinfo {author} {\bibfnamefont
  {M.}~\bibnamefont {Kocher}}, \bibinfo {author} {\bibfnamefont
  {S.}~\bibnamefont {Cholia}}, \bibinfo {author} {\bibfnamefont
  {D.}~\bibnamefont {Gunter}}, \bibinfo {author} {\bibfnamefont {V.~L.}\
  \bibnamefont {Chevrier}}, \bibinfo {author} {\bibfnamefont {K.~A.}\
  \bibnamefont {Persson}}, \ and\ \bibinfo {author} {\bibfnamefont
  {G.}~\bibnamefont {Ceder}},\ }\bibfield  {title} {\enquote {\bibinfo {title}
  {Python materials genomics (pymatgen): A robust, open-source python library
  for materials analysis},}\ }\href@noop {} {\bibfield  {journal} {\bibinfo
  {journal} {Computational Materials Science}\ }\textbf {\bibinfo {volume}
  {68}},\ \bibinfo {pages} {314--319} (\bibinfo {year} {2013})}\BibitemShut
  {NoStop}%
\bibitem [{\citenamefont {Larsen}\ \emph {et~al.}(2017)\citenamefont {Larsen},
  \citenamefont {Mortensen}, \citenamefont {Blomqvist}, \citenamefont
  {Castelli}, \citenamefont {Christensen}, \citenamefont {Du{\l}ak},
  \citenamefont {Friis}, \citenamefont {Groves}, \citenamefont {Hammer},
  \citenamefont {Hargus} \emph {et~al.}}]{larsen2017atomic}%
  \BibitemOpen
  \bibfield  {author} {\bibinfo {author} {\bibfnamefont {A.~H.}\ \bibnamefont
  {Larsen}}, \bibinfo {author} {\bibfnamefont {J.~J.}\ \bibnamefont
  {Mortensen}}, \bibinfo {author} {\bibfnamefont {J.}~\bibnamefont
  {Blomqvist}}, \bibinfo {author} {\bibfnamefont {I.~E.}\ \bibnamefont
  {Castelli}}, \bibinfo {author} {\bibfnamefont {R.}~\bibnamefont
  {Christensen}}, \bibinfo {author} {\bibfnamefont {M.}~\bibnamefont
  {Du{\l}ak}}, \bibinfo {author} {\bibfnamefont {J.}~\bibnamefont {Friis}},
  \bibinfo {author} {\bibfnamefont {M.~N.}\ \bibnamefont {Groves}}, \bibinfo
  {author} {\bibfnamefont {B.}~\bibnamefont {Hammer}}, \bibinfo {author}
  {\bibfnamefont {C.}~\bibnamefont {Hargus}},  \emph {et~al.},\ }\bibfield
  {title} {\enquote {\bibinfo {title} {The atomic simulation environment—a
  python library for working with atoms},}\ }\href@noop {} {\bibfield
  {journal} {\bibinfo  {journal} {Journal of Physics: Condensed Matter}\
  }\textbf {\bibinfo {volume} {29}},\ \bibinfo {pages} {273002} (\bibinfo
  {year} {2017})}\BibitemShut {NoStop}%
\bibitem [{\citenamefont {Joubert}(1999)}]{Joubert1999}%
  \BibitemOpen
  \bibfield  {author} {\bibinfo {author} {\bibfnamefont {D.}~\bibnamefont
  {Joubert}},\ }\bibfield  {title} {\enquote {\bibinfo {title} {{From ultrasoft
  pseudopotentials to the projector augmented-wave method}},}\ }\href {\doibase
  10.1103/PhysRevB.59.1758} {\bibfield  {journal} {\bibinfo  {journal}
  {Physical Review B - Condensed Matter and Materials Physics}\ }\textbf
  {\bibinfo {volume} {59}},\ \bibinfo {pages} {1758--1775} (\bibinfo {year}
  {1999})}\BibitemShut {NoStop}%
\bibitem [{\citenamefont {Kresse}\ and\ \citenamefont
  {Furthm{\"{u}}ller}(1996)}]{Kresse1996}%
  \BibitemOpen
  \bibfield  {author} {\bibinfo {author} {\bibfnamefont {G.}~\bibnamefont
  {Kresse}}\ and\ \bibinfo {author} {\bibfnamefont {J.}~\bibnamefont
  {Furthm{\"{u}}ller}},\ }\bibfield  {title} {\enquote {\bibinfo {title}
  {{Efficient iterative schemes for ab initio total-energy calculations using a
  plane-wave basis set}},}\ }\href {\doibase 10.1103/PhysRevB.54.11169}
  {\bibfield  {journal} {\bibinfo  {journal} {Physical Review B - Condensed
  Matter and Materials Physics}\ }\textbf {\bibinfo {volume} {54}},\ \bibinfo
  {pages} {11169--11186} (\bibinfo {year} {1996})}\BibitemShut {NoStop}%
\bibitem [{\citenamefont {Kresse}\ and\ \citenamefont
  {Hafner}(1993)}]{Kresse1993}%
  \BibitemOpen
  \bibfield  {author} {\bibinfo {author} {\bibfnamefont {G.}~\bibnamefont
  {Kresse}}\ and\ \bibinfo {author} {\bibfnamefont {J.}~\bibnamefont
  {Hafner}},\ }\bibfield  {title} {\enquote {\bibinfo {title} {{Ab initio
  molecular dynamics for open-shell transition metals}},}\ }\href {\doibase
  10.1103/PhysRevB.48.13115} {\bibfield  {journal} {\bibinfo  {journal}
  {Physical Review B}\ }\textbf {\bibinfo {volume} {48}},\ \bibinfo {pages}
  {13115--13118} (\bibinfo {year} {1993})}\BibitemShut {NoStop}%
\bibitem [{\citenamefont {Kresse}\ and\ \citenamefont
  {Hafner}(1994)}]{Kresse1994}%
  \BibitemOpen
  \bibfield  {author} {\bibinfo {author} {\bibfnamefont {G.}~\bibnamefont
  {Kresse}}\ and\ \bibinfo {author} {\bibfnamefont {J.}~\bibnamefont
  {Hafner}},\ }\bibfield  {title} {\enquote {\bibinfo {title} {{Ab initio
  molecular-dynamics simulation of the liquid-metalamorphous- semiconductor
  transition in germanium}},}\ }\href {\doibase 10.1103/PhysRevB.49.14251}
  {\bibfield  {journal} {\bibinfo  {journal} {Physical Review B}\ }\textbf
  {\bibinfo {volume} {49}},\ \bibinfo {pages} {14251--14269} (\bibinfo {year}
  {1994})}\BibitemShut {NoStop}%
\bibitem [{\citenamefont {Dull}\ \emph {et~al.}(2020)\citenamefont {Dull},
  \citenamefont {Wang}, \citenamefont {Johnson}, \citenamefont {Shayegan},
  \citenamefont {Shapiro}, \citenamefont {Priestley}, \citenamefont {Geerts},\
  and\ \citenamefont {Rand}}]{Dull2021}%
  \BibitemOpen
  \bibfield  {author} {\bibinfo {author} {\bibfnamefont {J.~T.}\ \bibnamefont
  {Dull}}, \bibinfo {author} {\bibfnamefont {Y.}~\bibnamefont {Wang}}, \bibinfo
  {author} {\bibfnamefont {H.}~\bibnamefont {Johnson}}, \bibinfo {author}
  {\bibfnamefont {K.}~\bibnamefont {Shayegan}}, \bibinfo {author}
  {\bibfnamefont {E.}~\bibnamefont {Shapiro}}, \bibinfo {author} {\bibfnamefont
  {R.~D.}\ \bibnamefont {Priestley}}, \bibinfo {author} {\bibfnamefont {Y.~H.}\
  \bibnamefont {Geerts}}, \ and\ \bibinfo {author} {\bibfnamefont {B.~P.}\
  \bibnamefont {Rand}},\ }\bibfield  {title} {\enquote {\bibinfo {title}
  {Thermal properties, molecular structure, and thin-film organic semiconductor
  crystallization},}\ }\href {\doibase 10.1021/acs.jpcc.0c09408} {\bibfield
  {journal} {\bibinfo  {journal} {The Journal of Physical Chemistry C}\
  }\textbf {\bibinfo {volume} {124}},\ \bibinfo {pages} {27213--27221}
  (\bibinfo {year} {2020})}\BibitemShut {NoStop}%
\bibitem [{\citenamefont {Zur}\ and\ \citenamefont {McGill}(1984)}]{Zur1984}%
  \BibitemOpen
  \bibfield  {author} {\bibinfo {author} {\bibfnamefont {A.}~\bibnamefont
  {Zur}}\ and\ \bibinfo {author} {\bibfnamefont {T.~C.}\ \bibnamefont
  {McGill}},\ }\bibfield  {title} {\enquote {\bibinfo {title} {{Lattice match:
  An application to heteroepitaxy}},}\ }\href {\doibase 10.1063/1.333084}
  {\bibfield  {journal} {\bibinfo  {journal} {Journal of Applied Physics}\
  }\textbf {\bibinfo {volume} {55}},\ \bibinfo {pages} {378--386} (\bibinfo
  {year} {1984})}\BibitemShut {NoStop}%
\bibitem [{\citenamefont {Sassella}, \citenamefont {Campione},\ and\
  \citenamefont {Borghesi}(2008)}]{Sassella2008}%
  \BibitemOpen
  \bibfield  {author} {\bibinfo {author} {\bibfnamefont {A.}~\bibnamefont
  {Sassella}}, \bibinfo {author} {\bibfnamefont {M.}~\bibnamefont {Campione}},
  \ and\ \bibinfo {author} {\bibfnamefont {A.}~\bibnamefont {Borghesi}},\
  }\bibfield  {title} {\enquote {\bibinfo {title} {{Organic epitaxy}},}\ }\href
  {\doibase 10.1393/NCR/I2008-10035-Y} {\bibfield  {journal} {\bibinfo
  {journal} {La Rivista del Nuovo Cimento}\ }\textbf {\bibinfo {volume} {31}},\
  \bibinfo {pages} {457--490} (\bibinfo {year} {2008})}\BibitemShut {NoStop}%
\bibitem [{\citenamefont {Scholz}\ and\ \citenamefont
  {Stirner}(2019)}]{Scholz2019}%
  \BibitemOpen
  \bibfield  {author} {\bibinfo {author} {\bibfnamefont {D.}~\bibnamefont
  {Scholz}}\ and\ \bibinfo {author} {\bibfnamefont {T.}~\bibnamefont
  {Stirner}},\ }\bibfield  {title} {\enquote {\bibinfo {title} {{Convergence of
  surface energy calculations for various methods: (0 0 1) hematite as
  benchmark}},}\ }\href {\doibase 10.1088/1361-648X/ab069d} {\bibfield
  {journal} {\bibinfo  {journal} {Journal of Physics Condensed Matter}\
  }\textbf {\bibinfo {volume} {31}},\ \bibinfo {pages} {195901} (\bibinfo
  {year} {2019})}\BibitemShut {NoStop}%
\bibitem [{\citenamefont {Antonijevi{\'{c}}}\ \emph {et~al.}(2019)\citenamefont
  {Antonijevi{\'{c}}}, \citenamefont {Malenov}, \citenamefont {Hall},\ and\
  \citenamefont {Zari{\'{c}}}}]{Antonijevic2019}%
  \BibitemOpen
  \bibfield  {author} {\bibinfo {author} {\bibfnamefont {I.~S.}\ \bibnamefont
  {Antonijevi{\'{c}}}}, \bibinfo {author} {\bibfnamefont {D.~P.}\ \bibnamefont
  {Malenov}}, \bibinfo {author} {\bibfnamefont {M.~B.}\ \bibnamefont {Hall}}, \
  and\ \bibinfo {author} {\bibfnamefont {S.~D.}\ \bibnamefont {Zari{\'{c}}}},\
  }\bibfield  {title} {\enquote {\bibinfo {title} {{Study of stacking
  interactions between two neutral tetrathiafulvalene molecules in Cambridge
  Structural Database crystal structures and by quantum chemical
  calculations}},}\ }\href {\doibase 10.1107/S2052520618015494} {\bibfield
  {journal} {\bibinfo  {journal} {Acta Crystallographica Section B}\ }\textbf
  {\bibinfo {volume} {75}},\ \bibinfo {pages} {1--7} (\bibinfo {year}
  {2019})}\BibitemShut {NoStop}%
\bibitem [{\citenamefont {Frazier}(2018{\natexlab{a}})}]{frazier2018tutorial}%
  \BibitemOpen
  \bibfield  {author} {\bibinfo {author} {\bibfnamefont {P.~I.}\ \bibnamefont
  {Frazier}},\ }\bibfield  {title} {\enquote {\bibinfo {title} {A tutorial on
  bayesian optimization},}\ }\href@noop {} {\bibfield  {journal} {\bibinfo
  {journal} {arXiv}\ ,\ \bibinfo {pages} {1807.02811}} (\bibinfo {year}
  {2018}{\natexlab{a}})}\BibitemShut {NoStop}%
\bibitem [{\citenamefont {Brochu}, \citenamefont {Cora},\ and\ \citenamefont
  {De~Freitas}(2010)}]{brochu2010tutorial}%
  \BibitemOpen
  \bibfield  {author} {\bibinfo {author} {\bibfnamefont {E.}~\bibnamefont
  {Brochu}}, \bibinfo {author} {\bibfnamefont {V.~M.}\ \bibnamefont {Cora}}, \
  and\ \bibinfo {author} {\bibfnamefont {N.}~\bibnamefont {De~Freitas}},\
  }\bibfield  {title} {\enquote {\bibinfo {title} {A tutorial on bayesian
  optimization of expensive cost functions, with application to active user
  modeling and hierarchical reinforcement learning},}\ }\href@noop {}
  {\bibfield  {journal} {\bibinfo  {journal} {arXiv}\ ,\ \bibinfo {pages}
  {1012.2599}} (\bibinfo {year} {2010})}\BibitemShut {NoStop}%
\bibitem [{\citenamefont {Frazier}(2018{\natexlab{b}})}]{Frazier2018}%
  \BibitemOpen
  \bibfield  {author} {\bibinfo {author} {\bibfnamefont {P.~I.}\ \bibnamefont
  {Frazier}},\ }\bibfield  {title} {\enquote {\bibinfo {title} {{A Tutorial on
  Bayesian Optimization}},}\ }\href {http://arxiv.org/abs/1807.02811}
  {\bibfield  {journal} {\bibinfo  {journal} {arXiv}\ ,\ \bibinfo {pages}
  {1807.02811}} (\bibinfo {year} {2018}{\natexlab{b}})}\BibitemShut {NoStop}%
\bibitem [{\citenamefont {Williams}\ and\ \citenamefont
  {Rasmussen}(2006)}]{williams2006gaussian}%
  \BibitemOpen
  \bibfield  {author} {\bibinfo {author} {\bibfnamefont {C.}~\bibnamefont
  {Williams}}\ and\ \bibinfo {author} {\bibfnamefont {C.~E.}\ \bibnamefont
  {Rasmussen}},\ }\bibfield  {title} {\enquote {\bibinfo {title} {Gaussian
  processes for machine learning, vol. 2},}\ }\href@noop {} {\bibfield
  {journal} {\bibinfo  {journal} {MIT press Cambridge, MA}\ }\textbf {\bibinfo
  {volume} {302}},\ \bibinfo {pages} {303} (\bibinfo {year}
  {2006})}\BibitemShut {NoStop}%
\bibitem [{\citenamefont {Nogueira}(2014)}]{Nogueira2014}%
  \BibitemOpen
  \bibfield  {author} {\bibinfo {author} {\bibfnamefont {F.}~\bibnamefont
  {Nogueira}},\ }\href {https://github.com/fmfn/BayesianOptimization} {\enquote
  {\bibinfo {title} {Bayesian optimization: Open source constrained global
  optimization tool for python},}\ } (\bibinfo {year} {2014})\BibitemShut
  {NoStop}%
\bibitem [{\citenamefont {Xiong}\ \emph {et~al.}(2017)\citenamefont {Xiong},
  \citenamefont {Liu}, \citenamefont {Zhang}, \citenamefont {Du},\ and\
  \citenamefont {Chen}}]{xiong2017first}%
  \BibitemOpen
  \bibfield  {author} {\bibinfo {author} {\bibfnamefont {H.}~\bibnamefont
  {Xiong}}, \bibinfo {author} {\bibfnamefont {Z.}~\bibnamefont {Liu}}, \bibinfo
  {author} {\bibfnamefont {H.}~\bibnamefont {Zhang}}, \bibinfo {author}
  {\bibfnamefont {Z.}~\bibnamefont {Du}}, \ and\ \bibinfo {author}
  {\bibfnamefont {C.}~\bibnamefont {Chen}},\ }\bibfield  {title} {\enquote
  {\bibinfo {title} {First principles calculation of interfacial stability,
  energy and electronic properties of {SiC/ZrB2} interface},}\ }\href@noop {}
  {\bibfield  {journal} {\bibinfo  {journal} {Journal of Physics and Chemistry
  of Solids}\ }\textbf {\bibinfo {volume} {107}},\ \bibinfo {pages} {162--169}
  (\bibinfo {year} {2017})}\BibitemShut {NoStop}%
\bibitem [{\citenamefont {Christensen}, \citenamefont {Dudiy},\ and\
  \citenamefont {Wahnstr{\"o}m}(2002)}]{christensen2002first}%
  \BibitemOpen
  \bibfield  {author} {\bibinfo {author} {\bibfnamefont {M.}~\bibnamefont
  {Christensen}}, \bibinfo {author} {\bibfnamefont {S.}~\bibnamefont {Dudiy}},
  \ and\ \bibinfo {author} {\bibfnamefont {G.}~\bibnamefont {Wahnstr{\"o}m}},\
  }\bibfield  {title} {\enquote {\bibinfo {title} {First-principles simulations
  of metal-ceramic interface adhesion: Co/wc versus co/tic},}\ }\href@noop {}
  {\bibfield  {journal} {\bibinfo  {journal} {Physical Review B}\ }\textbf
  {\bibinfo {volume} {65}},\ \bibinfo {pages} {045408} (\bibinfo {year}
  {2002})}\BibitemShut {NoStop}%
\bibitem [{\citenamefont {Li}\ \emph {et~al.}(2016)\citenamefont {Li},
  \citenamefont {Hui}, \citenamefont {Shao}, \citenamefont {Chen},
  \citenamefont {Wang}, \citenamefont {Jia}, \citenamefont {Li}, \citenamefont
  {Chen},\ and\ \citenamefont {Cheng}}]{li2016first}%
  \BibitemOpen
  \bibfield  {author} {\bibinfo {author} {\bibfnamefont {X.}~\bibnamefont
  {Li}}, \bibinfo {author} {\bibfnamefont {Q.}~\bibnamefont {Hui}}, \bibinfo
  {author} {\bibfnamefont {D.}~\bibnamefont {Shao}}, \bibinfo {author}
  {\bibfnamefont {J.}~\bibnamefont {Chen}}, \bibinfo {author} {\bibfnamefont
  {P.}~\bibnamefont {Wang}}, \bibinfo {author} {\bibfnamefont {Z.}~\bibnamefont
  {Jia}}, \bibinfo {author} {\bibfnamefont {C.}~\bibnamefont {Li}}, \bibinfo
  {author} {\bibfnamefont {Z.}~\bibnamefont {Chen}}, \ and\ \bibinfo {author}
  {\bibfnamefont {N.}~\bibnamefont {Cheng}},\ }\bibfield  {title} {\enquote
  {\bibinfo {title} {First-principles study on the stability and electronic
  structure of {Mg/ZrB2} interfaces},}\ }\href@noop {} {\bibfield  {journal}
  {\bibinfo  {journal} {Science China Materials}\ }\textbf {\bibinfo {volume}
  {59}},\ \bibinfo {pages} {28--37} (\bibinfo {year} {2016})}\BibitemShut
  {NoStop}%
\bibitem [{\citenamefont {Liu}, \citenamefont {Wang},\ and\ \citenamefont
  {Ye}(2004)}]{liu2004first}%
  \BibitemOpen
  \bibfield  {author} {\bibinfo {author} {\bibfnamefont {L.}~\bibnamefont
  {Liu}}, \bibinfo {author} {\bibfnamefont {S.}~\bibnamefont {Wang}}, \ and\
  \bibinfo {author} {\bibfnamefont {H.}~\bibnamefont {Ye}},\ }\bibfield
  {title} {\enquote {\bibinfo {title} {First-principles study of polar {Al/TiN}
  (1 1 1) interfaces},}\ }\href@noop {} {\bibfield  {journal} {\bibinfo
  {journal} {Acta materialia}\ }\textbf {\bibinfo {volume} {52}},\ \bibinfo
  {pages} {3681--3688} (\bibinfo {year} {2004})}\BibitemShut {NoStop}%
\bibitem [{\citenamefont {Zhuo}\ \emph {et~al.}(2018)\citenamefont {Zhuo},
  \citenamefont {Mao}, \citenamefont {Xu},\ and\ \citenamefont
  {Fu}}]{zhuo2018density}%
  \BibitemOpen
  \bibfield  {author} {\bibinfo {author} {\bibfnamefont {Z.}~\bibnamefont
  {Zhuo}}, \bibinfo {author} {\bibfnamefont {H.}~\bibnamefont {Mao}}, \bibinfo
  {author} {\bibfnamefont {H.}~\bibnamefont {Xu}}, \ and\ \bibinfo {author}
  {\bibfnamefont {Y.}~\bibnamefont {Fu}},\ }\bibfield  {title} {\enquote
  {\bibinfo {title} {Density functional theory study of {Al/NbB2} heterogeneous
  nucleation interface},}\ }\href@noop {} {\bibfield  {journal} {\bibinfo
  {journal} {Applied Surface Science}\ }\textbf {\bibinfo {volume} {456}},\
  \bibinfo {pages} {37--42} (\bibinfo {year} {2018})}\BibitemShut {NoStop}%
\bibitem [{\citenamefont {Wang}, \citenamefont {Li},\ and\ \citenamefont
  {Xu}(2020)}]{wang2020first}%
  \BibitemOpen
  \bibfield  {author} {\bibinfo {author} {\bibfnamefont {J.}~\bibnamefont
  {Wang}}, \bibinfo {author} {\bibfnamefont {Y.}~\bibnamefont {Li}}, \ and\
  \bibinfo {author} {\bibfnamefont {R.}~\bibnamefont {Xu}},\ }\bibfield
  {title} {\enquote {\bibinfo {title} {First-principles calculations on
  electronic structure and interfacial stability of {Mg/NbB2} heterogeneous
  nucleation interface},}\ }\href@noop {} {\bibfield  {journal} {\bibinfo
  {journal} {Surface Science}\ }\textbf {\bibinfo {volume} {691}},\ \bibinfo
  {pages} {121487} (\bibinfo {year} {2020})}\BibitemShut {NoStop}%
\bibitem [{\citenamefont {Park}\ \emph {et~al.}(2017)\citenamefont {Park},
  \citenamefont {Atalla}, \citenamefont {Smith},\ and\ \citenamefont
  {Yoon}}]{Park2017}%
  \BibitemOpen
  \bibfield  {author} {\bibinfo {author} {\bibfnamefont {C.}~\bibnamefont
  {Park}}, \bibinfo {author} {\bibfnamefont {V.}~\bibnamefont {Atalla}},
  \bibinfo {author} {\bibfnamefont {S.}~\bibnamefont {Smith}}, \ and\ \bibinfo
  {author} {\bibfnamefont {M.}~\bibnamefont {Yoon}},\ }\bibfield  {title}
  {\enquote {\bibinfo {title} {Understanding the charge transfer at the
  interface of electron donors and acceptors: {TTF–TCNQ} as an example},}\
  }\href {\doibase 10.1021/acsami.7b04148} {\bibfield  {journal} {\bibinfo
  {journal} {ACS Applied Materials \& Interfaces}\ }\textbf {\bibinfo {volume}
  {9}},\ \bibinfo {pages} {27266--27272} (\bibinfo {year} {2017})}\BibitemShut
  {NoStop}%
\bibitem [{\citenamefont {Wang}\ \emph {et~al.}(2019)\citenamefont {Wang},
  \citenamefont {Liu}, \citenamefont {Tom}, \citenamefont {Cook}, \citenamefont
  {Schatschneider},\ and\ \citenamefont {Marom}}]{Wang2019}%
  \BibitemOpen
  \bibfield  {author} {\bibinfo {author} {\bibfnamefont {X.}~\bibnamefont
  {Wang}}, \bibinfo {author} {\bibfnamefont {X.}~\bibnamefont {Liu}}, \bibinfo
  {author} {\bibfnamefont {R.}~\bibnamefont {Tom}}, \bibinfo {author}
  {\bibfnamefont {C.}~\bibnamefont {Cook}}, \bibinfo {author} {\bibfnamefont
  {B.}~\bibnamefont {Schatschneider}}, \ and\ \bibinfo {author} {\bibfnamefont
  {N.}~\bibnamefont {Marom}},\ }\bibfield  {title} {\enquote {\bibinfo {title}
  {Phenylated acene derivatives as candidates for intermolecular singlet
  fission},}\ }\href {\doibase 10.1021/acs.jpcc.8b12549} {\bibfield  {journal}
  {\bibinfo  {journal} {The Journal of Physical Chemistry C}\ }\textbf
  {\bibinfo {volume} {123}},\ \bibinfo {pages} {5890--5899} (\bibinfo {year}
  {2019})}\BibitemShut {NoStop}%
\bibitem [{\citenamefont {Kronik}\ and\ \citenamefont
  {K{\"u}mmel}(2014)}]{Kronik2014}%
  \BibitemOpen
  \bibfield  {author} {\bibinfo {author} {\bibfnamefont {L.}~\bibnamefont
  {Kronik}}\ and\ \bibinfo {author} {\bibfnamefont {S.}~\bibnamefont
  {K{\"u}mmel}},\ }\enquote {\bibinfo {title} {Gas-phase valence-electron
  photoemission spectroscopy using density functional theory},}\ in\ \href
  {\doibase 10.1007/128_2013_522} {\emph {\bibinfo {booktitle} {First
  Principles Approaches to Spectroscopic Properties of Complex Materials}}},\
  \bibinfo {editor} {edited by\ \bibinfo {editor} {\bibfnamefont
  {C.}~\bibnamefont {Di~Valentin}}, \bibinfo {editor} {\bibfnamefont
  {S.}~\bibnamefont {Botti}}, \ and\ \bibinfo {editor} {\bibfnamefont
  {M.}~\bibnamefont {Cococcioni}}}\ (\bibinfo  {publisher} {Springer Berlin
  Heidelberg},\ \bibinfo {address} {Berlin, Heidelberg},\ \bibinfo {year}
  {2014})\ pp.\ \bibinfo {pages} {137--191}\BibitemShut {NoStop}%
\bibitem [{\citenamefont {Ueno}\ and\ \citenamefont {Kera}(2008)}]{UENO2008}%
  \BibitemOpen
  \bibfield  {author} {\bibinfo {author} {\bibfnamefont {N.}~\bibnamefont
  {Ueno}}\ and\ \bibinfo {author} {\bibfnamefont {S.}~\bibnamefont {Kera}},\
  }\bibfield  {title} {\enquote {\bibinfo {title} {Electron spectroscopy of
  functional organic thin films: Deep insights into valence electronic
  structure in relation to charge transport property},}\ }\href {\doibase
  https://doi.org/10.1016/j.progsurf.2008.10.002} {\bibfield  {journal}
  {\bibinfo  {journal} {Progress in Surface Science}\ }\textbf {\bibinfo
  {volume} {83}},\ \bibinfo {pages} {490--557} (\bibinfo {year}
  {2008})}\BibitemShut {NoStop}%
\bibitem [{\citenamefont {Yang}, \citenamefont {Wu},\ and\ \citenamefont
  {Marom}(2020)}]{Yang2020}%
  \BibitemOpen
  \bibfield  {author} {\bibinfo {author} {\bibfnamefont {S.}~\bibnamefont
  {Yang}}, \bibinfo {author} {\bibfnamefont {C.}~\bibnamefont {Wu}}, \ and\
  \bibinfo {author} {\bibfnamefont {N.}~\bibnamefont {Marom}},\ }\bibfield
  {title} {\enquote {\bibinfo {title} {{Topological properties of SnSe/EuS and
  SnTe/CaTe interfaces}},}\ }\href {\doibase
  10.1103/PHYSREVMATERIALS.4.034203/FIGURES/6/MEDIUM} {\bibfield  {journal}
  {\bibinfo  {journal} {Physical Review Materials}\ }\textbf {\bibinfo {volume}
  {4}},\ \bibinfo {pages} {034203} (\bibinfo {year} {2020})}\BibitemShut
  {NoStop}%
\end{thebibliography}%

\end{document}